\documentclass[a4paper,fleqn]{cas-dc}
\usepackage[numbers]{natbib}

\def\tsc#1{\csdef{#1}{\textsc{\lowercase{#1}}\xspace}}
\tsc{WGM}
\tsc{QE}
\tsc{EP}
\tsc{PMS}
\tsc{BEC}
\tsc{DE}

\usepackage {hyperref}
\usepackage{booktabs}
\usepackage{colortbl}
\usepackage{rotating}
\usepackage{multirow}
\usepackage{pdfpages}

\usepackage{lscape}
\usepackage[caption=false,font=scriptsize,labelfont=sf,textfont=sf]{subfig}
\usepackage{graphicx}
\usepackage{placeins}
\begin{document}
\let\WriteBookmarks\relax
\def\floatpagepagefraction{1}
\def\textpagefraction{.001}
\shorttitle{GeMID: Generalizable Models for IoT Device Identification}
\shortauthors{Kostas et~al.}

\title [mode = title]{GeMID: Generalizable Models for IoT Device Identification}                      
\author[1,2]{Kahraman~Kostas}[type=editor,                        auid=001,bioid=1,                        orcid=0000-0002-4696-1857]
\cormark[1]

\author[3]{Rabia~Yasa~Kostas}[type=editor,                        auid=001,bioid=1,                        orcid=0000-0003-4023-1850]

\author[1]{Mike~Just}[type=editor,auid=000,bioid=1,orcid=0000-0002-9669-5067]

\author[1]{Michael~A.~Lones}[type=editor,auid=000,bioid=1,orcid=0000-0002-2745-9896]

\affiliation[1]{organization={Department of Computer Science, Heriot-Watt University},    postcode={EH14 4AS},        city={Edinburgh},country={UK}}

\affiliation[2]{organization={Ministry of		National Education},city={Anakara},                country={Türkiye}}

\affiliation[3]{organization={Gumushane University},city={Gumushane},country={Türkiye}}
\cortext[cor1]{Corresponding author}

\begin{abstract}
With the proliferation of devices on the Internet of Things (IoT), ensuring their security has become paramount. Device identification (DI), which distinguishes IoT devices based on their traffic patterns, plays a crucial role in both differentiating devices and identifying vulnerable ones, closing a serious security gap. 
However, existing approaches to DI that build machine learning models often overlook the challenge of model generalizability across diverse network environments.  In this study, we propose a novel framework to address this limitation and to evaluate the generalizability of DI models across data sets collected within different network environments. Our approach involves a two-step process: first, we develop a feature and model selection method that is more robust to generalization issues by using a genetic algorithm with external feedback and datasets from distinct environments to refine the selections. Second, the resulting DI models are then tested on further independent datasets to robustly assess their generalizability.
We demonstrate the effectiveness of our method by empirically comparing it to alternatives, highlighting how fundamental limitations of commonly employed techniques such as sliding window and flow statistics limit their generalizability. Moreover, we show that statistical methods, widely used in the literature, are unreliable for device identification due to their dependence on network-specific characteristics rather than device-intrinsic properties, challenging the validity of a significant portion of existing research.
Our findings advance research in IoT security and device identification, offering insight into improving model effectiveness and mitigating risks in IoT networks.

\end{abstract}


\maketitle

\section{Introduction}

The Internet of Things (IoT) seamlessly integrates our cyber world with the physical world and is becoming increasingly prevalent in daily life. According to published statistics, the number of IoT devices has exceeded 15 billion and is projected to reach approximately 30 billion by 2030~\cite{statista}.

Despite the rapid proliferation of devices and providers, security remains a critical issue. IoT devices can be more challenging to secure than conventional devices due to their diverse hardware, software, and varied manufacturer profiles~\cite{zarpelao2017survey}. Moreover, unlike conventional devices, many IoT devices have nonstandard interfaces that impair user interaction and make it difficult to mitigate against vulnerabilities~\cite{williams2017privacy}. The NETSCOUT Threat Intelligence Report~\cite{modi2019} indicates that a new IoT device on the network typically faces its first attack within 5 hours and becomes a specific attack target within 24 hours, and that most attacks exploit vulnerabilities in IoT devices~\cite{marzano2018evolution}. Beyond being targets in attacks, compromised IoT devices can serve as tools in botnet attacks, exemplified by Mirai, URSNIF, and BASHLITE~\cite{antonakakis2017understanding}, where captured devices are used to orchestrate high-volume Distributed Denial of Service (DDoS) attacks. 
While various methods have been developed to detect and mitigate ongoing attacks, primarily Intrusion Detection Systems (IDS), proactive measures such as device-specific updates, internet access restrictions or isolation can prevent these attacks before they happen. Given the variety, quantity and unfamiliar interfaces of IoT devices, it is impractical to rely on users to implement these solutions. As a result, Device identification (DI) systems have been developed to automatically identify devices based on their activity and make it possible to apply appropriate security measures to create a secure IoT ecosystem. \textcolor{black}{It is important to clarify that DI in this context aims to identify the device type (i.e., its make and model), and not to distinguish between unique instances of the same device type, which is a separate security challenge.} 

Most existing approaches to DI involve the construction of Machine Learning (ML) models. When developing ML models, it is important to ensure that they generalize beyond the specific environment in which they were trained. In DI, this means that models trained on data from one device should be able to detect other devices of the same make and model operating within different network environments. Achieving this requires accounting for the domain shifts and variations typical of network and IoT environments~\cite{zhou2022domain}. In this regard, a prominent limitation of previous work is that most studies have relied upon a single dataset collected in a single network environment for both developing and testing their models~\cite{meidan2017detection,bezawada2018behavioral,aksoy2019automated}. This has resulted in an incomplete, and potentially misleading, picture of DI generalizability. Where multiple datasets have been used, the focus has been on the generalizability of methodologies and feature sets rather than of model instances~\cite{chowdhury2020network,ortiz2019devicemien,kostas22IoTDevID}.

A critical issue in current DI research is the reliance on flow and window statistics, which capture network environment dynamics rather than device-specific behavior. This study demonstrates that such statistical methods fail to generalize across diverse settings, undermining their reliability. In contrast, features from individual packets offer a robust alternative, as they are less influenced by external variables. This distinction has profound implications, potentially questioning the validity of numerous existing studies.

In this study, we explicitly focus on developing DI model instances that are demonstrably generalizable across network environments. We measure this using multiple datasets containing the same devices operating in different network environments. We use a two-stage process to develop the models. The first stage involves feature and model selection, and focuses on identifying device-independent features that do not display over-dependence on their network environment. \textcolor{black}{Thus, while our work incorporates a sophisticated feature selection method, its purpose is to serve our primary goal: demonstrating a new, more robust methodology for creating and validating generalizable DI models.}

In the second stage, we use an independent dataset to train device-specific model instances. We then use a further dataset, containing the same devices operating in a different network environment, to evaluate the generalizability of these device-specific models.

We make the following contributions to the field of DI:
\begin{enumerate}

\item \textbf{A novel framework for generalizable DI}: \textcolor{black}{We propose and validate a new framework for building and evaluating machine learning-based DI models that are generalizable across diverse network environments. This framework introduces a rigorous two-stage, cross-dataset validation methodology to ensure models are robust to real-world domain shifts.}
	\item \textbf{Insight into feature selection and its impact on generalizability}: We demonstrate how the method of feature selection and construction critically impacts the generalizability of DI models. In particular, we show that features derived from individual packet characteristics lead to superior generalizability compared to those based on flow or window statistics. 
	\item \textbf{Validation of packet-based approaches}: Through empirical comparisons, we validate our packet-based method against other approaches, demonstrating that models built on individual packet features consistently outperform methods that use flow or window-based statistics in terms of generalizability.
	\item \textbf{Commitment to transparency and reproducibility}: To foster transparency and encourage further research, we openly share our code and analysis
  \footnote{To maintain review anonymity, the source code link will be provided after the review process.},
    providing the community with the resources to replicate and build upon our findings.
\end{enumerate}
This paper is organized as follows: Section~\ref{related_work} reviews related work, Section~\ref{mat&met} covers data selection and feature extraction,  Section~\ref{featuresel} details the process used for selecting and evaluating features, Section~\ref{modelev} presents model selection and the results of our model generalizability study, Section~\ref{limit} discusses limitations, and Section~\ref{con} concludes.

\section{Related Work} \label{related_work}

\begin{figure}[tb]
	\centering{
		\includegraphics[width=1\columnwidth]{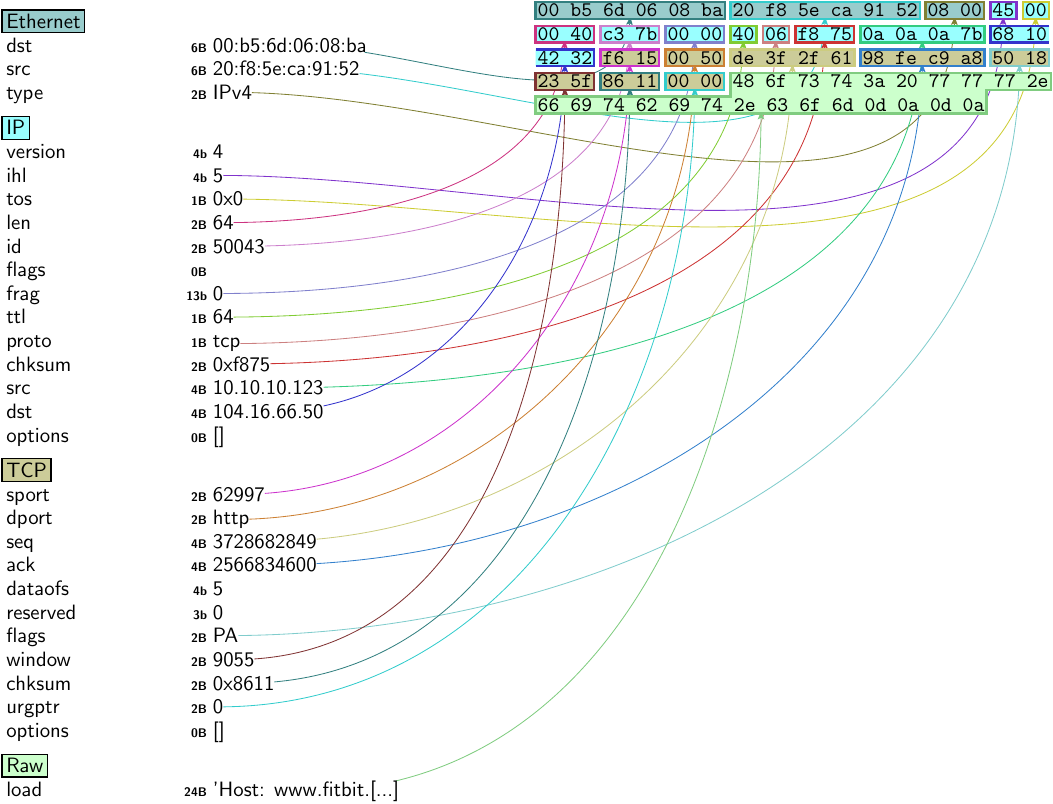}}
	\caption{Raw bytes and headers of a network packet from the Fitbit Aria WiFi enabled weighing device.}
	\label{fig:raw}
\end{figure}

The field of IoT DI has been developing in earnest since 2017. Among the pioneering works is IoTSentinel~\cite{miettinen2017iot}, which builds models using 23 features extracted from packet headers. Packet headers, analogous to envelopes for data traveling over a network, contain information for routing and processing the data. By analyzing these headers, various useful features for network tasks can be extracted. Figure~\ref{fig:raw} illustrates a sample network packet and lists the header features present in it. Extractable features from packet headers include source/destination IP/MAC addresses, protocol, Time-to-Live (TTL), flag information, port numbers, sequence/acknowledgment numbers, and checksum. A significant contribution of this study was the introduction of ~\href{https://research.aalto.fi/en/datasets/iot-devices-captures}{Aalto University IoT device captures}~\cite{aalto2017dataset}, one of the most widely used open-access datasets in the IoT DI domain. Subsequent studies, such as IoTSense~\cite{bezawada2018behavioral}, expanded the feature sets by incorporating payload features (such as payload size and entropy) alongside header-derived features. Concurrently, other research focused on statistics derived from headers, emphasizing the relationship between packets within a specified range (e.g., TTL~\cite{meidan2017detection}, packet size~\cite{kawai2017identification}, time~\cite{kawai2017identification}).

In IoTSentinel and IoTSense, features are derived from individual packet headers, but ML models are trained using features collected from multiple packets, known as fingerprints. These fingerprints consist of 12 packets in IoTSentinel and 20 packets in IoTSense. Despite using multiple packets, these fingerprints simply concatenate individual packet features, rather than generating inter-packet features. 
Alternatively, SysID~\cite{aksoy2019automated} and IoTDevID~\cite{kostas22IoTDevID} create fingerprints based on individual packet features, avoiding the need for merged data features. Obtaining fingerprints from an individual packet addresses the transfer problem that occurs when multiple devices share the same MAC address, a common issue due to the use of MAC addresses in the preassembly of packets~\cite{kostas22IoTDevID}.

Deriving features from flow data instead of packet headers is also common. A flow refers to a sequence of packets sent from a specific source to a specific destination, considered as part of a single session or connection. The size of the flow can be a predetermined number of packets or packets within a specific time/protocol interval. These features can include basic flow information (source, destination, protocol), time-related metrics (start-end time, duration), packet and byte counts (total, average, max, min sizes), rate metrics (packet, byte, flow rate), flag statistics, inter-arrival times (average, max, min), and flow direction indicators (packets/bytes sent between source and destination). Sivanathan et al.~\cite{sivanathan2018classifying}
used features obtained from network flows, such as flow volume, flow duration, and average flow rate. They also introduced the widely used UNSW-DI dataset. Many other DI studies have also used flow features~\cite{kostas23thesis,jmila2022survey}.

While flow-based features, such as flow volume and inter-arrival times, are prevalent~\cite{kostas23thesis,jmila2022survey}, they inherently encode network-specific traits---e.g., latency or bandwidth--- rather than device characteristics. Studies like~\cite{sivanathan2018classifying} and~\cite{msadek2019iot} exemplify this approach, yet their generalizability remains untested across diverse environments. Our analysis shows these methods are fragile, contrasting with packet-based features used in ~\cite{bezawada2018behavioral,miettinen2017iot}, which we validate as more reliable.

Features can also be obtained from packet headers or flows through sliding windows of various sizes. For example, features such as port ranges, packet size, packet quantity, availability time and inter-arrival time can be subjected to statistical analysis within the window to derive features such as maximum, minimum, mean and standard deviation~\cite{msadek2019iot}. This inherently involves features based on multiple packets. However, an issue with features derived from flow/network statistics and windowing methods is that they are likely to capture implicit characteristics of the network environment, rather than just the characteristics of individual devices. In this paper, we argue that this makes them fragile, and demonstrate that they are not a reliable basis for carrying out DI.


\textcolor{black}{Another approach moves away from hand-crafted features by using deep learning to automate the feature engineering process. A common technique is to use the raw bytes of a network packet directly, as illustrated in Figure~\ref{fig:raw}. In this method, the first n bytes are fed into a model—often a Convolutional Neural Network (CNN)—where each byte is treated as an image pixel~\cite{kotak2021iot,hoang2022data,hu2021robust,liu2022autonomous,yao2022cnn}. This data is truncated or padded to a fixed length n. This concept extends beyond raw packets to other data representations. For instance, DeviceMien~\cite{ortiz2019devicemien} uses stacked LSTM-Autoencoders on TCP-flow data, while Lopez-Martin et al.~\cite{lopez2017network} employ a hybrid CNN-RNN model to treat traffic flows as pseudo-images. Others focus on temporal patterns, such as using a CNN to learn long-term relationships from short-term packet arrival sequences~\cite{ma2020pinpointing}.}

However, using raw data for DI presents other problems that have not yet been adequately addressed. In studies where the entire raw packet data was used \cite{hoang2022data,liu2022autonomous}, this included headers, payloads, and metadata, making it difficult to filter out identifying information such as MAC/IP addresses or string identifiers, both of which can significantly impair model generalizability. 
Whilst this dependency can be reduced by solely using packet payloads~\cite{hu2021robust,kotak2022iot}, caution is needed because much of today's internet traffic is encrypted. In practice encrypted payloads are likely to be ineffective for device identification models as the data appears random and opaque, providing no useful information about the contents. Thus, the added complexity of raw data is unlikely to contribute to the model's performance and may even hinder it.

\textcolor{black}{
Beyond the methods that rely on flow statistics or concatenated packet headers, recent research has explored several other distinct avenues for IoT device identification.} 

\textcolor{black}{One prominent approach focuses on the analysis of periodic background traffic, which is often characteristic of IoT device behavior. AUDI~\cite{marchal2019audi} and BehavIoT~\cite{hu2023behaviot} exemplify this method. AUDI uses unsupervised learning, including Fourier transforms and signal autocorrelation, to model the periodic communication patterns of devices for identification. BehavIoT extends this by modeling not just periodic events but also user-initiated and aperiodic events to create comprehensive device and system behavior models from network traffic. These techniques capitalize on the insight that many IoT devices maintain constant, regular communication with backend servers, providing a stable basis for fingerprinting.}

\textcolor{black}{Another technique involves identifying fine-grained, packet-level signatures. The PINGPONG~\cite{trimananda2019pingpong} system demonstrates that short, unique sequences of packet lengths and directions can serve as reliable fingerprints for specific device events (e.g., a light turning on). These signatures often appear as request-reply pairs (``ping-pong'') between a device and a cloud server and can be extracted automatically from traffic without deep statistical analysis.}

\textcolor{black}{Other methods have leveraged metadata from different network protocols. For instance, IoTFinder~\cite{perdisci2020iotfinder} uses passive DNS traffic, modeling a device's identity based on the set of domain names it queries and the frequency of those queries. This DNS-based fingerprinting approach is designed to be scalable and effective even when devices are hidden behind Network Address Translation (NAT).}


\textcolor{black}{While these diverse techniques have proven effective within their respective evaluation contexts, a critical challenge that many early studies did not fully address is the generalizability of the resulting models. Work by Kolcun et al.~\cite{kolcun2021revisiting} highlighted this issue by demonstrating that the accuracy of various ML models for device identification degrades significantly over time when tested on data collected weeks or months after the training period, suggesting that static models are insufficient. Complementing this, Ahmed et al.~\cite{ahmed2022analyzing} conducted a comprehensive analysis across multiple datasets, showing that temporal, spatial, and data-collection-methodology differences all impact fingerprinting accuracy, and that features robust in one context may not be in another. These findings underscore the fundamental problem of model generalization, which is the primary focus of our work.}

A related issue with existing DI studies is the lack of cross-dataset validation, which involves evaluating a model trained on one dataset against a different dataset to assess its generalizability and robustness.
While some studies use multiple datasets~\cite{chowdhury2020network,ortiz2019devicemien,kostas22IoTDevID}, they typically apply their methods separately to each, training and testing within the same dataset (e.g., training on a portion of Dataset A and testing on the remainder).
This approach raises potential concerns about their performance in unseen environments. It also exacerbates the specific issues raised above surrounding the use of statistical features, since models using these features would---in effect---be tested within the same network environment as the one they were trained in. In this study, we demonstrate the importance of cross-dataset validation in developing generalizable models.

\textcolor{black}{Many of the studies cited in this section, such as \cite{bezawada2018behavioral,aksoy2019automated,chowdhury2020network,ortiz2019devicemien,kostas22IoTDevID,miettinen2017iot,kawai2017identification,sivanathan2018classifying,kostas23thesis,jmila2022survey,msadek2019iot,kotak2021iot,hoang2022data,hu2021robust,liu2022autonomous,yao2022cnn,kotak2022iot,ma2020pinpointing, trimananda2019pingpong, kolcun2021revisiting, hu2023behaviot, marchal2019audi, perdisci2020iotfinder, lopez2017network,  ortiz2019devicemien, ahmed2022analyzing} validate their approaches using training and testing sets drawn from the same dataset or environment. Our work diverges by focusing on the stricter challenge of model instance generalizability, where a specific model trained on data from one environment must perform accurately on data from a completely different environment. This evaluation distinction makes a direct F1-score comparison with these works an `apples-to-oranges' comparison and is the reason our study implements its own baselines to ensure a fair and controlled experiment.}

\section{Materials And Methods}\label{mat&met}

\subsection{Data Selection}

DI research typically relies on datasets comprising data collected from real devices. These datasets commonly feature over 20 devices and necessitate prolonged observation periods to gather sufficient data. However, the data collection process for IoT devices entails substantial resource allocation, including personnel and space, making it burdensome. Consequently, many researchers opt to utilize publicly available DI datasets, thereby enhancing reproducibility and transparency and facilitating comparative analysis.

In our study, we use
data from two dataset families, UNSW and MonIoTr. Each family consists of multiple datasets that are useful for our model building and evaluation. We use the UNSW datasets for feature and model selection, and the MonIoTr datasets to test the robustness and generalizability of the resulting models.
Additionally, we prioritized packet header features over flow or window statistics, as the latter are prone to network variability, reducing their utility for DI.

\begin{figure}[tb!]
	\centering{
		\includegraphics[width=1\columnwidth]{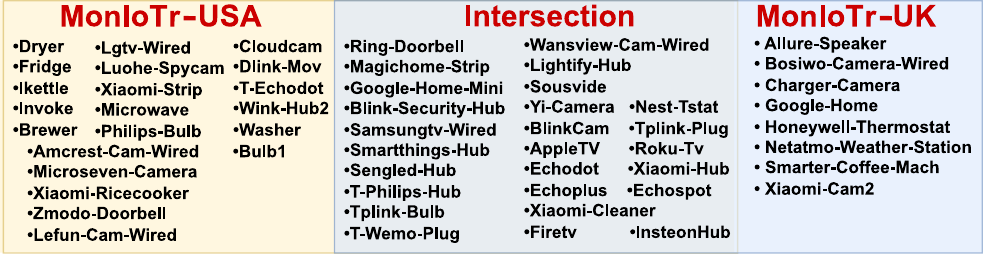}}
	\caption{Devices represented in the MonIoTr dataset from both UK (blue) and USA (yellow) sites, along with their intersection (grey). Adapted from~\cite{ren-imc19}.}
	\label{fig:mon-venn}
\end{figure}

\textcolor{black}{It is important to note that our framework uses these two dataset families for distinct stages (selection and evaluation) because they do not share a common set of device classes, as shown in Figures~\ref{fig:unsw-venn} and~\ref{fig:mon-venn}. Consequently, direct cross-testing, such as training a model on UNSW devices and testing on MonIoTr, is not possible as the classification labels are mutually exclusive.} 

\subsubsection{UNSW} \label{unsw_data}

For feature and model selection, we utilized the \href{https://iotanalytics.unsw.edu.au/attack-data.html}{IoT Attack Traces---ACM SOSR 2019} (UNSW-AD)~\cite{hamza2019detecting} and \href{https://iotanalytics.unsw.edu.au/iottraces.html}{IoT Traffic Traces---IEEE TMC 2018}  (UNSW-DI)~\cite{sivanathan2018classifying} datasets. The UNSW-DI dataset is tailored for DI tasks, comprising data from 24 benign devices collected over a 60-day period in 2016. 
In contrast, the UNSW-AD dataset, originally designed for anomaly detection (AD), contains 27 days of benign data and 17 days of both benign and malicious data from 28 devices collected over a 44-day period in 2018. For the UNSW-AD dataset, we used only the benign data.

\begin{figure}[tb!]
	\centering{
		\includegraphics[width=1\columnwidth]{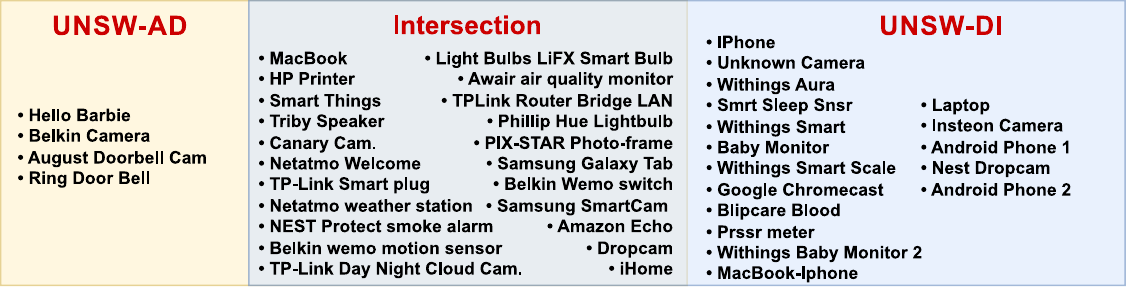}}
	\caption{Devices represented in the UNSW-DI (blue)  and  UNSW-AD (yellow) datasets, along with their intersection (grey).}
	\label{fig:unsw-venn}
\end{figure}

Notably, despite both datasets containing most devices in common (see Figure~\ref{fig:unsw-venn}), they were collected at different times, in distinct site environments with varying network characteristics, and for disparate purposes, likely by different users. 
We leverage these datasets for feature and model selection, employing UNSW-DI as training data and UNSW-AD as test data, and vice versa in different runs/iterations, to enhance model robustness and generalizability.

Each dataset contains a number of sessions with one session per day (UNSW-DI: 60 sessions, UNSW-AD: 27 sessions). The resultant datasets are very large, and some sessions lack certain devices. To address this we selected and merged some sessions to maximize the number of devices available, and then used these merged sessions. Specifically, in the DI dataset, sessions 16-10-03 and 16-11-22 were combined to form S1, while sessions 16-09-29 and 16-11-18 were combined to form S2 (names are in yy-mm-dd date format). In the AD dataset, sessions 18-10-13 and 18-06-14 were combined to form S1, and sessions 18-10-16 and 18-06-11 were combined to form S2.

\subsubsection{MonIoTr}

After selecting the features and models, to remove any bias, we evaluate their effectiveness on a different dataset, \href{https://moniotrlab.khoury.northeastern.edu/publications/imc19/}{IoT Information Exposure---IMC’19} (MonIoTr)~\cite{ren-imc19}. The MonIoTr dataset consists of 81 devices. An important characteristic of this dataset is that data collection took place at two separate sites, one in the UK (MonIoTr-UK) and the other in the USA (MonIoTr-USA). Of the 81 devices, 26 devices were present in both sites, with the USA site having 47 devices (including 21 unique devices) and the UK site having 34 devices (including 8 unique devices). Figure~\ref{fig:mon-venn} shows the distribution of devices by country.

Data collection in these sites spanned 112 hours, covering various scenarios:

\begin{description}
	\item[Idle] Data collected during an 8 hour period at night when devices were inactive.
	\item[Power] Data collected for 2 minutes immediately after device power-on, without interaction.
	\item[Interaction] Data collected through various interactions, including physical button presses, voice commands, phone app usage on the same network, communication with the device through a phone on a different network utilizing cloud infrastructure, and interaction with the device via Alexa Echo Spot.
\end{description}

\textcolor{black}{To facilitate our cross-environment evaluation, we define four distinct data partitions from this dataset based on the physical location of the device and its network routing. This was made possible as the original data collection included experiments where traffic was routed between sites using a VPN. The four partitions are:}

\begin{description}
    \item \textbf{MonIoTr-UK}: Data from devices physically located and operating on the UK network.
    \item \textbf{MonIoTr-USA}: Data from devices physically located and operating on the USA network.
    \item \textbf{MonIoTr-UK-VPN}: Data from devices physically located in the UK, but with their traffic routed to the USA site via a VPN before accessing the internet.
    \item \textbf{MonIoTr-USA-VPN}: Data from devices physically located in the USA, but with their traffic routed to the UK site via a VPN.
\end{description}


Many devices interact heavily with the local network, applications, tools, and cloud services during startup, so we merged power and interaction data into one \textit{active} class. In a previous study ~\cite{kostas2023externally}, we observed that when both active and idle data are available, active data is much more effective for DI. Consequently, we excluded idle data from this study.

\subsection{Feature Extraction} \label{FE}

For feature extraction, we used  \href{https://tshark.dev/}{Tshark}, a network protocol analyzer,  integrated with Python. We focused on features contained in the protocol headers of DNS, HTTP, ICMP, STUN, TCP, UDP, DHCP, EAPOL, IGMP, IP, NTP, and TLS. We eliminated features that consisted of string expressions or contained MAC and IP addresses. After this preliminary elimination, over 300 features remained from 4600 features.

\section{Feature Selection} \label{featuresel}

During the feature selection phase, we aimed to identify the most effective features for DI. For this, we used the four UNSW \textit{data partitions} described in Section \ref{unsw_data} to guide feature selection. These comprise two sessions (S1 and S2) from each of UNSW-AD and UNSW-DI (referred to hereafter according to their dataset-session number: \textit{AD-S1}, \textit{AD-S2}, \textit{DI-S1}, and \textit{DI-S2}).

The generalizability of features is assessed in the following three ways, in order of increasing strictness:
\begin{description}
	\item[5-fold cross-validation (CV)] within a partition, which is common in the literature, e.g., within \textit{AD-S1}.
	\item[Session versus session (SS)] where generalizability is measured between two partitions of the same dataset, e.g., \textit{AD-S1} vs \textit{AD-S2}.
	\item[Dataset versus dataset (DD)] where generalizability is measured between partitions of different datasets, e.g., \textit{AD-S1} vs \textit{DI-S2}.
\end{description}

\subsection{Predictive Features}

\begin{figure}[t]
	\centering{
		\includegraphics[width=.75\columnwidth]{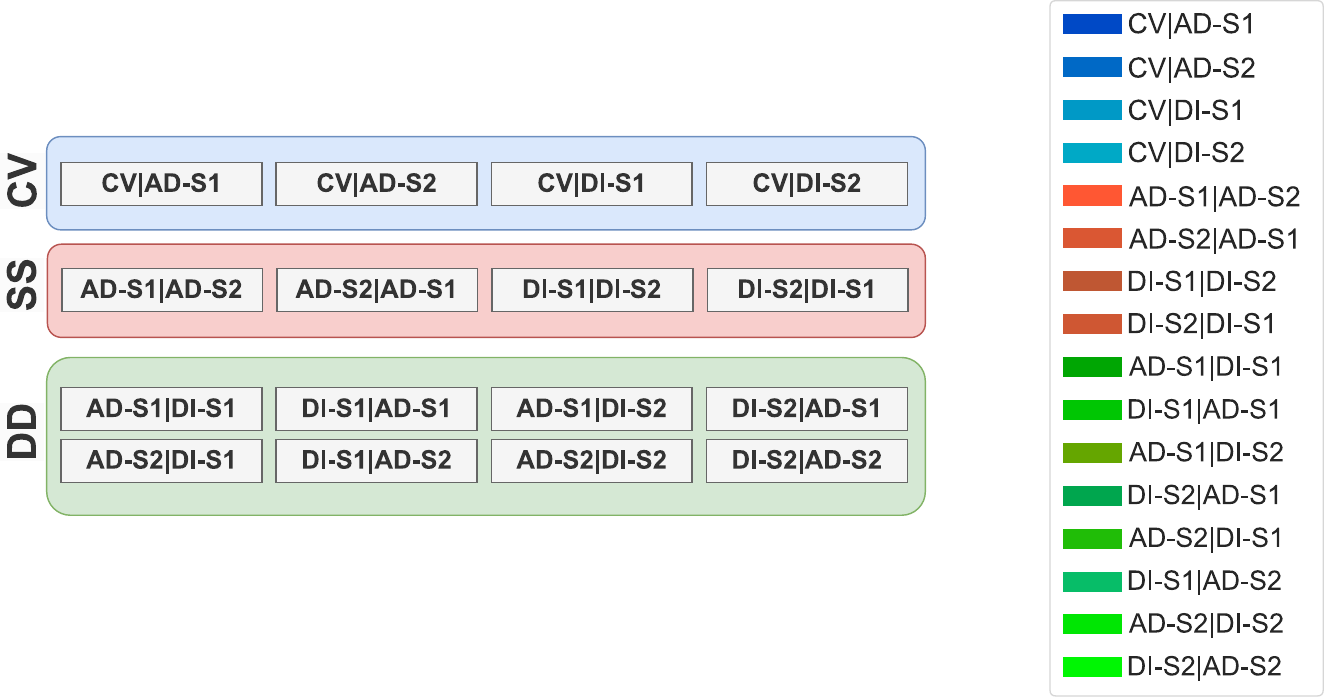}} 
	\caption{(Left) Data usage from UNSW datasets across different evaluation contexts. CV, SS, and DD steps are visualized in shades of blue, red, and green, respectively, throughout the paper for clarity. In the nomenclature, \textit{CV\textbar AD-S1} denotes cross-validation on the AD dataset session 1. For cases other than CV, the first dataset is used for training and the second for testing. For example, \textit{AD-S1\textbar DI-S2} means the first session of the UNSW-AD dataset is used for training and the second session of the UNSW-DI dataset is used for testing. (Right) The legend displayed is shared across Figures~\ref{fig:cizgi}-\ref{fig:vote8} for consistency and clarity.}
	\label{fig:sessions}
\end{figure}

\begin{figure}[t]
	\centering{
		\includegraphics[width=1\columnwidth]{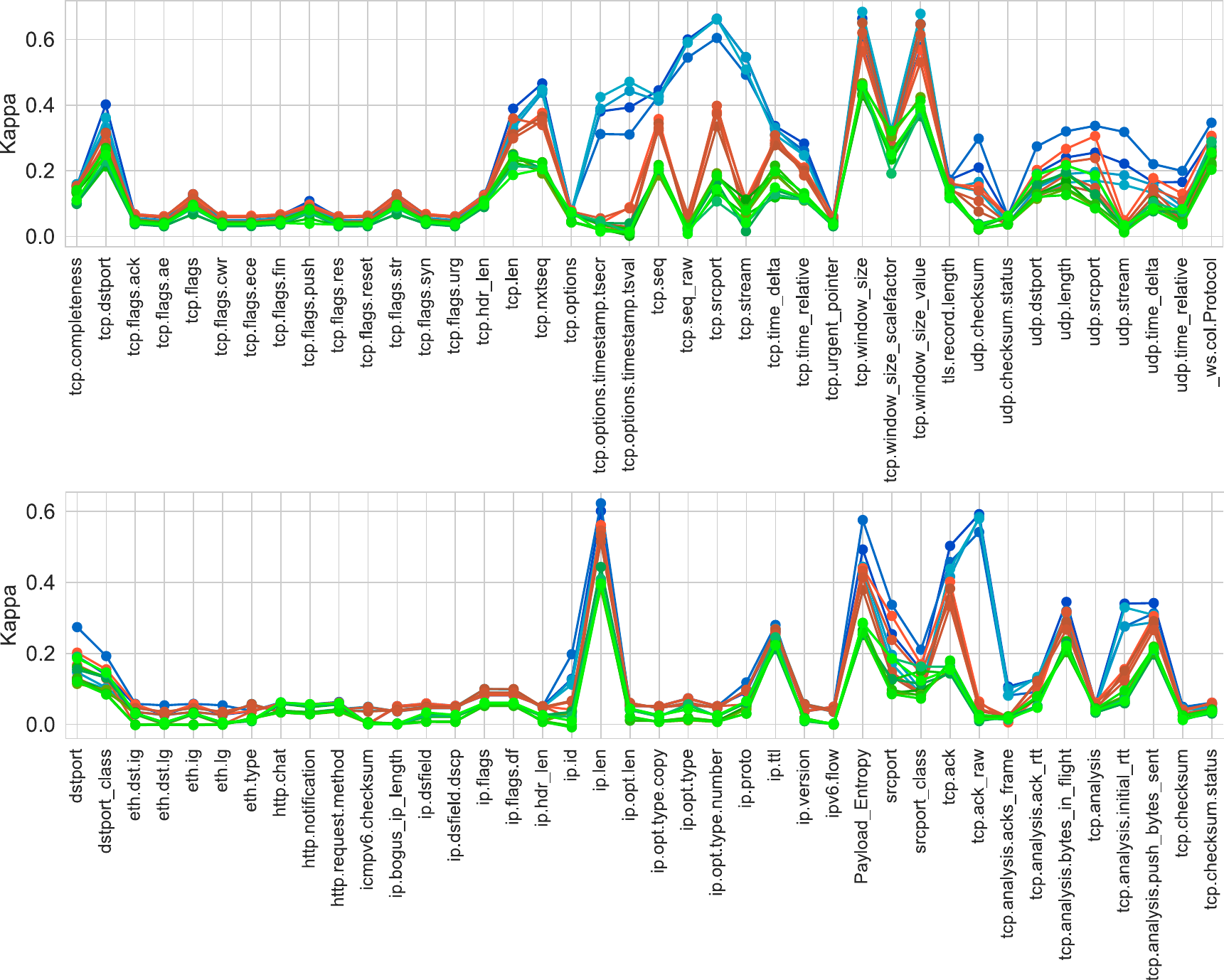}}
	\caption{Comparison of feature utility in UNSW datasets measured using CV (blue) and isolated methods, SS (red) and DD (green). CV tends to overestimate feature utility, with higher scores for many attributes. SS and DD produce more realistic evaluations.  The discrepancy highlights the potential for information leakage in cross-validation and the importance of using isolated validation methods for assessing feature utility in ML-based DI models. For the legend, please refer to Figure~\ref{fig:sessions}.}
	\label{fig:cizgi}
\end{figure}

\begin{figure}[h!]
	\centering{
		
		\includegraphics[width=1\columnwidth]{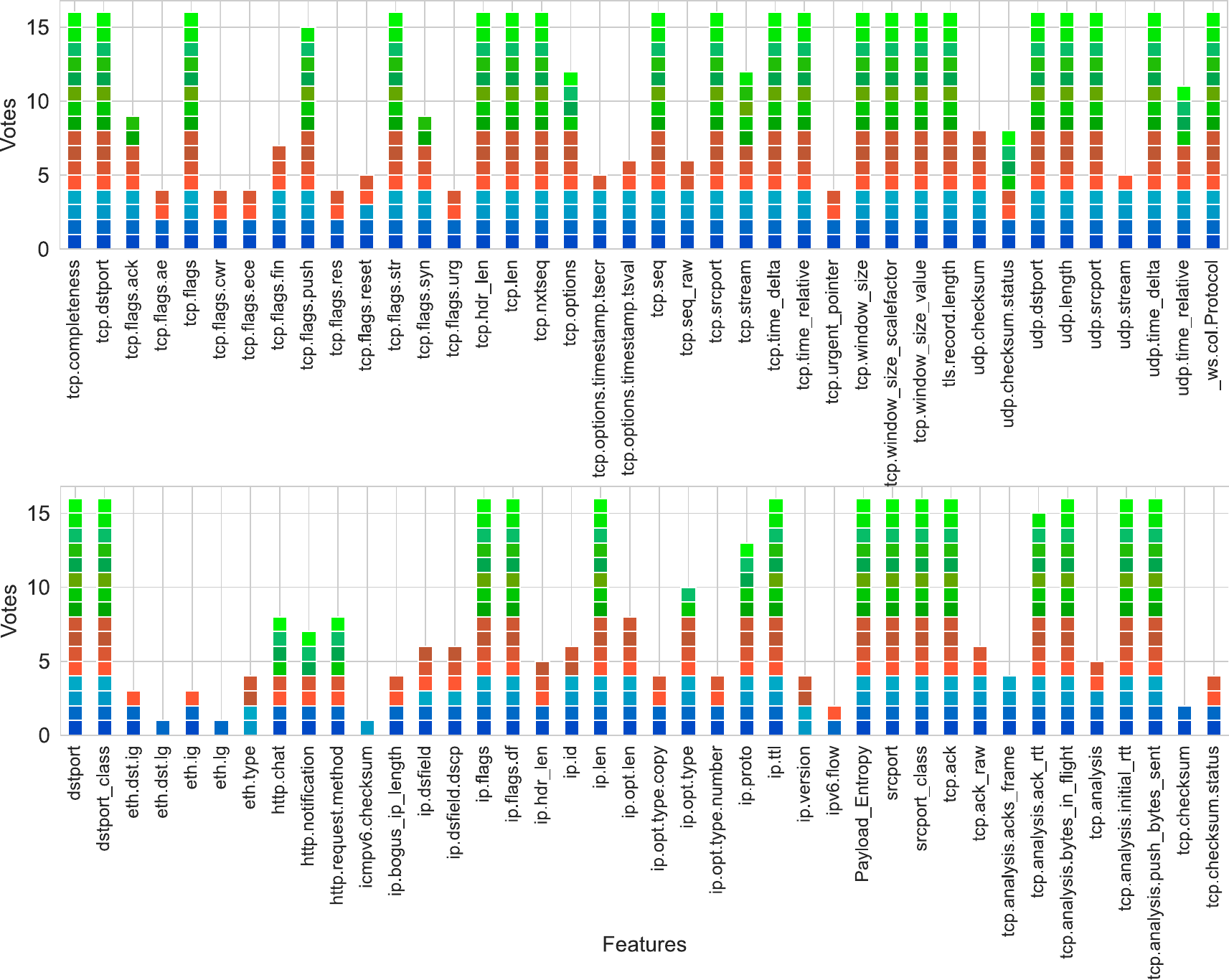}}
	\caption{Features identified as being predictive within one or more evaluation contexts (CV: blue, SS: red, DD: green), based on non-zero kappa scores of DT models. Note that the CV votes are not used to select features, but are shown here for completeness. For the legend, please refer to Figure~\ref{fig:sessions}.}
	\label{fig:vote86}
\end{figure}

First, each feature is evaluated individually using each relevant combination of partitions, leading to 16 distinct evaluation contexts (see Figure~\ref{fig:sessions}).~\textcolor{black}{This process began with an initial pool of 332 features extracted as described in Section~\ref{FE}. Each of these 332 features was evaluated individually. For clarity, Figures~\ref{fig:cizgi} and~\ref{fig:vote86} visualize only the subset of features (approximately 80) that demonstrated a non-zero kappa score in at least one of the 16 evaluation contexts. The remaining ~250 features showed no predictive ability in any context and were thus eliminated from further consideration. From this reduced set of potentially useful features, we then applied our stricter voting mechanism to select the 46 most promising candidates for the GA-based interaction analysis.} 
Generalizability is assessed by training a decision tree (DT) model using the feature, with DT selected for speed and explainability. \textcolor{black}{This selection step, particularly when embedded within a GA, requires thousands of individual model training runs. Using a computationally inexpensive model like a DT was therefore essential to make the search process feasible. A more complex model (such as a random forest) would have been prohibitively time-consuming in this phase.}  
\textcolor{black}{The utility of each feature is measured using the \textit{kappa}  metric. The kappa statistic is a measure of inter-rater reliability that evaluates the agreement between a classifier and the ground truth, correcting for the probability that agreement could occur by chance.} 

Figure~\ref{fig:cizgi} shows the resulting utility values. Notably, the scores measured under CV are much higher for some features than those measured under DD and SS scores. This indicates that information leaks may be occurring due to train and test folds being taken from the same partition, and suggests that CV is not a reliable basis for selecting features. Hence we only use DD and SS from now on.

A voting system was used for feature selection, with each non-zero kappa value (with a tolerance of $\pm5\%$, so $\ge0.05$), indicating the feature had a positive contribution, resulting in a vote. Figure~\ref{fig:vote86} illustrates this system. Features receiving four or more votes from either DD (green) or SS (red) categories, with at least one DD vote, advanced to the next stage. Features not meeting this criterion were eliminated. The requirement for at least one DD vote is based on its greater strictness in assessing generalizability.
As a result of voting, we identified 46 of the 332 features as being individually predictive. 

\begin{figure}[h!]
	\centering{
		\includegraphics[width=1\columnwidth]{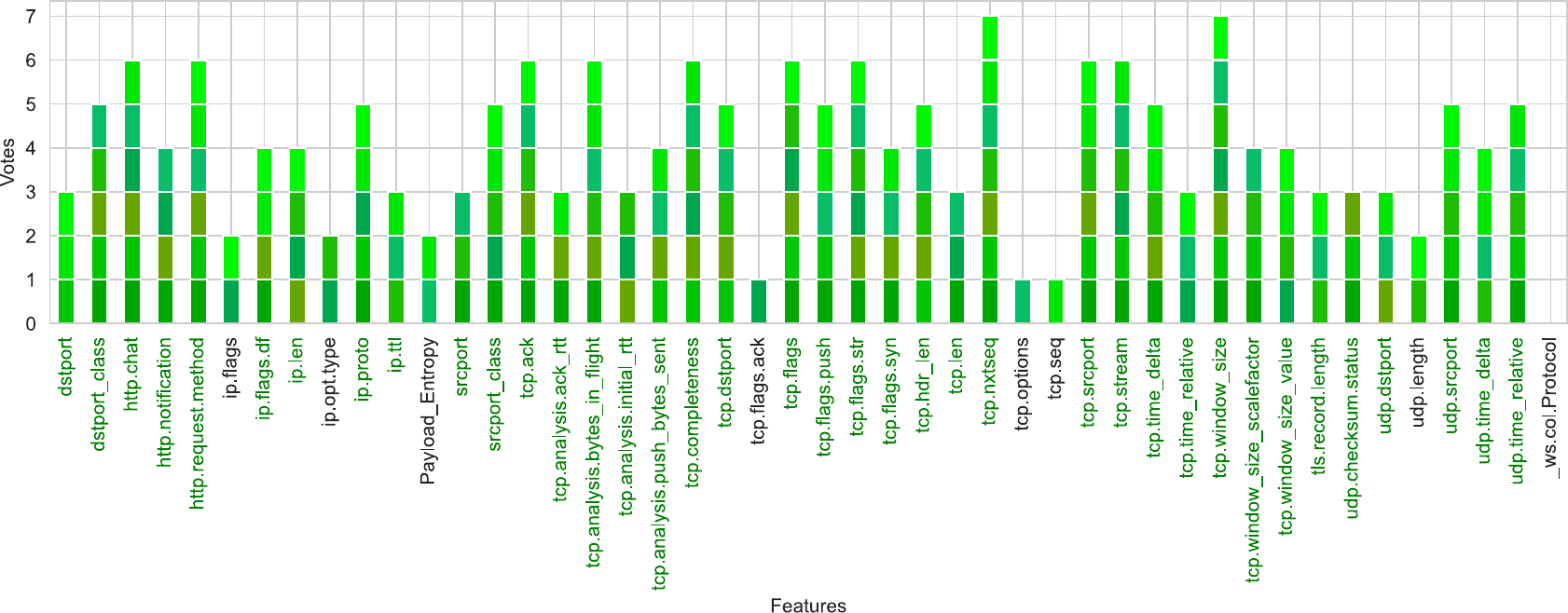}}
	\caption{List of intersecting features identified across eight dataset versus dataset (DD) cases using Genetic Algorithm (GA). Feature names highlighted in dark green were selected for model building after further post hoc analysis. For the legend, please refer to Figure~\ref{fig:sessions}.}
	\label{fig:vote8}
\end{figure}

\subsection{Feature Interactions}

\textcolor{black}{We then considered feature interactions, using a wrapper method to find the optimal combination of these individually predictive features. We chose a Genetic Algorithm (GA) for this purpose, as its global search characteristic is effective at navigating large feature spaces to find robust solutions. A detailed description of the GA implementation, including the specific algorithm, hyperparameters, and evaluation criteria, is provided in Appendix Section~\ref{ga-exp}, Table~\ref{tab:ga_implementation}.} 

 However, in general there is no guarantee of a GA finding an optimal subset, meaning it may include uninformative features in the feature set. Additionally, GA-selected features may develop a bias towards the data used for selection, potentially reducing their performance on other datasets.

To address this, all 46 features that passed individual voting were combined into a single feature set for selection using a genetic algorithm (GA). A decision tree (DT) model was employed for evaluation, with the fitness of each generation assessed using the F1 score across 8 additional DD datasets (see Figure~\ref{fig:sessions}). This external validation provides  feedback to the GA, ensuring that feature selection is based on success across different datasets and not dependent on only one. For each of the eight DD cases, the GA produced slightly different feature sets, so we then identified the intersections between them. Figure~\ref{fig:vote8} depicts this, showing that some features (e.g., tcp.window\_size, tcp.ack) were chosen multiple times, some (e.g., tcp.seq, tcp.flags.ack) only once.

\subsection{Deriving a Final Feature Set}

\begin{figure}[h!]
	\centering{
		\includegraphics[width=1\columnwidth]{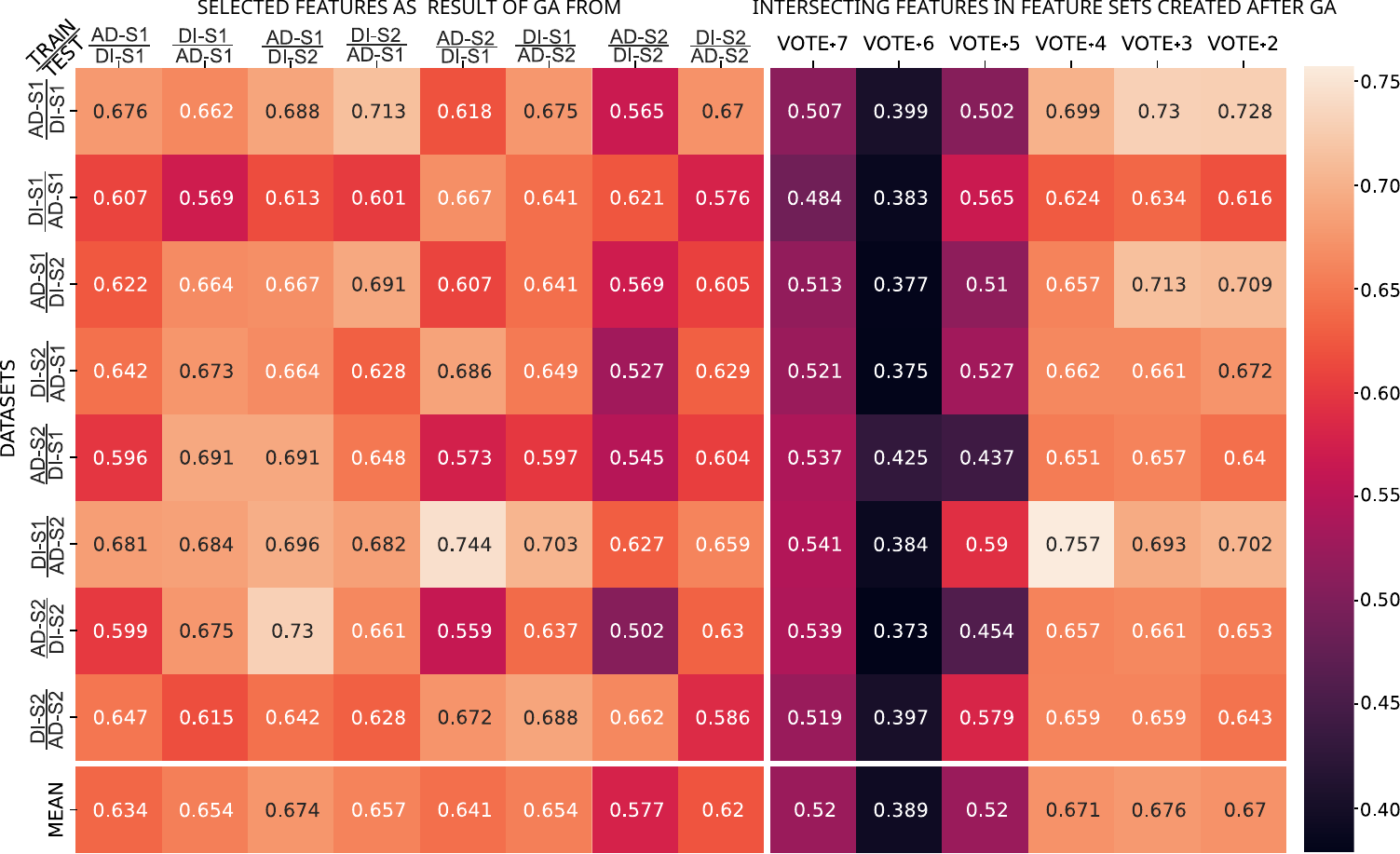}}
	\caption{Comparison of feature set performance across dataset versus dataset (DD) cases. The left side of the heatmap displays the results of applying feature sets obtained from the first step of the Genetic Algorithm (GA) to all DD data, yielding 64 results. On the right, the performance of the features grouped according to their frequency, focusing on the intersection of features obtained from the output of the GA algorithm.}
	\label{fig:cm}
\end{figure}

To identify the most effective combination of features from the GA results, we conducted two post hoc analyses. First, the feature set derived from each DD case (each GA run) is reevaluated on each other DD case. F1 scores of the resulting DT models are presented in the left part of Figure~\ref{fig:cm}. These show how well the feature sets from each GA run generalize.

The second analysis uses a voting mechanism based on intersections of the eight GA feature sets (Figure~\ref{fig:vote8}), with a feature receiving votes proportional to how often it appears. Features are then grouped based on thresholds, i.e., whether they appear in 2 or more feature sets (Vote+2), 3 or more (Vote+3) etc., and DT models are trained using these feature groups. The results are shown on the right side of Figure~\ref{fig:cm}.

From the mean F1 scores shown in the final row of Figure~\ref{fig:cm}, it can be seen that feature sets based on grouping generally lead to better performance, at least for voting thresholds up to 4. For larger thresholds, the feature sets become very small, explaining the lower scores. Generally, the feature sets from single runs lead to models with lower F1 scores, suggesting that voting adds more robustness. Consequently, we use the Vote+3 feature set---which has the highest F1 score---in the next section. The selected features are highlighted in Figure~\ref{fig:vote8}.

\section{Model Evaluation}\label{modelev}

Next, we construct and evaluate ML models using the selected features and compare their performance against established baselines using an independent dataset.

\subsection{Model Selection}

We experimented with various ML modeling approaches commonly used in DI to identify the most effective one for this purpose, in terms of both predictive success and inference time---the latter being an important consideration when scanning network packets. The models we considered were Logistic Regression (LR), Decision Trees (DT), Naive Bayes (NB), Support Vector Machines (SVM), Random Forest (RF), Extreme Gradient Boosting (XGB) Multi-Layer Perceptron (MLP), K-Nearest Neighbors (KNN), Convolutional Neural Networks (CNN), Long Short-Term Memory (LSTM), and Bidirectional Encoder Representations from Transformers (BERT).

\begin{figure}[h!]
	\centering{
		\includegraphics[width=.95\columnwidth]{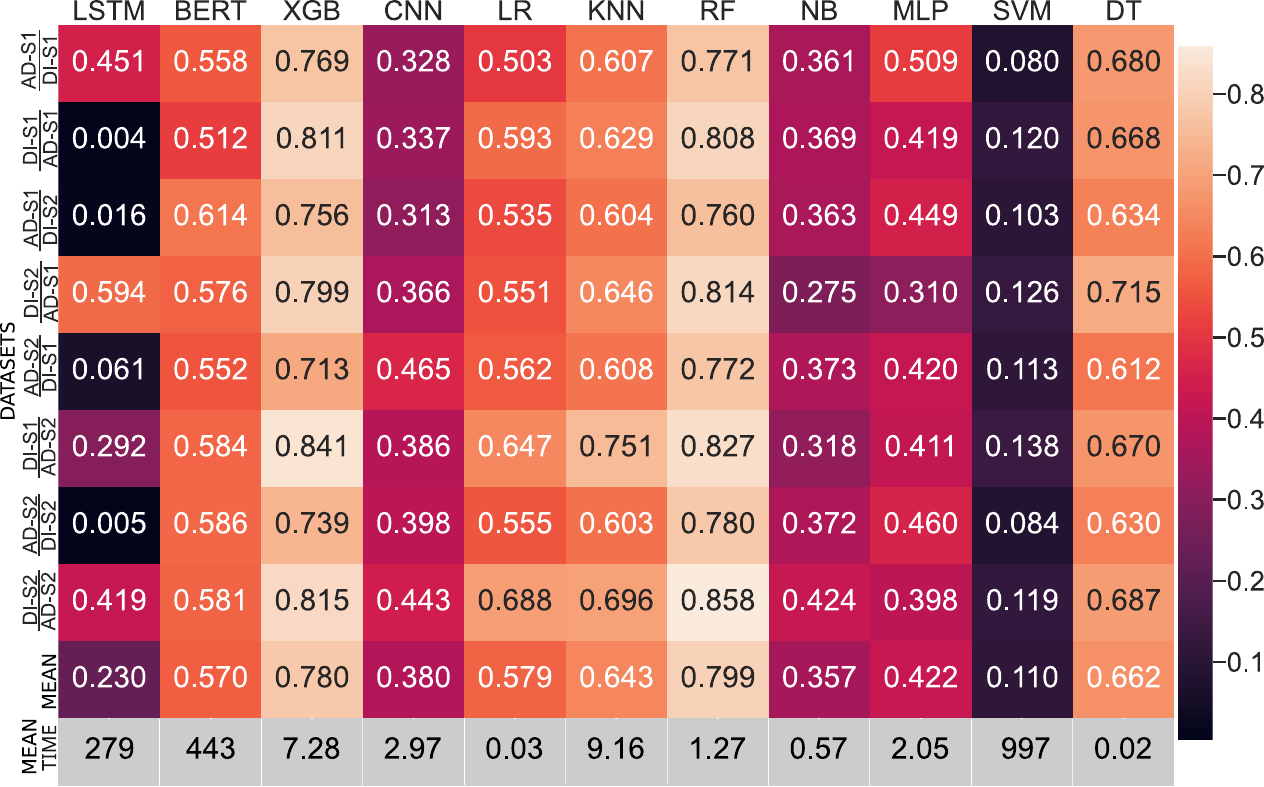}}
	\caption{F1 scores and average inference times of various ML algorithms applied to DD datasets. RF and XGB demonstrate the highest F1 scores of 0.799 and 0.780, respectively, while DT show the fastest inference time.}
	\label{fig:cmML}
\end{figure}

We used random search for hyperparameter optimization and applied each model to the DD datasets---see Appendix Section~\ref{Hyperparameter} for details. The F1 scores and average inference times are shown in Figure~\ref{fig:cmML}. From this, it is clear that the most successful models are RF and XGB, with F1 scores of 0.799 and 0.780, respectively. While their predictive performance is similar, RF runs about 6 times faster than XGB, so we chose this as our preferred model. Notably, DT has the fastest inference time, but its predictive performance lags behind RF by about 10 percentage points. 

\textcolor{black}{This performance difference is characteristic of the trade-off between single learners and ensemble models. A single DT is computationally simple and therefore very fast, but it can be prone to overfitting the training data. RF mitigates this by building a multitude of decision trees on different subsets of the data and features, and then averaging their predictions. This ensemble approach reduces the model's overall variance and improves its ability to generalize to unseen data, resulting in a higher F1 score at the cost of increased computational overhead for inference.}

\subsection{Evaluating Model Generalizability}

In this section, we measure the generalizability of our method (GeMID) and other methods using the MonIoTr dataset, using the USA, USA-VPN, UK, and UK-VPN partitions of the MonIoTr dataset within three different evaluation contexts. In CV, we carry out cross-validation within a single partition. In SS, we train models using non-VPN data and test them using VPN data, and vice versa. In DD, we use data collected from one geographical site to train models, and data collected from the other geographical site to test them.


\textcolor{black}{To test our hypothesis about the fragility of statistical features, we needed to compare our packet-based approach against strong, representative baselines for flow-based and window-based methods. Rather than attempting to replicate individual studies, which often rely on different evaluation criteria and settings, we chose a more fundamental approach. For this purpose, we selected CICFlowmeter~\cite{flowmeter} and Kitsune~\cite{mirsky2018kitsune} not as published methods to compare against, but as well-established tools for extracting broad sets of flow- and window-based features, respectively (see Table~\ref{tab:results}). These tools generate feature sets that cover the majority of statistical features commonly used across the DI literature, making them suitable for constructing baseline models that fairly represent these entire classes of methods. We then applied our own rigorous feature selection and evaluation pipeline to these feature sets to ensure a controlled, fair, and direct comparison of the feature types themselves. We also compare against IoTDevID~\cite{kostas22IoTDevID}, another packet header-based DI method that aimed to build generalizable DI models and demonstrated improved performance over alternative methods (see Appendix Sections~\ref{CICFlowmeter} and~\ref{Kitsune} for results).}

\begin{table}[tb]
	\centering
	\caption{Comparison of DI methods on MonIoTr dataset,  in F1 scores. \textcolor{black}{The final section of the table quantifies the average generalization gap ($\Delta F1$) between the different evaluation contexts (Cross-Validation, Session-vs-Session, and Dataset-vs-Dataset), where a smaller value indicates better generalizability.}} 
	\resizebox{\columnwidth}{!}{ \begin{tabular}{clcccc}
			\toprule
			\multicolumn{1}{l}{\textbf{}} & \textbf{Dataset} & \multicolumn{1}{l}{\textbf{GeMID}} & \multicolumn{1}{l}{\textbf{CICFlwM}} & \multicolumn{1}{l}{\textbf{IoTDevID}} & \multicolumn{1}{l}{\textbf{Kitsune}} \\
			\midrule
			\multirow{5}[2]{*}{\begin{sideways}\textbf{CV}\end{sideways}} & UK    & 0.960 & 0.857 & 0.900 & 0.931 \\
			& UKVPN & 0.965 & 0.853 & 0.908 & 0.924 \\
			& USA    & 0.965 & 0.904 & 0.910 & 0.903 \\
			& USAVPN & 0.954 & 0.904 & 0.923 & 0.944 \\
			& \cellcolor[rgb]{ .851,  .851,  .851}Mean & \cellcolor[rgb]{ .851,  .851,  .851}\textbf{0.961} & \cellcolor[rgb]{ .851,  .851,  .851}0.879 & \cellcolor[rgb]{ .851,  .851,  .851}0.910 & \cellcolor[rgb]{ .851,  .851,  .851}0.925 \\
			\midrule
			\multirow{5}[2]{*}{\begin{sideways}\textbf{SS}\end{sideways}} & UK\textbar UKVPN & 0.895 & 0.644 & 0.824 & 0.741 \\
			& UKVPN\textbar UK & 0.877 & 0.555 & 0.802 & 0.710 \\
			& USA\textbar USAVPN & 0.861 & 0.606 & 0.821 & 0.712 \\
			& USAVPN\textbar USA & 0.894 & 0.616 & 0.839 & 0.687 \\
			& \cellcolor[rgb]{ .851,  .851,  .851}Mean & \cellcolor[rgb]{ .851,  .851,  .851}\textbf{0.882} & \cellcolor[rgb]{ .851,  .851,  .851}0.605 & \cellcolor[rgb]{ .851,  .851,  .851}0.822 & \cellcolor[rgb]{ .851,  .851,  .851}0.712 \\
			\midrule
			\multirow{9}[2]{*}{\begin{sideways}\textbf{DD}\end{sideways}} & UK\textbar USA & 0.760 & 0.475 & 0.703 & 0.464 \\
			& UK\textbar USAVPN & 0.769 & 0.490 & 0.666 & 0.435 \\
			& UKVPN\textbar USA & 0.748 & 0.516 & 0.654 & 0.392 \\
			& UKVPN\textbar USAVPN & 0.746 & 0.454 & 0.709 & 0.444 \\
			& USA\textbar UK & 0.778 & 0.454 & 0.752 & 0.541 \\
			& USA\textbar UKVPN & 0.780 & 0.517 & 0.694 & 0.455 \\
			& USAVPN\textbar UK & 0.811 & 0.515 & 0.710 & 0.439 \\
			& USAVPN\textbar UKVPN & 0.816 & 0.464 & 0.731 & 0.521 \\
			& \cellcolor[rgb]{ .851,  .851,  .851}Mean & \cellcolor[rgb]{ .851,  .851,  .851}\textbf{0.776} & \cellcolor[rgb]{ .851,  .851,  .851}0.486 & \cellcolor[rgb]{ .851,  .851,  .851}0.702 & \cellcolor[rgb]{ .851,  .851,  .851}0.461\\
                \midrule
    \multicolumn{1}{c}{\multirow{3}[2]{*}{\begin{sideways}$\Delta$F1\end{sideways}}} & CV-SS & 0.079 & 0.274 & 0.088 & 0.213 \\
          & CV-DD & 0.185 & 0.393 & 0.208 & 0.464 \\
          & SS-DD & 0.106 & 0.119 & 0.12  & 0.251 \\

			\bottomrule
	\end{tabular}}%
	\label{tab:results}%
\end{table}%

The results are shown in Table~\ref{tab:results}. All methods achieved high scores for CV, close to or above 0.90. 
However, a decrease is observed in all methods for SS and DD, with this decrease being particularly dramatic for the statistics-based methods (CICFlowMeter and Kitsune). For DD, GeMID maintains 81\% of its CV F1 score and IoTDevID 77\%, compared to only 55\% for CIC and 50\% for Kitsune. This significant drop supports our claim that flow or window-based statistical methods, which focus on relationships between devices within the network rather than individual device activities, are not suitable for building generalizable DI models due to their poor transferability across heterogeneous network environments. 

\textcolor{black}{The performance drop between evaluation contexts can be more formally analyzed by examining the generalization gap ($\Delta F1$), summarized in the final section of Table~\ref{tab:results}. This metric quantifies the loss in F1 score as the testing conditions become more challenging. The most critical gap is between the lenient cross-validation and the strict dataset-versus-dataset context (CV-DD). Here, the packet-header-based methods, GeMID and IoTDevID, show superior resilience with gaps of 0.185 and 0.208, respectively. In contrast, the statistics-based methods fail to generalize effectively, with CICFlowMeter exhibiting a gap of 0.393 and Kitsune a gap of 0.464—more than double that of GeMID. This quantitatively demonstrates that statistical features are not robust to changes in the network environment. A more detailed, per-case analysis of this generalization gap is available in Appendix  Table~\ref{tab:deltaf1}.}

The two methods---GeMID and IoTDevID---based on packet headers both perform significantly better, with GeMID outperforming IoTDevID within all evaluation contexts. Table~\ref{tab:class} shows the per-device results for GeMID within the DD evaluation context. It can be seen that most of the devices are identified with a high level of accuracy. A notable exception is the \textit{echo} family of devices, which, presumably due to underlying similarities, are easily misclassified with each other (see Appendix Figures~\ref{fig:mon_data_uk2us} and~\ref{fig:mon_data_us2uk}  for confusion matrices). There is also a degree of misclassification between devices from different manufacturers that perform the same function, e.g., between \textit{roku-tv} and \textit{samsungtv-wired}, and between \textit{yi-camera} and \textit{wansview-cam-wired}. A third category of poor performance is due to data paucity, e.g., for \textit{sousvide}, there are only 70 USA samples, compared to 15,000 for UK. The challenges in classifying these devices are not specific to GeMID but are also observed in other methods (IoTDevID, CICFlowMeter, and Kitsune). These methods also encounter similar difficulties due to the inherent complexity of distinguishing between devices with similar characteristics.

\begin{table}[tb]
	\centering
	\caption{F1 scores per device in all 8 DD cases for MonIoTr data, with green for higher and red for lower success.}
	\resizebox{\columnwidth}{!}{ \begin{tabular}{llrrrrrrrr}
			\toprule
			\multirow{2}[2]{*}{Devices} & \multicolumn{1}{r}{Train→} & \multicolumn{1}{c}{UK} & \multicolumn{1}{c}{UK} & \multicolumn{1}{c}{UKVPN} & \multicolumn{1}{c}{UKVPN} & \multicolumn{1}{c}{USA} & \multicolumn{1}{c}{USA} & \multicolumn{1}{c}{USAVPN} & \multicolumn{1}{c}{USAVPN} \\
			& \multicolumn{1}{r}{Test→} & \multicolumn{1}{c}{USA} & \multicolumn{1}{c}{USAVPN} & \multicolumn{1}{c}{USA} & \multicolumn{1}{c}{USAVPN} & \multicolumn{1}{c}{UK} & \multicolumn{1}{c}{UKVPN} & \multicolumn{1}{c}{UK} & \multicolumn{1}{c}{UKVPN} \\
			\midrule

			\multicolumn{2}{l}{appletv} & \cellcolor[rgb]{ .875,  .945,  .902}0.868 & \cellcolor[rgb]{ .871,  .941,  .898}0.879 & \cellcolor[rgb]{ .988,  .988,  1}0.851 & \cellcolor[rgb]{ .984,  .961,  .973}0.830 & \cellcolor[rgb]{ .98,  .812,  .824}0.658 & \cellcolor[rgb]{ .984,  .988,  .996}0.858 & \cellcolor[rgb]{ .976,  .616,  .627}0.614 & \cellcolor[rgb]{ .984,  .882,  .89}0.802 \\
			\multicolumn{2}{l}{blink-camera} & \cellcolor[rgb]{ .518,  .8,  .592}0.963 & \cellcolor[rgb]{ .451,  .773,  .537}0.983 & \cellcolor[rgb]{ .573,  .82,  .643}0.952 & \cellcolor[rgb]{ .451,  .773,  .537}0.985 & \cellcolor[rgb]{ .553,  .812,  .627}0.951 & \cellcolor[rgb]{ .671,  .859,  .725}0.929 & \cellcolor[rgb]{ .745,  .89,  .792}0.945 & \cellcolor[rgb]{ .839,  .929,  .871}0.932 \\
			\multicolumn{2}{l}{blink-security-hub} & \cellcolor[rgb]{ .722,  .882,  .769}0.909 & \cellcolor[rgb]{ .827,  .925,  .863}0.889 & \cellcolor[rgb]{ .98,  .714,  .722}0.498 & \cellcolor[rgb]{ .984,  .922,  .933}0.775 & \cellcolor[rgb]{ .973,  .412,  .42}0.259 & \cellcolor[rgb]{ .973,  .506,  .514}0.213 & \cellcolor[rgb]{ .976,  .663,  .671}0.649 & \cellcolor[rgb]{ .976,  .686,  .694}0.599 \\
			\multicolumn{2}{l}{echodot} & \cellcolor[rgb]{ .973,  .412,  .42}0.161 & \cellcolor[rgb]{ .973,  .412,  .42}0.093 & \cellcolor[rgb]{ .973,  .412,  .42}0.106 & \cellcolor[rgb]{ .973,  .412,  .42}0.053 & \cellcolor[rgb]{ .976,  .675,  .682}0.521 & \cellcolor[rgb]{ .973,  .412,  .42}0.083 & \cellcolor[rgb]{ .976,  .569,  .576}0.575 & \cellcolor[rgb]{ .973,  .412,  .42}0.313 \\
			\multicolumn{2}{l}{echoplus} & \cellcolor[rgb]{ .973,  .498,  .506}0.263 & \cellcolor[rgb]{ .976,  .635,  .643}0.387 & \cellcolor[rgb]{ .976,  .659,  .671}0.430 & \cellcolor[rgb]{ .973,  .545,  .553}0.242 & \cellcolor[rgb]{ .973,  .553,  .561}0.402 & \cellcolor[rgb]{ .98,  .788,  .796}0.590 & \cellcolor[rgb]{ .973,  .451,  .459}0.479 & \cellcolor[rgb]{ .973,  .506,  .514}0.413 \\
			\multicolumn{2}{l}{echospot} & \cellcolor[rgb]{ .98,  .808,  .82}0.629 & \cellcolor[rgb]{ .98,  .753,  .761}0.542 & \cellcolor[rgb]{ .98,  .816,  .827}0.633 & \cellcolor[rgb]{ .98,  .749,  .761}0.532 & \cellcolor[rgb]{ .984,  .859,  .867}0.704 & \cellcolor[rgb]{ .984,  .941,  .953}0.798 & \cellcolor[rgb]{ .98,  .8,  .812}0.760 & \cellcolor[rgb]{ .98,  .82,  .831}0.739 \\
			\multicolumn{2}{l}{firetv} & \cellcolor[rgb]{ .98,  .808,  .82}0.629 & \cellcolor[rgb]{ .984,  .871,  .878}0.695 & \cellcolor[rgb]{ .98,  .827,  .839}0.646 & \cellcolor[rgb]{ .984,  .863,  .875}0.694 & \cellcolor[rgb]{ .98,  .792,  .8}0.638 & \cellcolor[rgb]{ .984,  .953,  .961}0.809 & \cellcolor[rgb]{ .976,  .663,  .675}0.651 & \cellcolor[rgb]{ .984,  .882,  .894}0.805 \\
			\multicolumn{2}{l}{google-home-mini} & \cellcolor[rgb]{ .984,  .961,  .973}0.806 & \cellcolor[rgb]{ .843,  .933,  .878}0.885 & \cellcolor[rgb]{ .984,  .918,  .929}0.761 & \cellcolor[rgb]{ .957,  .976,  .973}0.873 & \cellcolor[rgb]{ .839,  .929,  .871}0.873 & \cellcolor[rgb]{ .82,  .922,  .855}0.895 & \cellcolor[rgb]{ .988,  .988,  1}0.909 & \cellcolor[rgb]{ .984,  .98,  .992}0.903 \\
			\multicolumn{2}{l}{insteon-hub} & \cellcolor[rgb]{ .722,  .882,  .769}0.909 & \cellcolor[rgb]{ .984,  .98,  .992}0.842 & \cellcolor[rgb]{ .98,  .827,  .839}0.648 & \cellcolor[rgb]{ .98,  .753,  .765}0.537 & \cellcolor[rgb]{ .984,  .961,  .973}0.806 & \cellcolor[rgb]{ .984,  .878,  .89}0.712 & \cellcolor[rgb]{ .659,  .855,  .718}0.958 & \cellcolor[rgb]{ .624,  .839,  .686}0.962 \\
			\multicolumn{2}{l}{lightify-hub} & \cellcolor[rgb]{ .561,  .816,  .631}0.951 & \cellcolor[rgb]{ .62,  .839,  .682}0.941 & \cellcolor[rgb]{ .596,  .831,  .663}0.946 & \cellcolor[rgb]{ .671,  .863,  .729}0.936 & \cellcolor[rgb]{ .965,  .98,  .98}0.839 & \cellcolor[rgb]{ .69,  .871,  .745}0.924 & \cellcolor[rgb]{ .58,  .824,  .651}0.969 & \cellcolor[rgb]{ .498,  .792,  .576}0.979 \\
			\multicolumn{2}{l}{magichome-strip} & \cellcolor[rgb]{ .988,  .988,  1}0.837 & \cellcolor[rgb]{ .953,  .976,  .973}0.857 & \cellcolor[rgb]{ .769,  .902,  .812}0.905 & \cellcolor[rgb]{ .812,  .918,  .847}0.905 & \cellcolor[rgb]{ .459,  .776,  .541}0.977 & \cellcolor[rgb]{ .596,  .831,  .663}0.945 & \cellcolor[rgb]{ .525,  .804,  .6}0.978 & \cellcolor[rgb]{ .694,  .871,  .745}0.952 \\
			\multicolumn{2}{l}{nest-tstat} & \cellcolor[rgb]{ .835,  .925,  .867}0.879 & \cellcolor[rgb]{ .984,  .953,  .965}0.803 & \cellcolor[rgb]{ .91,  .957,  .933}0.870 & \cellcolor[rgb]{ .988,  .988,  1}0.866 & \cellcolor[rgb]{ .525,  .8,  .6}0.959 & \cellcolor[rgb]{ .6,  .831,  .663}0.945 & \cellcolor[rgb]{ .667,  .859,  .722}0.957 & \cellcolor[rgb]{ .663,  .859,  .718}0.956 \\
			\multicolumn{2}{l}{ring-doorbell} & \cellcolor[rgb]{ .388,  .745,  .482}0.997 & \cellcolor[rgb]{ .388,  .745,  .482}0.998 & \cellcolor[rgb]{ .388,  .745,  .482}0.997 & \cellcolor[rgb]{ .388,  .745,  .482}0.999 & \cellcolor[rgb]{ .388,  .745,  .482}0.996 & \cellcolor[rgb]{ .388,  .745,  .482}0.991 & \cellcolor[rgb]{ .408,  .753,  .498}0.995 & \cellcolor[rgb]{ .408,  .757,  .502}0.991 \\
			\multicolumn{2}{l}{roku-tv} & \cellcolor[rgb]{ .976,  .561,  .569}0.337 & \cellcolor[rgb]{ .98,  .714,  .722}0.490 & \cellcolor[rgb]{ .98,  .702,  .71}0.482 & \cellcolor[rgb]{ .98,  .733,  .741}0.507 & \cellcolor[rgb]{ .984,  .847,  .855}0.692 & \cellcolor[rgb]{ .984,  .863,  .875}0.693 & \cellcolor[rgb]{ .973,  .412,  .42}0.448 & \cellcolor[rgb]{ .976,  .651,  .659}0.563 \\
			\multicolumn{2}{l}{samsungtv-wired} & \cellcolor[rgb]{ .984,  .984,  .996}0.834 & \cellcolor[rgb]{ .988,  .988,  1}0.848 & \cellcolor[rgb]{ .62,  .839,  .682}0.941 & \cellcolor[rgb]{ .8,  .914,  .835}0.908 & \cellcolor[rgb]{ .984,  .961,  .973}0.808 & \cellcolor[rgb]{ .984,  .957,  .969}0.820 & \cellcolor[rgb]{ .976,  .675,  .682}0.659 & \cellcolor[rgb]{ .98,  .729,  .741}0.645 \\
			\multicolumn{2}{l}{sengled-hub} & \cellcolor[rgb]{ .984,  .843,  .855}0.670 & \cellcolor[rgb]{ .984,  .918,  .929}0.758 & \cellcolor[rgb]{ .565,  .82,  .635}0.954 & \cellcolor[rgb]{ .737,  .886,  .784}0.922 & \cellcolor[rgb]{ .976,  .678,  .686}0.525 & \cellcolor[rgb]{ .976,  .686,  .698}0.455 & \cellcolor[rgb]{ .973,  .518,  .525}0.533 & \cellcolor[rgb]{ .976,  .639,  .651}0.552 \\
			\multicolumn{2}{l}{smartthings-hub} & \cellcolor[rgb]{ .933,  .969,  .953}0.853 & \cellcolor[rgb]{ .863,  .937,  .89}0.880 & \cellcolor[rgb]{ .745,  .89,  .792}0.910 & \cellcolor[rgb]{ .804,  .914,  .839}0.907 & \cellcolor[rgb]{ .624,  .843,  .686}0.932 & \cellcolor[rgb]{ .78,  .906,  .82}0.904 & \cellcolor[rgb]{ .824,  .922,  .859}0.934 & \cellcolor[rgb]{ .988,  .988,  1}0.911 \\
			\multicolumn{2}{l}{sousvide} & \cellcolor[rgb]{ .976,  .631,  .643}0.422 & \cellcolor[rgb]{ .976,  .639,  .647}0.394 & \cellcolor[rgb]{ .973,  .427,  .435}0.128 & \cellcolor[rgb]{ .973,  .439,  .447}0.092 & \cellcolor[rgb]{ .404,  .753,  .494}0.992 & \cellcolor[rgb]{ .431,  .765,  .518}0.982 & \cellcolor[rgb]{ .388,  .745,  .482}0.997 & \cellcolor[rgb]{ .388,  .745,  .482}0.994 \\
			\multicolumn{2}{l}{t-philips-hub} & \cellcolor[rgb]{ .427,  .761,  .514}0.987 & \cellcolor[rgb]{ .431,  .765,  .522}0.988 & \cellcolor[rgb]{ .431,  .765,  .522}0.987 & \cellcolor[rgb]{ .443,  .769,  .529}0.987 & \cellcolor[rgb]{ .533,  .804,  .608}0.956 & \cellcolor[rgb]{ .984,  .965,  .976}0.826 & \cellcolor[rgb]{ .529,  .804,  .604}0.977 & \cellcolor[rgb]{ .553,  .812,  .627}0.971 \\
			\multicolumn{2}{l}{t-wemo-plug} & \cellcolor[rgb]{ .624,  .839,  .686}0.935 & \cellcolor[rgb]{ .984,  .984,  .996}0.844 & \cellcolor[rgb]{ .631,  .843,  .69}0.938 & \cellcolor[rgb]{ .984,  .969,  .98}0.841 & \cellcolor[rgb]{ .529,  .804,  .604}0.957 & \cellcolor[rgb]{ .443,  .769,  .529}0.979 & \cellcolor[rgb]{ .663,  .859,  .722}0.957 & \cellcolor[rgb]{ .478,  .784,  .561}0.982 \\
			\multicolumn{2}{l}{tplink-bulb} & \cellcolor[rgb]{ .6,  .831,  .667}0.941 & \cellcolor[rgb]{ .475,  .78,  .557}0.977 & \cellcolor[rgb]{ .984,  .988,  .996}0.852 & \cellcolor[rgb]{ .627,  .843,  .69}0.946 & \cellcolor[rgb]{ .725,  .882,  .773}0.904 & \cellcolor[rgb]{ .988,  .988,  1}0.857 & \cellcolor[rgb]{ .416,  .757,  .506}0.994 & \cellcolor[rgb]{ .443,  .769,  .529}0.987 \\
			\multicolumn{2}{l}{tplink-plug} & \cellcolor[rgb]{ .984,  .941,  .953}0.783 & \cellcolor[rgb]{ .765,  .898,  .808}0.905 & \cellcolor[rgb]{ .984,  .976,  .984}0.836 & \cellcolor[rgb]{ .706,  .875,  .757}0.929 & \cellcolor[rgb]{ .988,  .988,  1}0.832 & \cellcolor[rgb]{ .984,  .984,  .996}0.855 & \cellcolor[rgb]{ .984,  .859,  .871}0.808 & \cellcolor[rgb]{ .984,  .961,  .973}0.886 \\
			\multicolumn{2}{l}{wansview-cam-wired} & \cellcolor[rgb]{ .847,  .933,  .878}0.875 & \cellcolor[rgb]{ .984,  .965,  .976}0.822 & \cellcolor[rgb]{ .886,  .949,  .914}0.876 & \cellcolor[rgb]{ .984,  .969,  .98}0.841 & \cellcolor[rgb]{ .69,  .871,  .745}0.914 & \cellcolor[rgb]{ .722,  .882,  .773}0.917 & \cellcolor[rgb]{ .984,  .984,  .996}0.908 & \cellcolor[rgb]{ .973,  .984,  .988}0.913 \\
			\multicolumn{2}{l}{xiaomi-hub} & \cellcolor[rgb]{ .984,  .976,  .988}0.825 & \cellcolor[rgb]{ .867,  .941,  .894}0.879 & \cellcolor[rgb]{ .984,  .969,  .98}0.830 & \cellcolor[rgb]{ .969,  .98,  .984}0.870 & \cellcolor[rgb]{ .984,  .906,  .918}0.753 & \cellcolor[rgb]{ .69,  .867,  .741}0.924 & \cellcolor[rgb]{ .471,  .78,  .553}0.986 & \cellcolor[rgb]{ .424,  .761,  .514}0.989 \\
			\multicolumn{2}{l}{yi-camera} & \cellcolor[rgb]{ .984,  .894,  .906}0.729 & \cellcolor[rgb]{ .98,  .839,  .847}0.654 & \cellcolor[rgb]{ .984,  .882,  .894}0.719 & \cellcolor[rgb]{ .984,  .851,  .859}0.673 & \cellcolor[rgb]{ .98,  .749,  .761}0.596 & \cellcolor[rgb]{ .98,  .788,  .796}0.589 & \cellcolor[rgb]{ .976,  .655,  .667}0.646 & \cellcolor[rgb]{ .98,  .741,  .753}0.657 \\
			\midrule
			\multicolumn{2}{l}{Mean accuracy} & 0.857 & 0.825 & 0.861 & 0.830 & 0.893 & 0.888 & 0.898 & 0.889 \\
			\multicolumn{2}{l}{Mean F1 score} & 0.760 & 0.769 & 0.748 & 0.746 & 0.778 & 0.780 & 0.811 & 0.816 \\
			\bottomrule
			
	\end{tabular}}%
	\label{tab:class}%
\end{table}%

\subsection{Feature Usage}

GeMID and IoTDevID demonstrate better generalization in model performance compared to those that use statistical features. Of these two, GeMID is approximately 8 percentage points higher in accuracy. Given that the model building stages are similar, this is presumably due in part to a more rigorous feature selection process, using two data sets rather than just one to ensure that the selected features are generalizable.

Notably, IoTDevID's feature set is rich in features related to BOOTP, DNS, and EAPOL protocols, whereas GeMID lacks these. This may be due to a dataset-specific bias, owing to the reliance on a single dataset---Aalto---for IoTDevID's feature selection process. This is supported by the observation that while GeMID outperforms IoTDevID in experiments using the UNSW and MonIoTr datasets, IoTDevID shows markedly better performance when applied to the Aalto dataset~\cite{kostas22IoTDevID} (see Appendix Table~\ref{tab:aalto}). 

However, it is worth noting that both share the  features dstport\_class, ip.flags.df, ip.len, ip.ttl, and tcp.window\_size, suggesting that these are particularly important for DI.

Another reason for the difference in performance may be the larger initial pool of extracted features in GeMID (332 features) compared to IoTDevID (112 features). Although both studies generally use the same protocols, GeMID captures more sub-features within each protocol, providing finer-grained information in the feature set. In terms of selected features, GeMID offers more comprehensive protocol coverage, starting with a richer set of HTTP features and extending to broader coverage of TCP and UDP protocols. This includes features such as sequence numbers, timing metrics, and window size information, which enhance its ability to capture more detailed network behavior (see Figure~\ref{fig:vote8}).
\textcolor{black}{Unlike in the IoTDevID study, we retained features containing session-based identifiers (such as TCP acknowledgment, TCP sequence, and source/destination port numbers). Whilst these can lead to over-optimistic metric scores~\cite{kostas2024individual}, this is only the case when sessions are split between test and train sets, e.g., when using CV. When this is controlled for using stricter evaluation contexts (as in this study), the inclusion of these features can enhance model performance by capturing meaningful relationships between packets.}

In practice, the accuracy of packet-level classification methods can be improved by aggregating predictions from individual packets. The IoTDevID study introduced an aggregation algorithm that enhances performance by combining multiple packet-level predictions. This approach is particularly effective for packet-level methods such as GeMID, IoTDevID, and Kitsune (despite Kitsune focusing on the relationship between the window system and packets, as it still relies on packet-based labels). In contrast, CICFlowMeter operates at the flow level, where such aggregation is not applicable. 
Our results in Appendix Tables~\ref{tab:aggre1} and~\ref{tab:aggre2} show the impact of this approach, with GeMID achieving a mean F1 score of 0.892 in the DD evaluation context.

\section{Limitations and Practical Considerations}\label{limit}

 \textcolor{black}{\subsection{Practical Considerations and Deployment Scenarios}} 
While the core of this research is methodological, it is important to consider the practical path to deployment for a DI system built on our framework.

\noindent
 \textcolor{black}{\textbf{Deployment Environment:} We envision the GeMID classifier operating at the edge of the local network, for instance, on a home router or an enterprise gateway. This location is ideal as it has visibility into all traffic from local IoT devices and allows for device-specific policies (e.g., isolation, access control) to be enforced locally. Deployment at the ISP level, while possible, would present significant privacy and scalability challenges.}

 \noindent
\textcolor{black}{\textbf{Model Scalability and Management:} The multi-class models developed in this paper were aimed at assessing the effectiveness of our feature set across a predefined list of devices. In a real-world, dynamic environment, a more practical approach would be a combination of binary, one-vs-all models. When a new device is added to the network, the system could either attempt to classify it against its known models or flag it as unknown. If a new model for this device type becomes available, it can be added to the system without retraining the entire multi-class classifier. This binary model approach inherently solves the issue of adding or removing devices from a household.}

\noindent
\textcolor{black}{\textbf{Model Training and Distribution:} The responsibility for training would likely fall to device manufacturers, who are best positioned to generate traffic data for their products under controlled conditions. These pre-trained, signed models could be made available through a central repository. A network controller, perhaps using Software-Defined Networking (SDN) principles, could then automatically detect a new, unclassified device, query the repository for a corresponding model, and integrate it into the local DI system.}

\noindent
\textcolor{black}{\textbf{Model Size and Efficiency:} A key concern for deployment on resource-constrained hardware like a home router is model efficiency. Our framework directly addresses this. The final GeMID model utilizes a Random Forest classifier with a small, optimized feature set. As shown in our evaluation (Figure~\ref{fig:cmML}), this model has a very fast average inference time, making it highly efficient and suitable for online, real-time deployment without significant computational overhead.}

\subsection{Study Limitations and Future Work}

The primary limitation of this work is the availability and quality of public datasets. Although we used well-regarded datasets and cross-dataset validation, there will always be a limit in terms of the diversity, representativeness and sampling biases of any dataset. 

\textcolor{black}{A further consideration is the generalizability of the feature selection process itself. Our methodology deliberately used the UNSW dataset family for feature and model selection and the entirely independent MonIoTr dataset family for final validation. This design ensures that our final performance metrics are not biased by the selection process. While we believe that our process, and the datasets used, provide strong evidence for the robustness of our framework, we encourage others to investigate enhancements to our model, such as by studying other datasets for feature selection and testing. Though our UNSW feature-based selection and testing on MonIoTr demonstrate that our model selected fundamental, device-intrinsic properties (and are not dataset-specific), it is likely that the exact features used would vary if selected on a different device population, due to both the different network behaviors of the devices and the stochastic nature of the genetic algorithm used. Nevertheless, our framework is designed to be a robust process for identifying the most stable features in any given context, and we anticipate that a core set of highly predictive features would remain consistent across datasets.}

\textcolor{black}{
A second limitation relates to GeMID's dependence on packet header features. Our method relies on features extracted from unencrypted headers such as those from IP, TCP, and UDP. This approach remains effective for a majority of today's IoT traffic where the payload is encrypted by a protocol like TLS but the transport headers remain visible. However, the increasing adoption of newer protocols, most notably QUIC, which encrypts most of the transport layer header information, presents a future challenge. The effectiveness of GeMID's current feature set would likely be diminished on networks with heavy QUIC adoption, and future work should investigate features that are robust to transport-layer encryption.}

\textcolor{black}{
Third, it is important to discuss the computational overhead of our framework. The GA used for feature selection is a computationally intensive process. However, this is a one-time, offline cost incurred during the model design and training phase. It is not part of the operational, real-time device identification system. The final GeMID model, which uses a small, optimized feature set with a Random Forest classifier, is highly efficient and suitable for online deployment, as evidenced by the fast inference times shown in Figure~\ref{fig:cmML}. Therefore, while the development phase requires significant resources, the deployment phase is lightweight.}

\textcolor{black}{
Finally, the scalability of GeMID to environments with a significantly larger number of distinct device types has not yet been validated. Our experiments were conducted on datasets with several dozen unique devices. Scaling to networks with hundreds or thousands of device classes would make the multi-class classification task inherently more complex. This could potentially reduce classification accuracy, especially among devices with very similar functionalities and traffic patterns. Therefore, while our framework is designed to produce generalizable models, its performance limits at a very large scale remain an open question for future research.}

Future work in DI would benefit from larger datasets designed specifically for this purpose, containing device samples that reflect realistic usage situations. Whilst we already consider devices situated in two different network environments, collecting data from the same devices operating in a broader range of networks would inevitably provide a broader perspective on generalizability. There is also a need for more data from non-IP protocols such as ZigBee and Z-Wave to test how well DI models of non-IP devices generalize.

Whilst this study (in common with previous studies) focuses on benign data, future DI studies would also benefit from using attack or malicious data, since this would be more representative of the actual environments within which DI models are deployed.

\section{Conclusions}\label{con}

Our work presents a novel approach to improving the generalizability of IoT DI models. Using a two-stage framework involving feature and algorithm selection followed by cross-dataset validation, we show that models trained on datasets in one environment can effectively identify devices in another environment. This addresses an important gap in existing methodologies, which often do not validate across different datasets.

Comparative analysis of DI methods showed that packet header-based features performed better than network statistics and window methods for building generalizable DI models. Packet features offer a more consistent and reliable basis for model training because they are less influenced by network-specific variables. \textcolor{black}{This claim is most strongly supported by our generalization gap analysis (Table~\ref{tab:results}), where the performance of statistical methods like CICFlowMeter and Kitsune degraded by over 100\% more than our packet-based GeMID method when moving from single-environment validation to cross-environment testing (CV-DD $\Delta$F1 of 0.393 and 0.464 vs. 0.185). This demonstrates that packet features offer a more consistent and reliable basis for model training because they capture device-intrinsic attributes rather than network-specific variables.} 
This finding highlights the importance of selecting features that inherently capture device-specific attributes.

Our results also highlight the limitations of relying only on single datasets for model training and validation. Such approaches run the risk of overfitting and may fail to generalize to different network environments. In contrast, our methodology provides robustness and adaptability, which are crucial for practical IoT security applications.

Our findings underscore that statistical methods, despite their prevalence, are unreliable for DI due to their entanglement with network conditions. This challenges nearly half of existing research~\cite{kostas23thesis,jmila2022survey}, urging a re-evaluation of studies reliant on flow or window statistics. Future work should prioritize packet-based or similarly device-centric features to ensure generalizability, reshaping IoT security practices.

\bibliographystyle{IEEEtran}
\bibliography{references}
\vspace{-30pt}

\onecolumn
\appendix

\section{Appendices}\label{app}

\section{CICFlowmeter Feature Selection Process and Results}~\label{CICFlowmeter}
\begin{figure}[h!]
	\centering{
		\includegraphics[width=1\columnwidth]{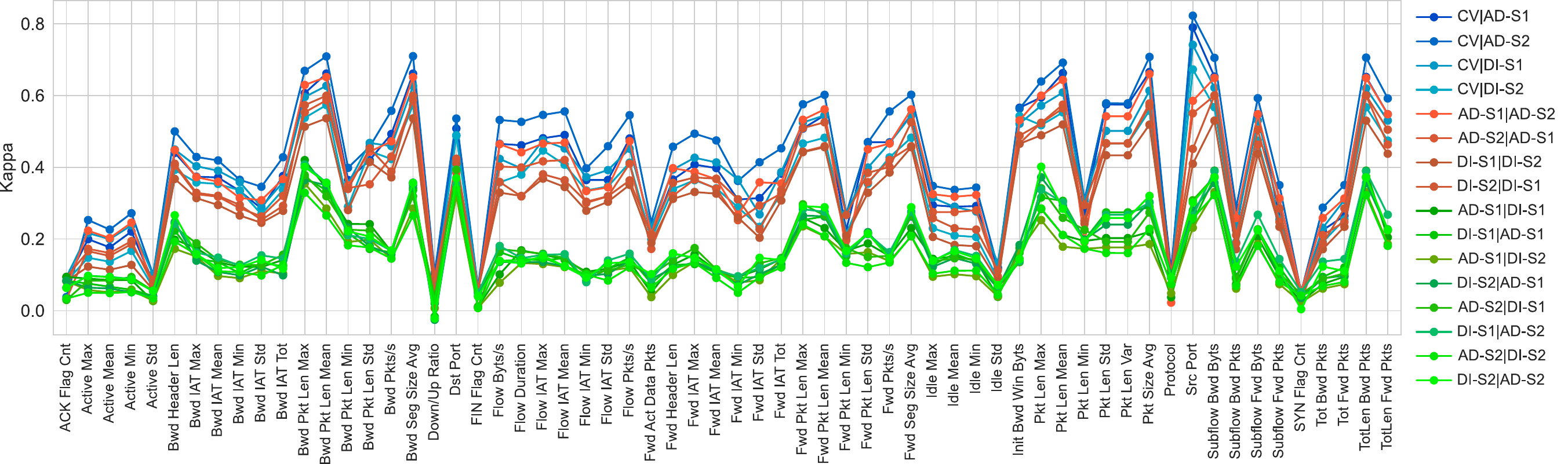}}
	\caption{Comparison of kappa scores for CICFlowmeter features in UNSW datasets using CV (blue) and isolated methods, SS (red) and DD (green). The CV method tends to overestimate feature utility, showing higher scores for many attributes, while SS and DD produce more realistic evaluations.}
	\label{fig:cizgi-cic}
\end{figure}

\begin{figure}[h!]
	\centering{	
	\includegraphics[width=1\columnwidth]{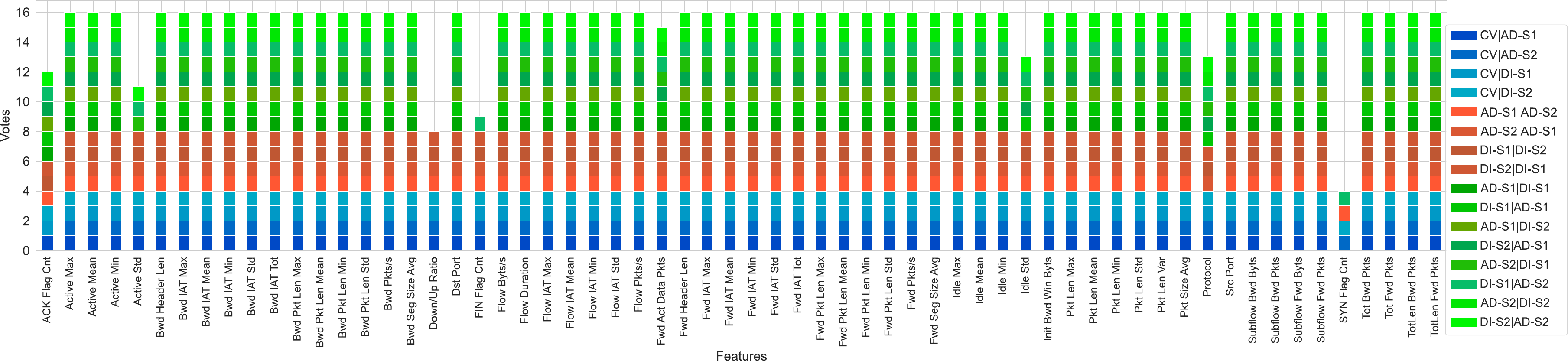}}
	\caption{A voting system for CICFlowmeter features that gives one vote to each feature with a kappa value different from 0 that is individually predictive within CV (blue), SS (red) \& DD (green).}
	\label{fig:vote-cic}
\end{figure}

\begin{figure}[h!]
	\centering{
		\includegraphics[width=1\columnwidth]{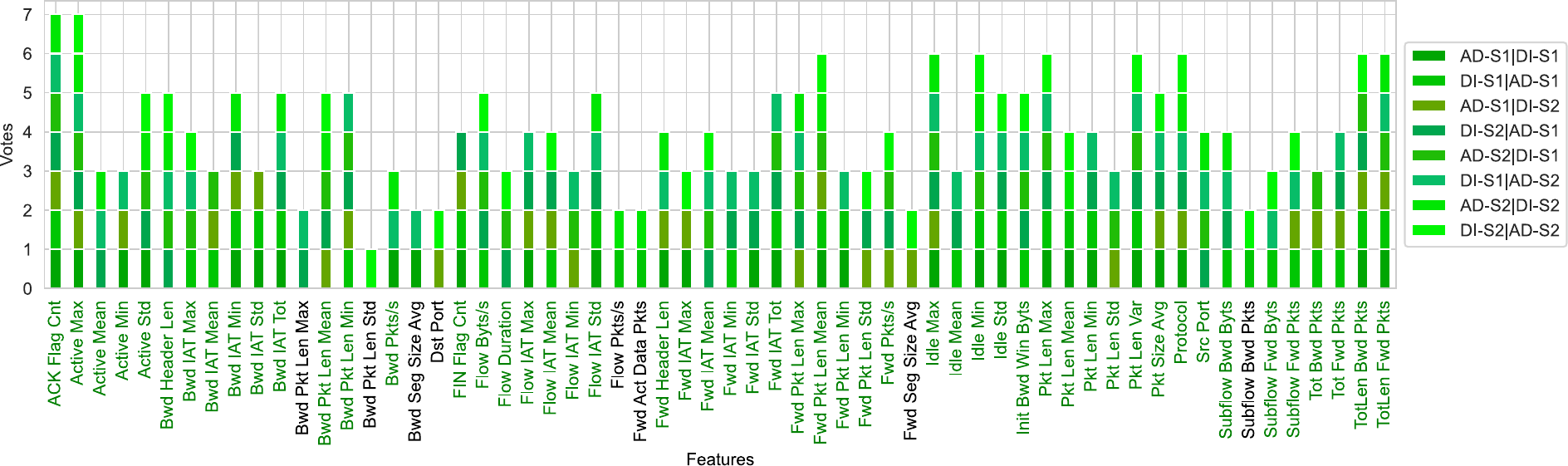}}
	\caption{ List of intersecting CICFlowmeter features identified across eight dataset versus dataset (DD) cases using Genetic Algorithm (GA).}
	\label{fig:v-viv-g}
\end{figure}

\begin{figure}[h!]
	\centering{
		\includegraphics[width=0.8\columnwidth]{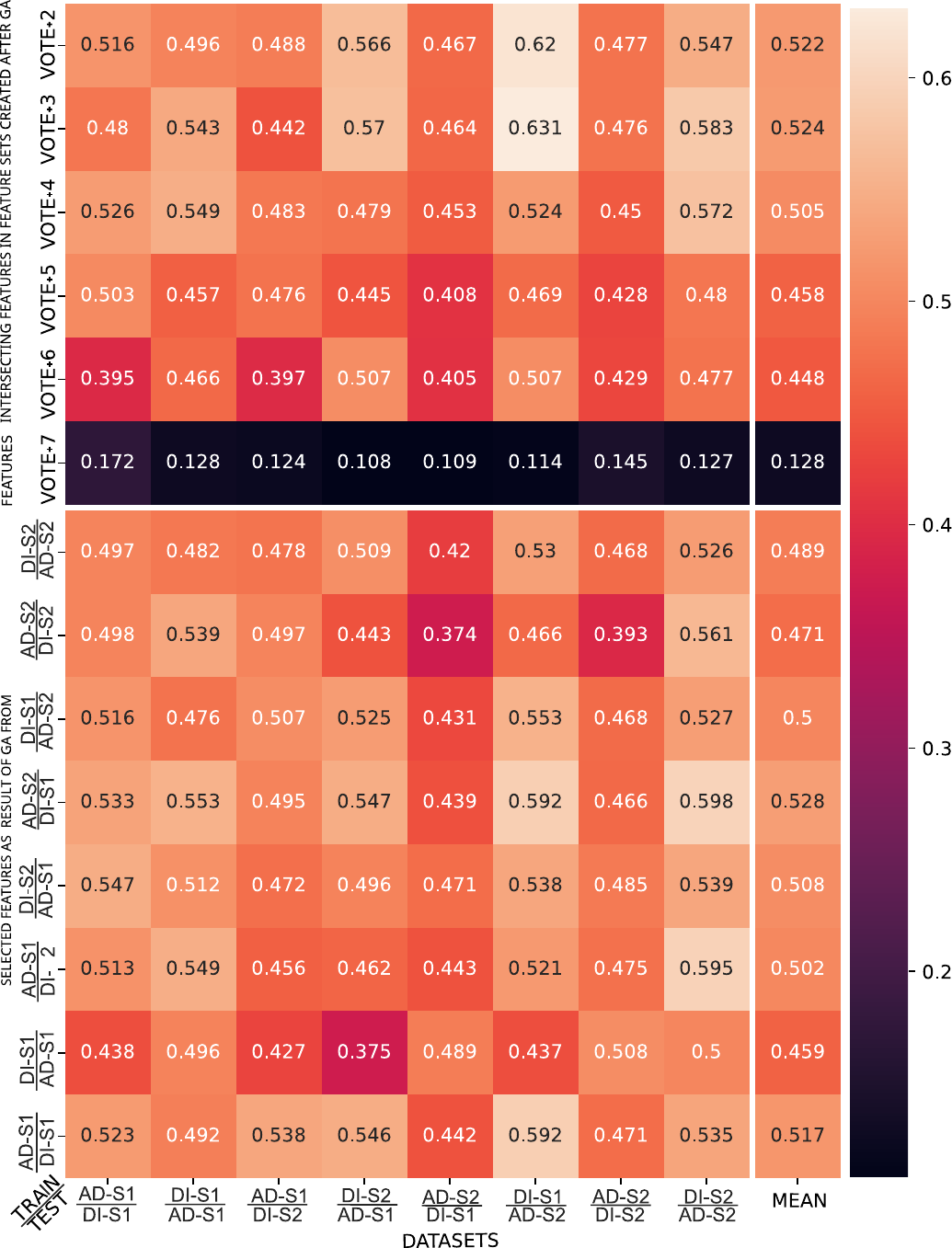}}
	\caption{Comparison of CICFlowmeter feature set performance across dataset versus dataset (DD) cases. The left side of the heatmap displays the results of applying feature sets obtained from the first step of the Genetic Algorithm (GA) to all DD data, yielding 64 results. On the right, the performance of the features grouped according to their frequency, focusing on the intersection of features obtained from the output of the GA algorithm.}
	\label{fig:CIC-f-cm}
\end{figure}

\begin{figure}[h!]
	\centering{
		\includegraphics[width=0.8\columnwidth]{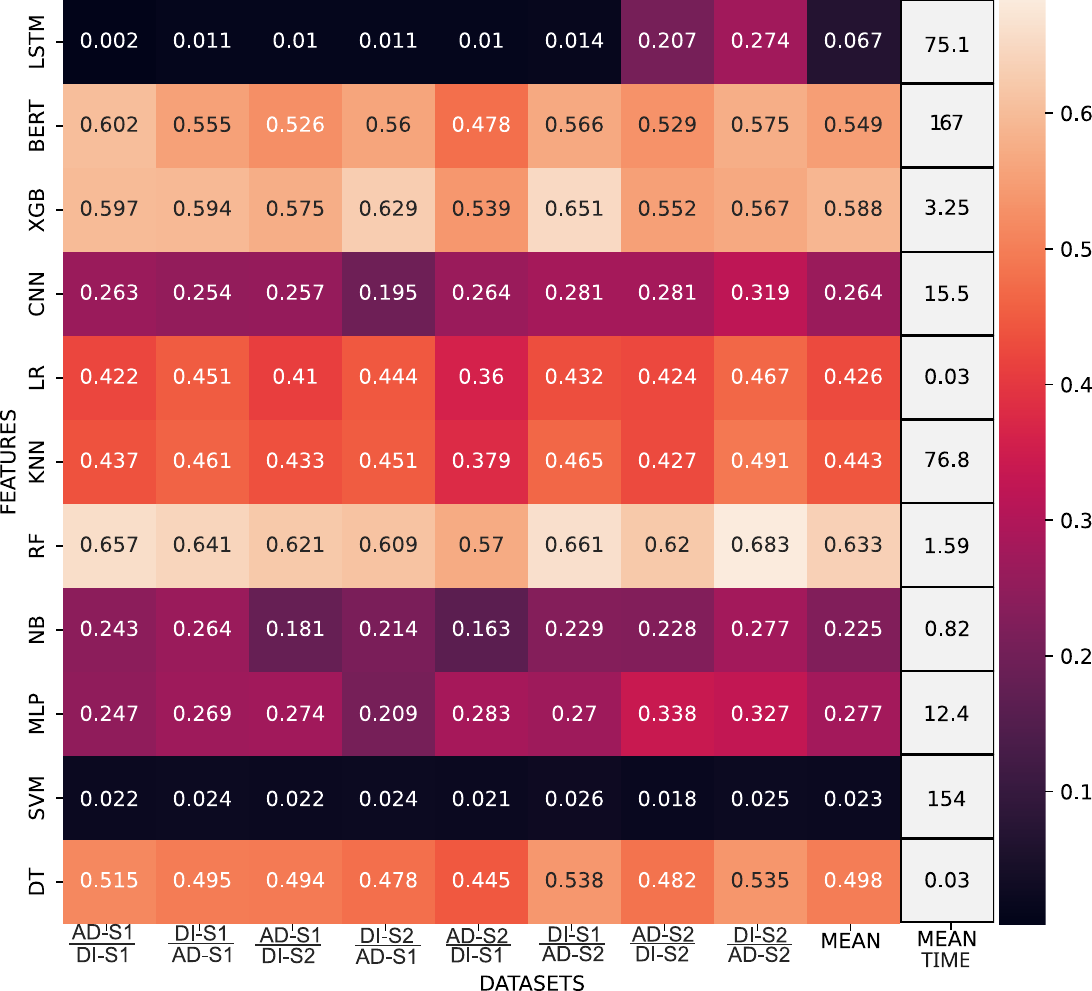}}
	\caption{F1 scores and average inference times of various ML algorithms applied to DD datasets with Kitsune features. The Random Forest (RF) and XGB methods demonstrate the highest F1 scores of 0.633 and 0.588, respectively, while Decision Trees (DT) show the fastest inference time.}
	\label{fig:CICcmML}
\end{figure}

In this section, we apply the same feature selection process to the CICFlowMeter features as we did with GeMID. Table~\ref{tab:feature-cic} lists all features included in the CICFlowMeter method. Figure~\ref{fig:cizgi-cic} compares feature utility in the UNSW datasets, using both cross-validation and isolated evaluation methods (SS and DD). The voting process for selecting the most suitable features, based on the utility of each feature, is illustrated in Figure~\ref{fig:vote-cic}. Candidate sub-features generated by applying a genetic algorithm to each dataset series are presented in Figure~\ref{fig:v-viv-g}. The testing results for these features and their intersections across all UNSW-DD datasets are shown in Figure~\ref{fig:CIC-f-cm}. The best-performing feature set, as determined by these tests, is detailed in Table~\ref{tab:cic-final-features}. The results of testing the final features with various machine learning and neural network-based techniques to evaluate model performance are shown in Figure~\ref{fig:CICcmML}.
\begin{table}[h!]
	\centering
	\caption{Features obtained after applying the CICFlowmeter tool to the pcap file.}
	\begin{tabular}{rlr}
		\toprule
		\multicolumn{3}{c}{CICFlowmeter Full Feature List} \\
		\midrule
		\multicolumn{1}{l}{Idle Min} & total Length of Fwd Packet & \multicolumn{1}{l}{Flow Bytes/s} \\
		\multicolumn{1}{l}{Idle Max} & total Length of Bwd Packet & \multicolumn{1}{l}{Flow IAT Std} \\
		\multicolumn{1}{l}{Idle Std} & Packet Length Variance   & \multicolumn{1}{l}{Flow IAT Max} \\
		\multicolumn{1}{l}{Idle Mean} & Fwd Packet Length Min  & \multicolumn{1}{l}{Flow IAT Min} \\
		\multicolumn{1}{l}{Active Min} & Fwd Packet Length Max  & \multicolumn{1}{l}{Fwd IAT Mean} \\
		\multicolumn{1}{l}{Active Max} & Fwd Packet Length Mean & \multicolumn{1}{l}{Bwd IAT Mean} \\
		\multicolumn{1}{l}{Active Std} & Bwd Packet Length Mean & \multicolumn{1}{l}{Flow duration} \\
		\multicolumn{1}{l}{Fwd IAT Min} & Fwd Packet Length Std & \multicolumn{1}{l}{Flow IAT Mean} \\
		\multicolumn{1}{l}{Fwd IAT Max} & Bwd Packet Length Min & \multicolumn{1}{l}{Bwd IAT Total} \\
		\multicolumn{1}{l}{Fwd IAT Std} & Bwd Packet Length Max & \multicolumn{1}{l}{Fwd PSH flags} \\
		\multicolumn{1}{l}{Bwd IAT Min} & Bwd Packet Length Std & \multicolumn{1}{l}{Bwd PSH Flags} \\
		\multicolumn{1}{l}{Bwd IAT Max} & Fwd Segment Size Avg  & \multicolumn{1}{l}{Fwd URG Flags} \\
		\multicolumn{1}{l}{Bwd IAT Std} & Bwd Segment Size Avg  & \multicolumn{1}{l}{Bwd URG Flags} \\
		\multicolumn{1}{l}{Active Mean} & Average Packet Size  & \multicolumn{1}{l}{FWD Packets/s} \\
		\multicolumn{1}{l}{Fwd Header Length} & Bwd Packet/Bulk Avg  & \multicolumn{1}{l}{Bwd Packets/s} \\
		\multicolumn{1}{l}{Bwd Header Length} & Packet Length Mean  & \multicolumn{1}{l}{down/Up Ratio} \\
		\multicolumn{1}{l}{Packet Length Max} & Fwd Packet/Bulk Avg & \multicolumn{1}{l}{Flow Packets/s} \\
		\multicolumn{1}{l}{Packet Length Std} & Subflow Fwd Packets & \multicolumn{1}{l}{FIN Flag Count } \\
		\multicolumn{1}{l}{Bwd Bulk Rate Avg} & Subflow Bwd Packets & \multicolumn{1}{l}{SYN Flag Count } \\
		\multicolumn{1}{l}{Subflow Fwd Bytes} & Packet Length Min  & \multicolumn{1}{l}{RST Flag Count } \\
		\multicolumn{1}{l}{Subflow Bwd Bytes} & Fwd Bytes/Bulk Avg & \multicolumn{1}{l}{PSH Flag Count } \\
		\multicolumn{1}{l}{Fwd Act Data Pkts} & Fwd Bulk Rate Avg  & \multicolumn{1}{l}{ACK Flag Count } \\
		\multicolumn{1}{l}{total Fwd Packet} & Bwd Bytes/Bulk Avg & \multicolumn{1}{l}{URG Flag Count } \\
		\multicolumn{1}{l}{Fwd IAT Total   } & Fwd Init Win bytes & \multicolumn{1}{l}{CWR Flag Count } \\
		\multicolumn{1}{l}{Fwd Seg Size Min} & Bwd Init Win bytes & \multicolumn{1}{l}{ECE Flag Count } \\
		& total Bwd packets &  \\
		\bottomrule
	\end{tabular}%
	\label{tab:feature-cic}%
\end{table}%



\begin{table}[h!]
	\centering
	\caption{Final list of features obtained as a result of multi-step feature selection.}
	\begin{tabular}{lll}
		\toprule
		\multicolumn{3}{c}{CICFlowmeter Final  Feature List} \\
		\midrule
    ACK Flag Cnt & Flow IAT Mean & \multicolumn{1}{l}{Init Bwd Win Byts} \\
    Active Max & Flow IAT Min & \multicolumn{1}{l}{Pkt Len Max} \\
    Active Mean & Flow IAT Std & \multicolumn{1}{l}{Pkt Len Mean} \\
    Active Min & Fwd Header Len & \multicolumn{1}{l}{Pkt Len Min} \\
    Active Std & Fwd IAT Max & \multicolumn{1}{l}{Pkt Len Std} \\
    Bwd Header Len & Fwd IAT Mean & \multicolumn{1}{l}{Pkt Len Var} \\
    Bwd IAT Max & Fwd IAT Min & \multicolumn{1}{l}{Pkt Size Avg} \\
    Bwd IAT Mean & Fwd IAT Std & \multicolumn{1}{l}{Protocol} \\
    Bwd IAT Min & Fwd IAT Tot & \multicolumn{1}{l}{Src Port} \\
    Bwd IAT Std & Fwd Pkt Len Max & \multicolumn{1}{l}{Subflow Bwd Byts} \\
    Bwd IAT Tot & Fwd Pkt Len Mean & \multicolumn{1}{l}{Subflow Fwd Byts} \\
    Bwd Pkt Len Mean & Fwd Pkt Len Min & \multicolumn{1}{l}{Subflow Fwd Pkts} \\
    Bwd Pkt Len Min & Fwd Pkt Len Std & \multicolumn{1}{l}{Tot Bwd Pkts} \\
    Bwd Pkts/s & Fwd Pkts/s & \multicolumn{1}{l}{Tot Fwd Pkts} \\
    FIN Flag Cnt & Idle Max & \multicolumn{1}{l}{TotLen Bwd Pkts} \\
    Flow Byts/s & Idle Mean & \multicolumn{1}{l}{TotLen Fwd Pkts} \\
    Flow Duration & Idle Min &  \\
    Flow IAT Max & Idle Std &  \\
		\bottomrule
	\end{tabular}%
	\label{tab:cic-final-features}%
\end{table}%

\newpage
\section{Kitsune Feature Selection Process and Results}~\label{Kitsune}
\begin{figure}[h!]
	\centering{
		\includegraphics[width=1\columnwidth]{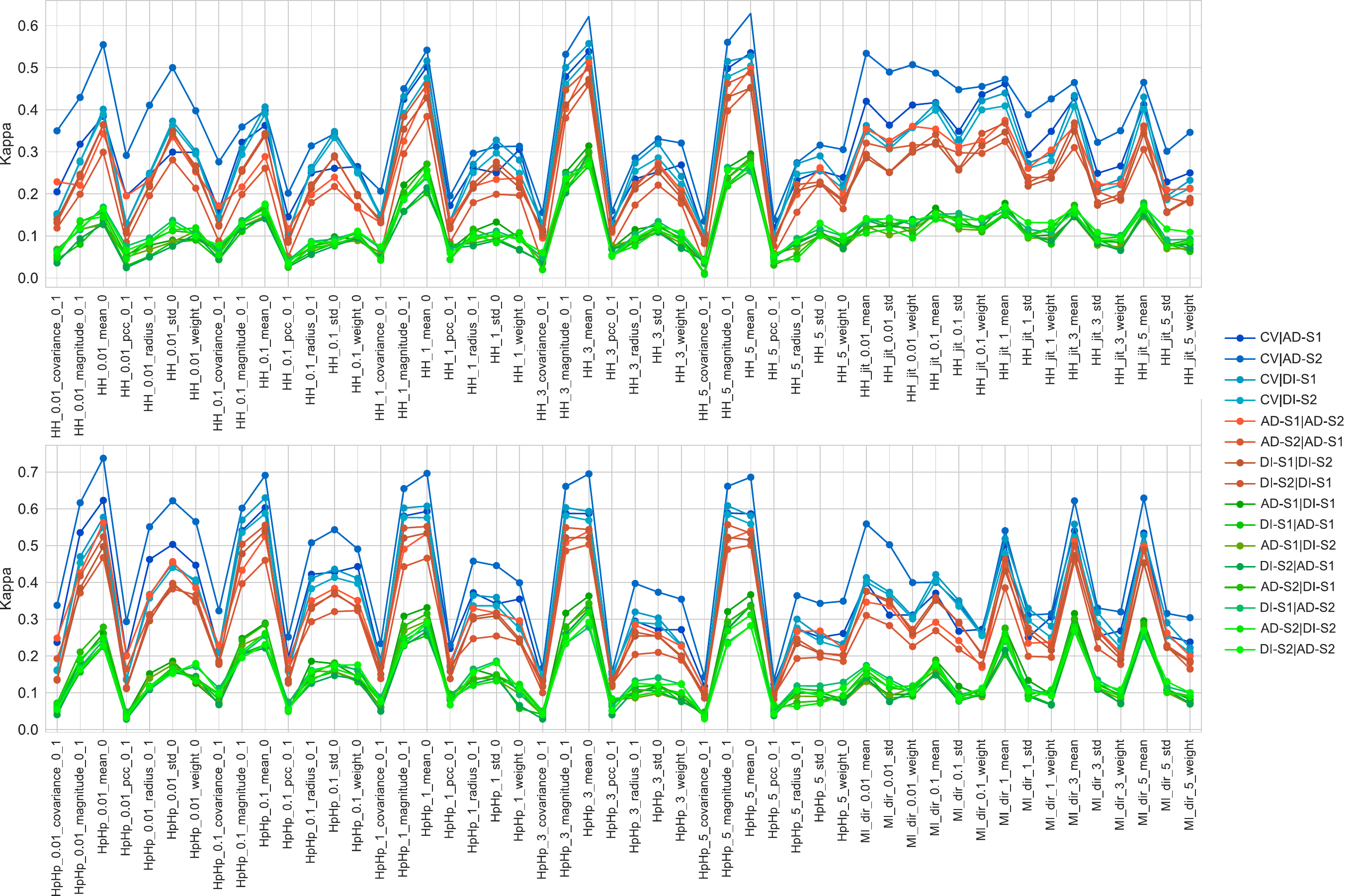}}
	\caption{Comparison of kappa scores for Kitsune features in UNSW datasets using CV (blue) and isolated methods, SS (red) and DD (green). The CV method tends to overestimate feature utility, showing higher scores for many attributes, while SS and DD produce more realistic evaluations.  This discrepancy highlights the potential for information leakage in cross-validation and the importance of using isolated validation methods for assessing feature utility in ML-based DI models.}
	\label{fig:cizgi-kit}
\end{figure}

\begin{figure}[h!]
	\centering{	
		\includegraphics[width=1\columnwidth]{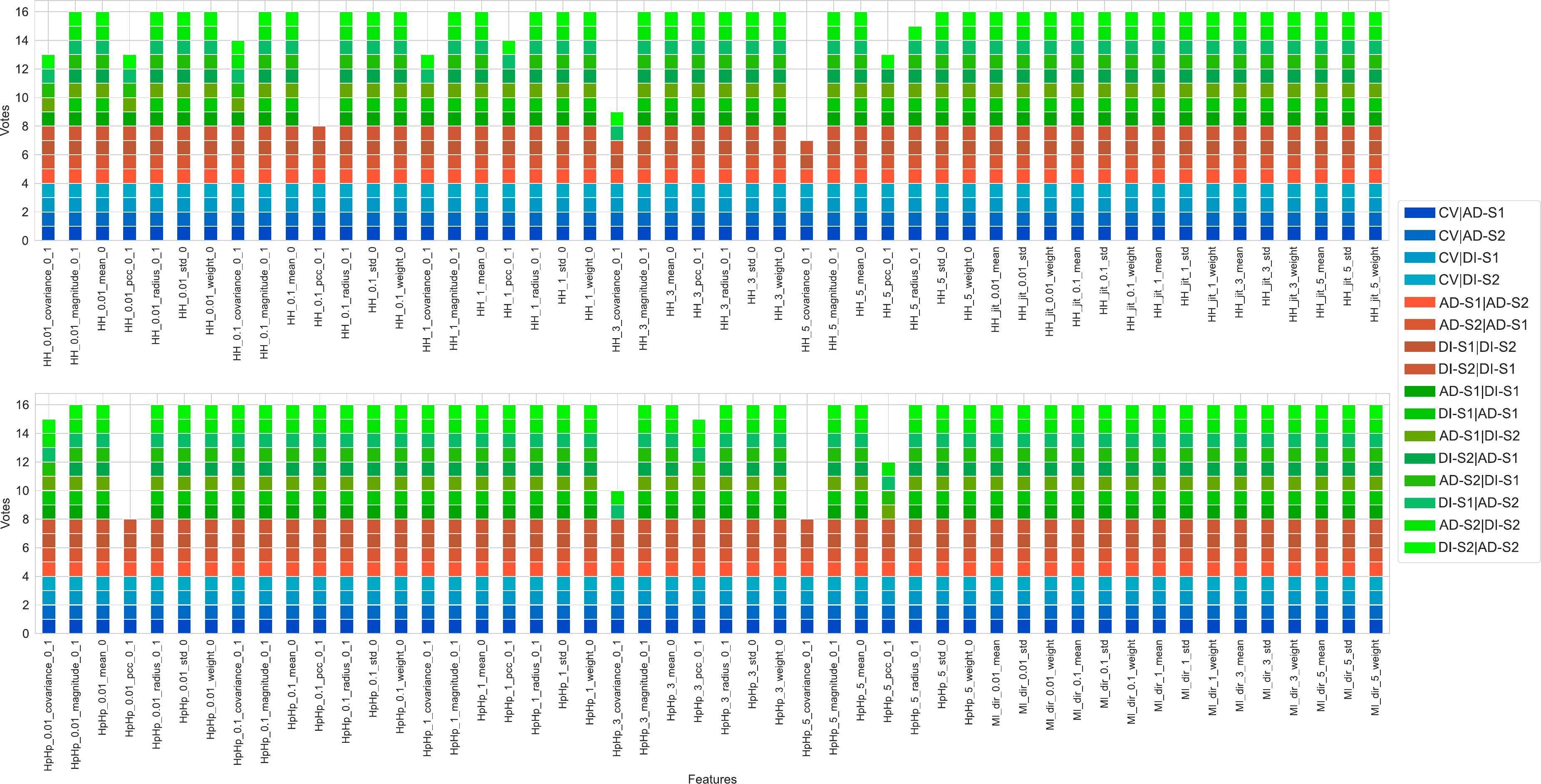}}
	\caption{A voting system for Kitsune features that gives one vote to each feature with a kappa value different from 0 that is individually predictive within CV (blue), SS (red) \& DD (green).}
	\label{fig:vote-kit}
\end{figure}

\begin{figure}[h!]
	\centering{
		\includegraphics[width=1\columnwidth]{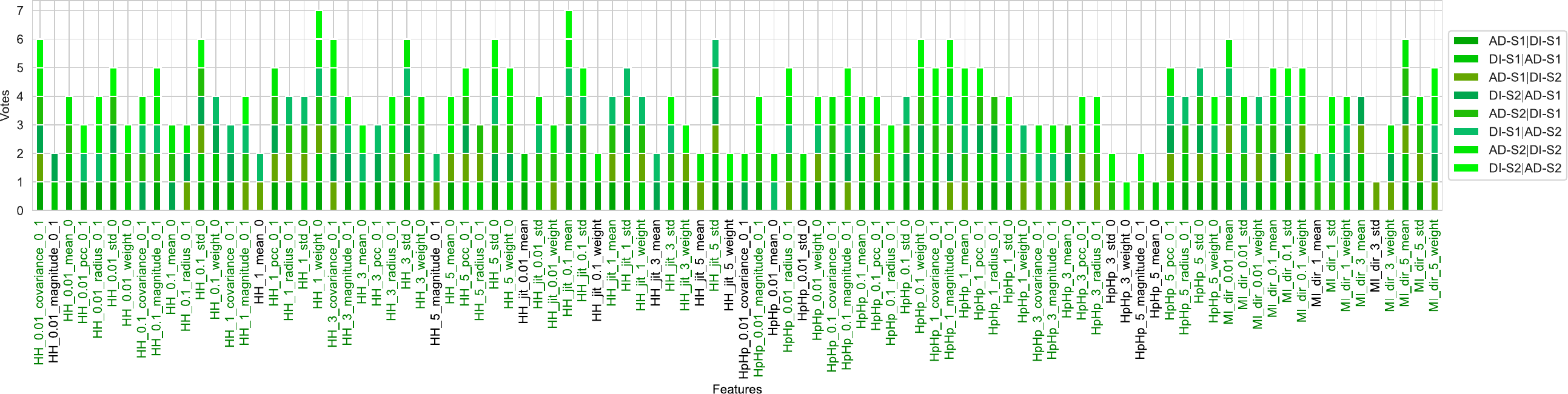}}
	\caption{ List of intersecting Kitsune features identified across eight dataset versus dataset (DD) cases using Genetic Algorithm (GA).}
	\label{fig:v-g-kit}
\end{figure}

\begin{figure}[h!]
	\centering{
		\includegraphics[width=0.8\columnwidth]{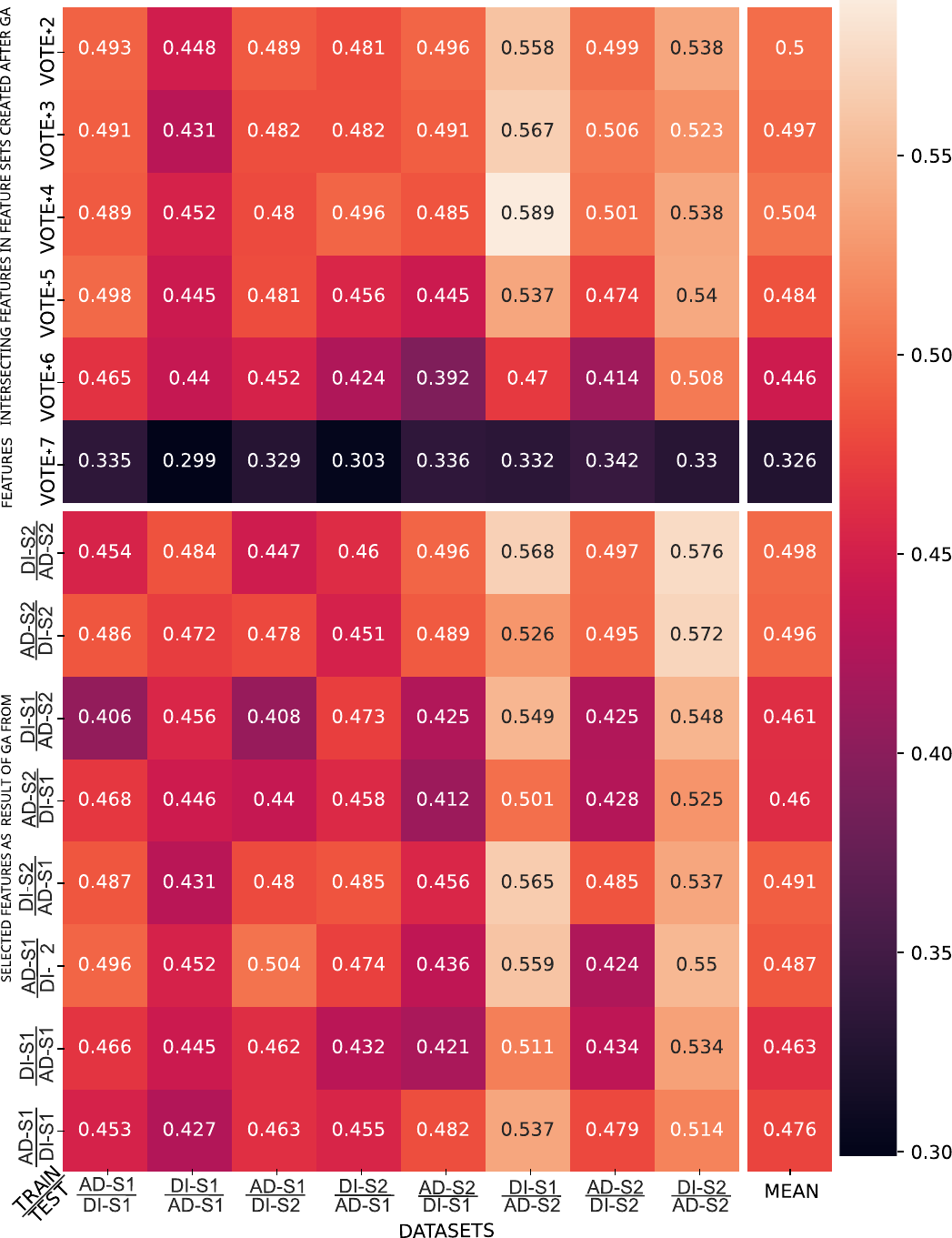}}
	\caption{Comparison of Kitsune feature set performance across dataset versus dataset (DD) cases. The left side of the heatmap displays the results of applying feature sets obtained from the first step of the Genetic Algorithm (GA) to all DD data, yielding 64 results. On the right, the performance of the features grouped according to their frequency, focusing on the intersection of features obtained from the output of the GA algorithm.}
	\label{fig:kit-f-cm}
\end{figure}

\begin{figure}[h!]
	\centering{
		\includegraphics[width=0.8\columnwidth]{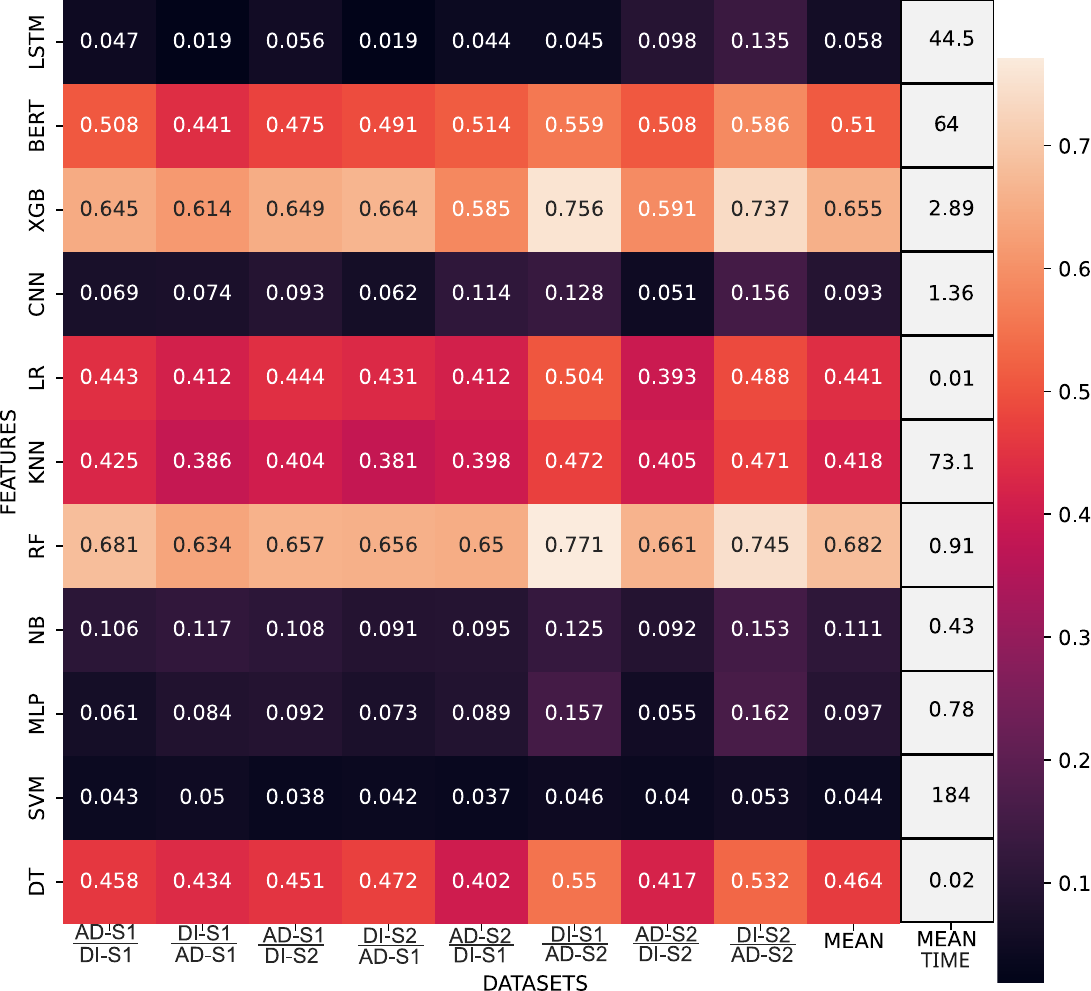}}
	\caption{F1 scores and average inference times of various ML algorithms applied to DD datasets with Kitsune features. The Random Forest (RF) and XGB methods demonstrate the highest F1 scores of 0.682 and 0.655, respectively, while Decision Trees (DT) show the fastest inference time.}
	\label{fig:kitcmML}
\end{figure}

In this section, we apply the same feature selection process to the Kitsune features as we did with GeMID. Table~\ref{tab:feature-kit} lists all features included in the Kitsune method. Figure~\ref{fig:cizgi-kit} compares feature utility in the UNSW datasets, using both cross-validation and isolated evaluation methods (SS and DD). The voting process for selecting the most suitable features, based on the utility of each feature, is illustrated in Figure~\ref{fig:vote-kit}. Candidate sub-features generated by applying a genetic algorithm to each dataset series are presented in Figure~\ref{fig:v-g-kit}. The testing results for these features and their intersections across all UNSW-DD datasets are shown in Figure~\ref{fig:kit-f-cm}. The best-performing feature set, as determined by these tests, is detailed in Table~\ref{tab:kit-final-features}. The results of testing the final features with various machine learning and neural network-based techniques to evaluate model performance are shown in Figure~\ref{fig:kitcmML}.

\begin{table}[h!]
	\centering
	\caption{Features obtained after applying the Kitsune tool to the pcap file.}
	\begin{tabular}{rlr}
		\toprule
		\multicolumn{3}{c}{Features obtained after applying the Kitsune tool to the pcap file.} \\
		\midrule
		\multicolumn{1}{l}{HH\_5\_std\_0} & HpHp\_0.01\_pcc\_0\_1' & \multicolumn{1}{l}{HpHp\_0.01\_mean\_0} \\
		\multicolumn{1}{l}{HH\_3\_std\_0} & HH\_5\_covariance\_0\_1 & \multicolumn{1}{l}{MI\_dir\_0.1\_weight} \\
		\multicolumn{1}{l}{HH\_1\_std\_0} & HH\_3\_covariance\_0\_1 & \multicolumn{1}{l}{HH\_0.1\_radius\_0\_1} \\
		\multicolumn{1}{l}{HH\_5\_mean\_0} & HH\_1\_covariance\_0\_1 & \multicolumn{1}{l}{HH\_jit\_0.1\_weight} \\
		\multicolumn{1}{l}{HH\_3\_mean\_0} & HpHp\_0.1\_radius\_0\_1 & \multicolumn{1}{l}{HpHp\_5\_radius\_0\_1} \\
		\multicolumn{1}{l}{HH\_1\_mean\_0} & HH\_0.1\_magnitude\_0\_1 & \multicolumn{1}{l}{HpHp\_3\_radius\_0\_1} \\
		\multicolumn{1}{l}{MI\_dir\_5\_std} & HpHp\_5\_magnitude\_0\_1 & \multicolumn{1}{l}{HpHp\_1\_radius\_0\_1} \\
		\multicolumn{1}{l}{MI\_dir\_3\_std} & HpHp\_3\_magnitude\_0\_1 & \multicolumn{1}{l}{HpHp\_0.1\_weight\_0} \\
		\multicolumn{1}{l}{MI\_dir\_1\_std} & HpHp\_1\_magnitude\_0\_1 & \multicolumn{1}{l}{MI\_dir\_0.01\_weight} \\
		\multicolumn{1}{l}{HH\_5\_pcc\_0\_1} & HpHp\_0.01\_radius\_0\_1 & \multicolumn{1}{l}{HH\_5\_magnitude\_0\_1} \\
		\multicolumn{1}{l}{HH\_3\_pcc\_0\_1} & HH\_0.1\_covariance\_0\_1 & \multicolumn{1}{l}{HH\_3\_magnitude\_0\_1} \\
		\multicolumn{1}{l}{HH\_1\_pcc\_0\_1} & HH\_0.01\_magnitude\_0\_1 & \multicolumn{1}{l}{HH\_1\_magnitude\_0\_1} \\
		\multicolumn{1}{l}{HH\_0.1\_std\_0} & HpHp\_5\_covariance\_0\_1 & \multicolumn{1}{l}{HH\_0.01\_radius\_0\_1} \\
		\multicolumn{1}{l}{HH\_jit\_5\_std} & HpHp\_3\_covariance\_0\_1 & \multicolumn{1}{l}{HH\_jit\_0.01\_weight} \\
		\multicolumn{1}{l}{HH\_jit\_3\_std} & HpHp\_1\_covariance\_0\_1 & \multicolumn{1}{l}{HpHp\_0.01\_weight\_0} \\
		\multicolumn{1}{l}{HH\_jit\_1\_std} & HH\_0.01\_covariance\_0\_1 & \multicolumn{1}{l}{HH\_0.01\_pcc\_0\_1} \\
		\multicolumn{1}{l}{HpHp\_5\_std\_0} & HpHp\_0.1\_magnitude\_0\_1 & \multicolumn{1}{l}{HH\_jit\_5\_weight} \\
		\multicolumn{1}{l}{HpHp\_3\_std\_0} & HpHp\_0.1\_covariance\_0\_1 & \multicolumn{1}{l}{HH\_jit\_3\_weight} \\
		\multicolumn{1}{l}{HpHp\_1\_std\_0} & HpHp\_0.01\_magnitude\_0\_1 & \multicolumn{1}{l}{HH\_jit\_1\_weight} \\
		\multicolumn{1}{l}{MI\_dir\_5\_mean} & HpHp\_0.01\_covariance\_0\_1 & \multicolumn{1}{l}{HH\_jit\_0.1\_mean} \\
		\multicolumn{1}{l}{MI\_dir\_3\_mean} & HpHp\_0.01\_std\_0 & \multicolumn{1}{l}{HH\_jit\_0.01\_std} \\
		\multicolumn{1}{l}{MI\_dir\_1\_mean} & MI\_dir\_0.01\_mean & \multicolumn{1}{l}{HpHp\_5\_weight\_0} \\
		\multicolumn{1}{l}{HH\_5\_weight\_0} & HH\_0.01\_weight\_0 & \multicolumn{1}{l}{HpHp\_3\_weight\_0} \\
		\multicolumn{1}{l}{HH\_3\_weight\_0} & HH\_jit\_0.01\_mean & \multicolumn{1}{l}{HpHp\_1\_weight\_0} \\
		\multicolumn{1}{l}{HH\_1\_weight\_0} & HpHp\_0.1\_pcc\_0\_1 & \multicolumn{1}{l}{HpHp\_0.1\_mean\_0} \\
		\multicolumn{1}{l}{HH\_0.1\_mean\_0} & MI\_dir\_0.1\_std & \multicolumn{1}{l}{MI\_dir\_5\_weight} \\
		\multicolumn{1}{l}{HH\_0.01\_std\_0} & HH\_0.1\_pcc\_0\_1 & \multicolumn{1}{l}{MI\_dir\_3\_weight} \\
		\multicolumn{1}{l}{HH\_jit\_5\_mean} & HH\_0.01\_mean\_0 & \multicolumn{1}{l}{MI\_dir\_1\_weight} \\
		\multicolumn{1}{l}{HH\_jit\_3\_mean} & HH\_jit\_0.1\_std & \multicolumn{1}{l}{MI\_dir\_0.1\_mean} \\
		\multicolumn{1}{l}{HH\_jit\_1\_mean} & HpHp\_5\_pcc\_0\_1 & \multicolumn{1}{l}{MI\_dir\_0.01\_std} \\
		\multicolumn{1}{l}{HpHp\_5\_mean\_0} & HpHp\_3\_pcc\_0\_1 & \multicolumn{1}{l}{HH\_5\_radius\_0\_1} \\
		\multicolumn{1}{l}{HpHp\_3\_mean\_0} & HpHp\_1\_pcc\_0\_1 & \multicolumn{1}{l}{HH\_3\_radius\_0\_1} \\
		\multicolumn{1}{l}{HpHp\_1\_mean\_0} & HpHp\_0.1\_std\_0 & \multicolumn{1}{l}{HH\_1\_radius\_0\_1} \\
		& HH\_0.1\_weight\_0 &  \\
		\bottomrule
	\end{tabular}%
	\label{tab:feature-kit}%
\end{table}%

\begin{table}[h!]
	\centering
	\caption{Final list of features obtained as a result of multi-step feature selection.}
	\begin{tabular}{lll}
		\toprule
		\multicolumn{3}{c}{Kitsune Final Feature List} \\
		\midrule
		    \multicolumn{1}{l}{HH\_0.01\_covariance\_0\_1} & HH\_5\_pcc\_0\_1 & \multicolumn{1}{l}{HpHp\_1\_pcc\_0\_1} \\
    \multicolumn{1}{l}{HH\_0.01\_mean\_0} & HH\_5\_radius\_0\_1 & \multicolumn{1}{l}{HpHp\_1\_radius\_0\_1} \\
    \multicolumn{1}{l}{HH\_0.01\_pcc\_0\_1} & HH\_5\_std\_0 & \multicolumn{1}{l}{HpHp\_1\_std\_0} \\
    \multicolumn{1}{l}{HH\_0.01\_radius\_0\_1} & HH\_5\_weight\_0 & \multicolumn{1}{l}{HpHp\_1\_weight\_0} \\
    \multicolumn{1}{l}{HH\_0.01\_std\_0} & HH\_jit\_0.01\_std & \multicolumn{1}{l}{HpHp\_3\_covariance\_0\_1} \\
    \multicolumn{1}{l}{HH\_0.01\_weight\_0} & HH\_jit\_0.01\_weight & \multicolumn{1}{l}{HpHp\_3\_magnitude\_0\_1} \\
    \multicolumn{1}{l}{HH\_0.1\_covariance\_0\_1} & HH\_jit\_0.1\_mean & \multicolumn{1}{l}{HpHp\_3\_mean\_0} \\
    \multicolumn{1}{l}{HH\_0.1\_magnitude\_0\_1} & HH\_jit\_0.1\_std & \multicolumn{1}{l}{HpHp\_3\_pcc\_0\_1} \\
    \multicolumn{1}{l}{HH\_0.1\_mean\_0} & HH\_jit\_1\_mean & \multicolumn{1}{l}{HpHp\_3\_radius\_0\_1} \\
    \multicolumn{1}{l}{HH\_0.1\_radius\_0\_1} & HH\_jit\_1\_std & \multicolumn{1}{l}{HpHp\_5\_pcc\_0\_1} \\
    \multicolumn{1}{l}{HH\_0.1\_std\_0} & HH\_jit\_1\_weight & \multicolumn{1}{l}{HpHp\_5\_radius\_0\_1} \\
    \multicolumn{1}{l}{HH\_0.1\_weight\_0} & HH\_jit\_3\_std & \multicolumn{1}{l}{HpHp\_5\_std\_0} \\
    \multicolumn{1}{l}{HH\_1\_covariance\_0\_1} & HH\_jit\_3\_weight & \multicolumn{1}{l}{HpHp\_5\_weight\_0} \\
    \multicolumn{1}{l}{HH\_1\_magnitude\_0\_1} & HH\_jit\_5\_std & \multicolumn{1}{l}{MI\_dir\_0.01\_mean} \\
    \multicolumn{1}{l}{HH\_1\_pcc\_0\_1} & HpHp\_0.01\_magnitude\_0\_1 & \multicolumn{1}{l}{MI\_dir\_0.01\_std} \\
    \multicolumn{1}{l}{HH\_1\_radius\_0\_1} & HpHp\_0.01\_radius\_0\_1 & \multicolumn{1}{l}{MI\_dir\_0.01\_weight} \\
    \multicolumn{1}{l}{HH\_1\_std\_0} & HpHp\_0.01\_weight\_0 & \multicolumn{1}{l}{MI\_dir\_0.1\_mean} \\
    \multicolumn{1}{l}{HH\_1\_weight\_0} & HpHp\_0.1\_covariance\_0\_1 & \multicolumn{1}{l}{MI\_dir\_0.1\_std} \\
    \multicolumn{1}{l}{HH\_3\_covariance\_0\_1} & HpHp\_0.1\_magnitude\_0\_1 & \multicolumn{1}{l}{MI\_dir\_0.1\_weight} \\
    \multicolumn{1}{l}{HH\_3\_magnitude\_0\_1} & HpHp\_0.1\_mean\_0 & \multicolumn{1}{l}{MI\_dir\_1\_std} \\
    \multicolumn{1}{l}{HH\_3\_mean\_0} & HpHp\_0.1\_pcc\_0\_1 & \multicolumn{1}{l}{MI\_dir\_1\_weight} \\
    \multicolumn{1}{l}{HH\_3\_pcc\_0\_1} & HpHp\_0.1\_radius\_0\_1 & \multicolumn{1}{l}{MI\_dir\_3\_mean} \\
    \multicolumn{1}{l}{HH\_3\_radius\_0\_1} & HpHp\_0.1\_std\_0 & \multicolumn{1}{l}{MI\_dir\_3\_weight} \\
    \multicolumn{1}{l}{HH\_3\_std\_0} & HpHp\_0.1\_weight\_0 & \multicolumn{1}{l}{MI\_dir\_5\_mean} \\
    \multicolumn{1}{l}{HH\_3\_weight\_0} & HpHp\_1\_covariance\_0\_1 & \multicolumn{1}{l}{MI\_dir\_5\_std} \\
    \multicolumn{1}{l}{HH\_5\_mean\_0} & HpHp\_1\_magnitude\_0\_1 & \multicolumn{1}{l}{MI\_dir\_5\_weight} \\
          & HpHp\_1\_mean\_0 &  \\
		\bottomrule
	\end{tabular}%

\label{tab:kit-final-features}%

\end{table}%

\newpage
\section{IoTDevID Results}
\begin{figure}[h!]
	\centering{
		\includegraphics[width=.80\columnwidth]{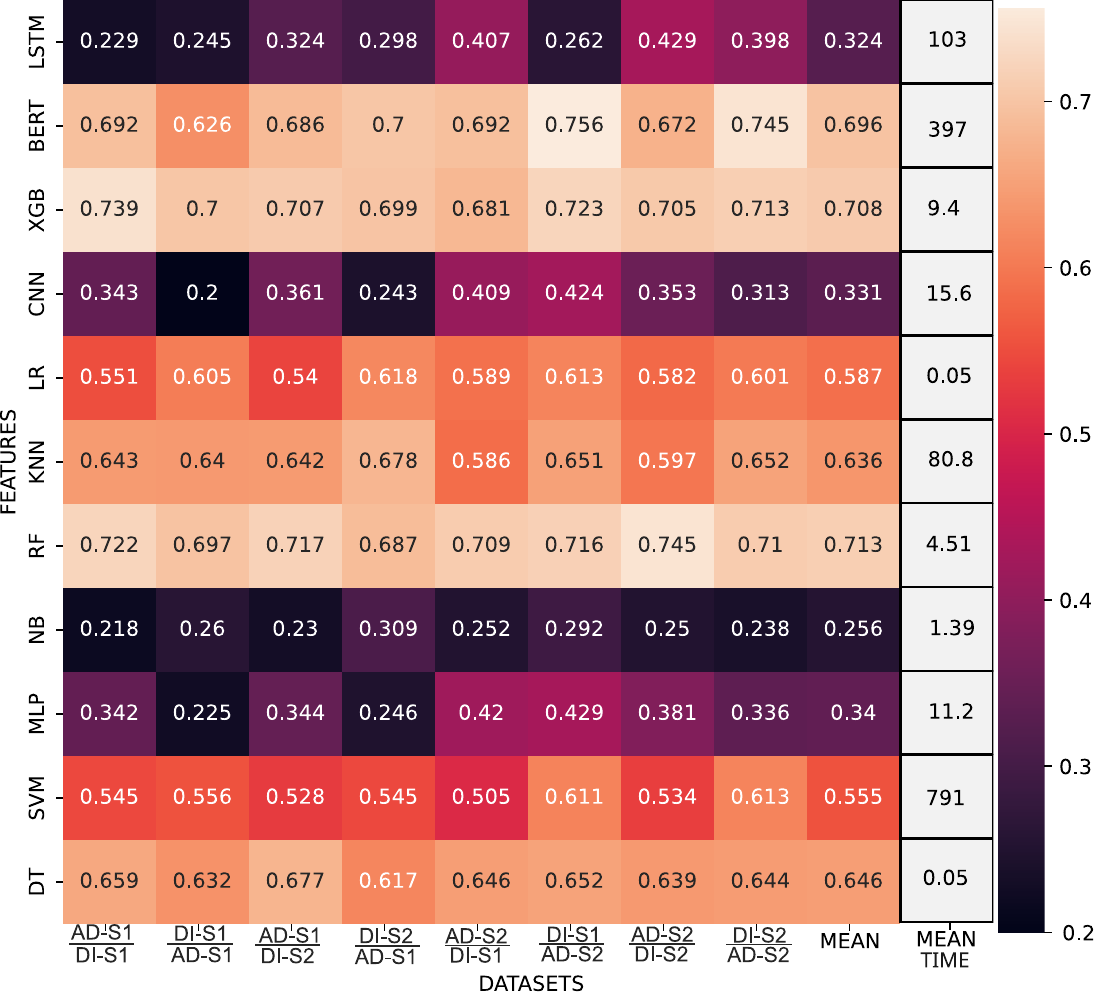}}
	\caption{F1 scores and average inference times of various ML algorithms applied to DD datasets with IoTDevID features. The Random Forest (RF) and XGB methods demonstrate the highest F1 scores of 0.713 and 0.708, respectively, while Decision Trees (DT) show the fastest inference time.}
	\label{fig:DevcmML}
\end{figure}

Since the original IoTDevID study included its own feature selection process, we did not perform an additional feature selection in this analysis. Instead, we used the features selected in the IoTDevID study for model selection. Therefore, unlike other methods, this section does not include the feature selection steps. The scores obtained during model selection are presented in Figure~\ref{fig:DevcmML}, along with the models and datasets used.


\section{GeMID Method Additional Tables and Figures}
The following tables and figures provide additional insights into the GeMID methodology and results, detailing feature selection, evaluation metrics, and model performance.

-----------------------------
\subsection{Final Evaluation Confusion Matrices}

This section presents the confusion matrices for the individual results shared in Table~\ref{tab:class} of the main paper.

\begin{figure}[htbp]
	\centering
	\subfloat[\label{fig:uk-us}UK$|$USA]{\includegraphics[width=80mm]{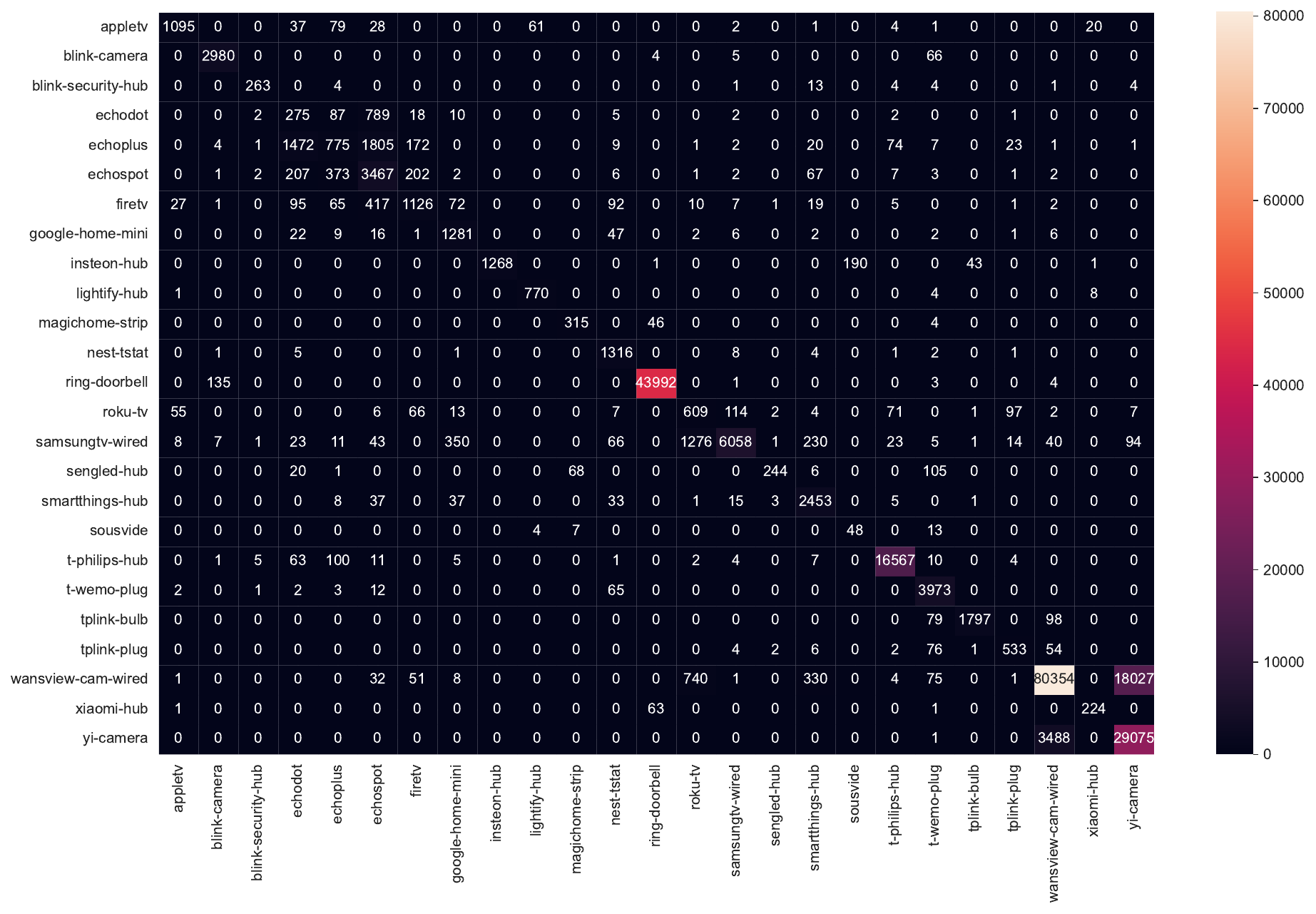}}
	\subfloat[\label{fig:uk-usv}UK$|$USA-VPN]{\includegraphics[width=80mm]{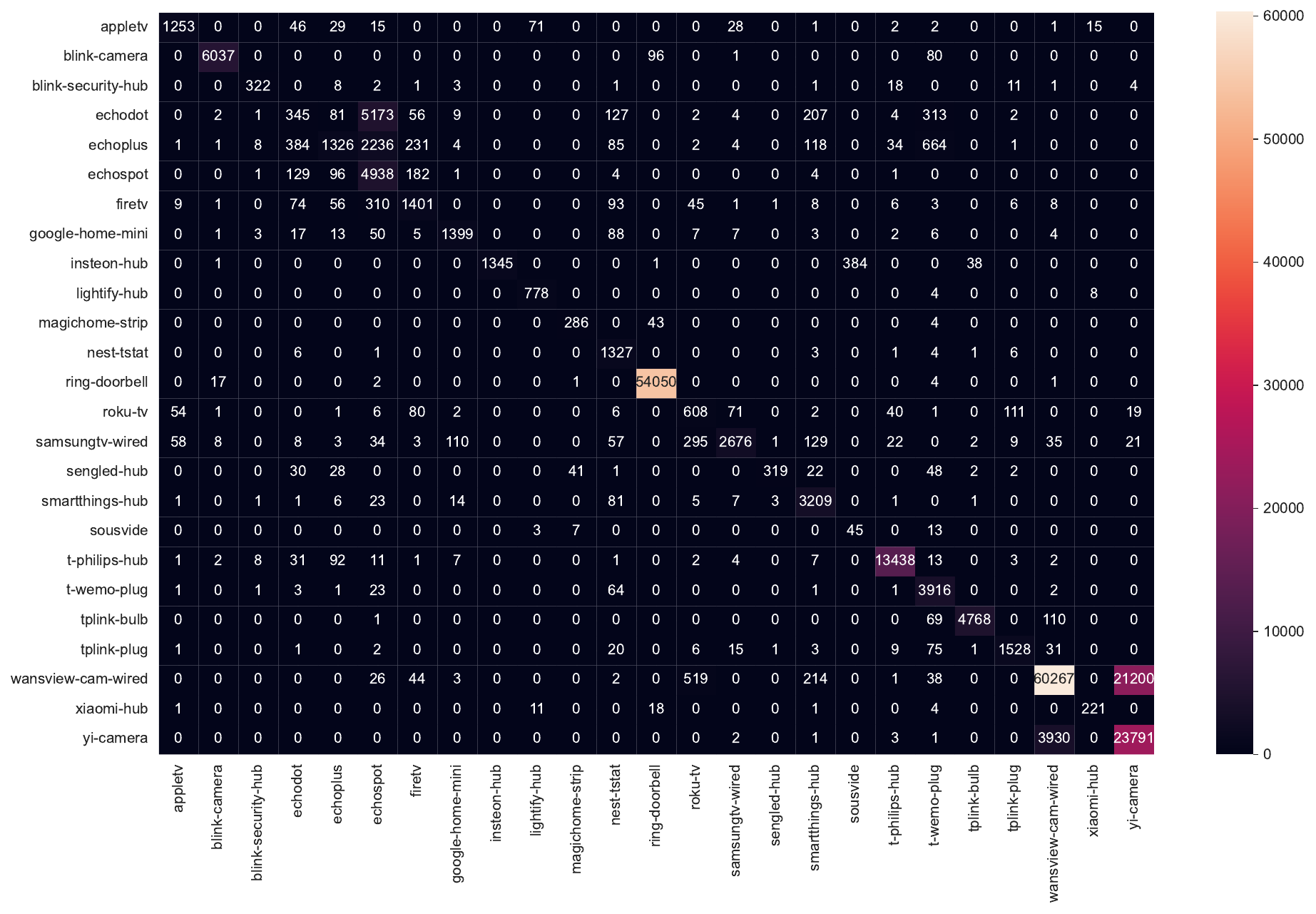}}\\
	\subfloat[\label{fig:ukv-us}UK-VPN$|$USA]{\includegraphics[width=80mm]{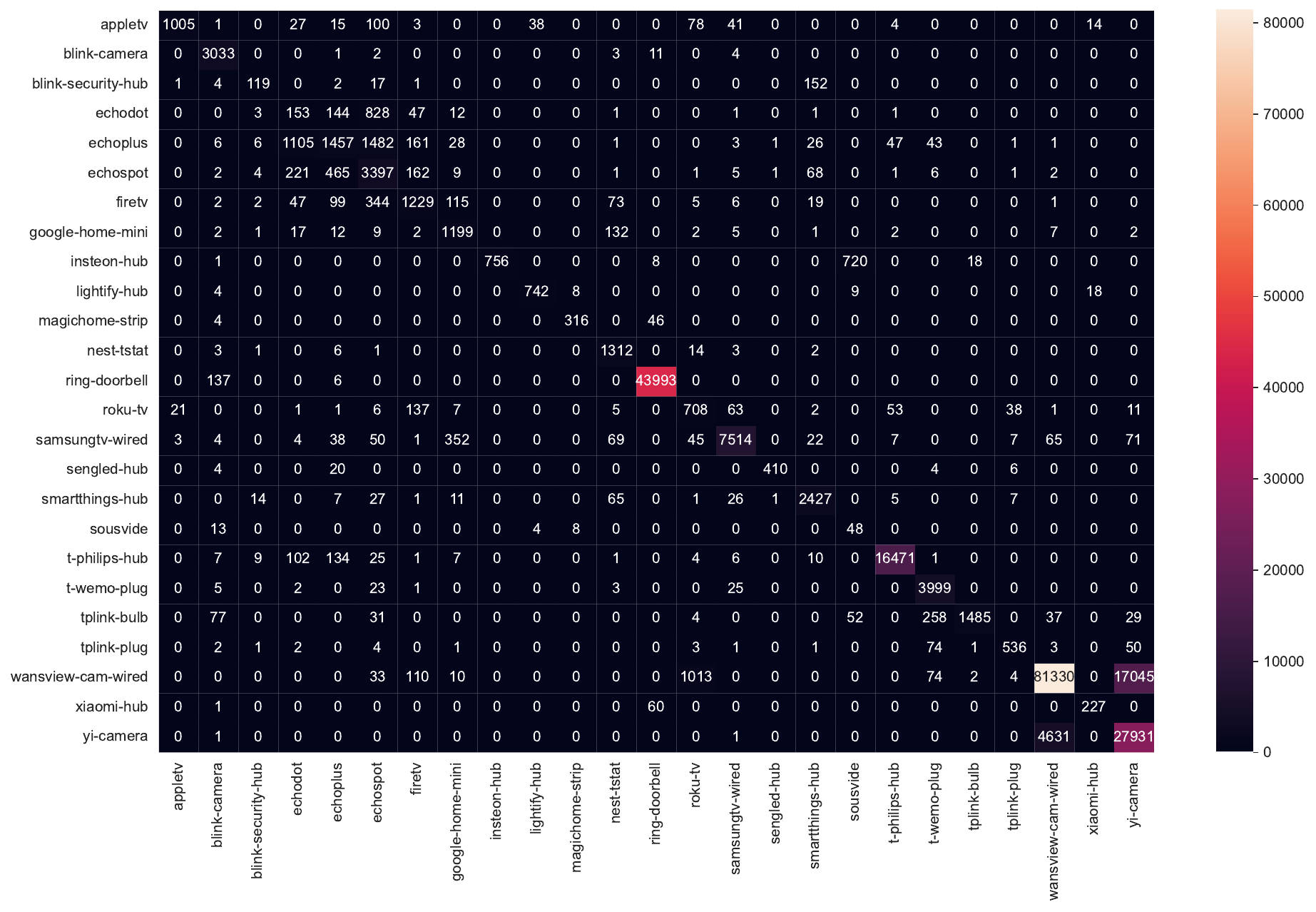}}
	\subfloat[\label{fig:ukv-usv}UK-VPN$|$USA-VPN]{\includegraphics[width=80mm]{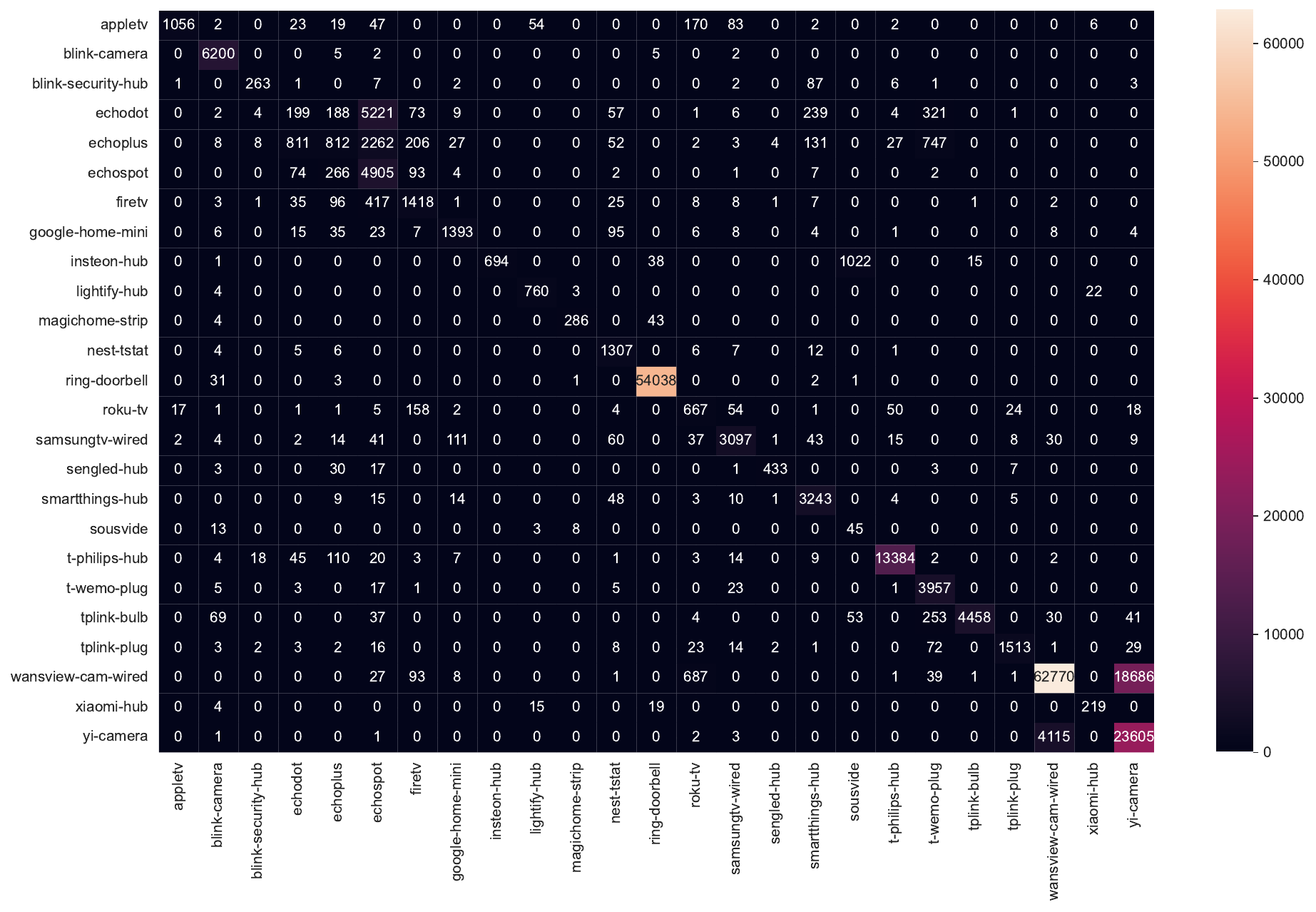}}\\
	\caption{Confusion matrices depict scenarios where distinct laboratories were employed in the MonIoTr dataset. UK data is utilized for training, while USA data is utilized for testing.}
	\label{fig:mon_data_uk2us}
\end{figure}

\begin{figure}[htbp]
	\centering
	\subfloat[\label{fig:us-uk}USA$|$UK]{\includegraphics[width=80mm]{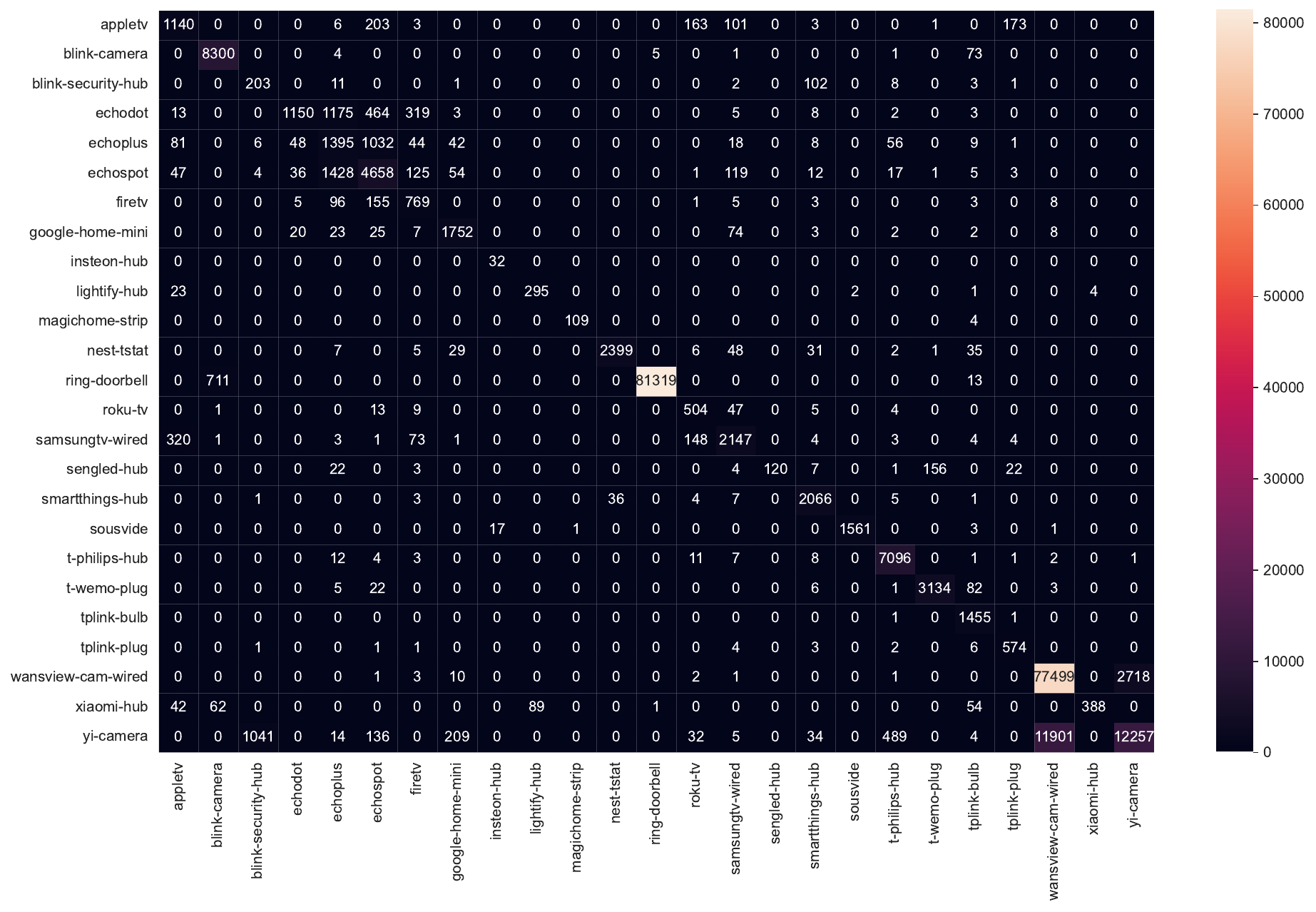}}
	\subfloat[\label{fig:us-ukv}USA$|$UK-VPN]{\includegraphics[width=80mm]{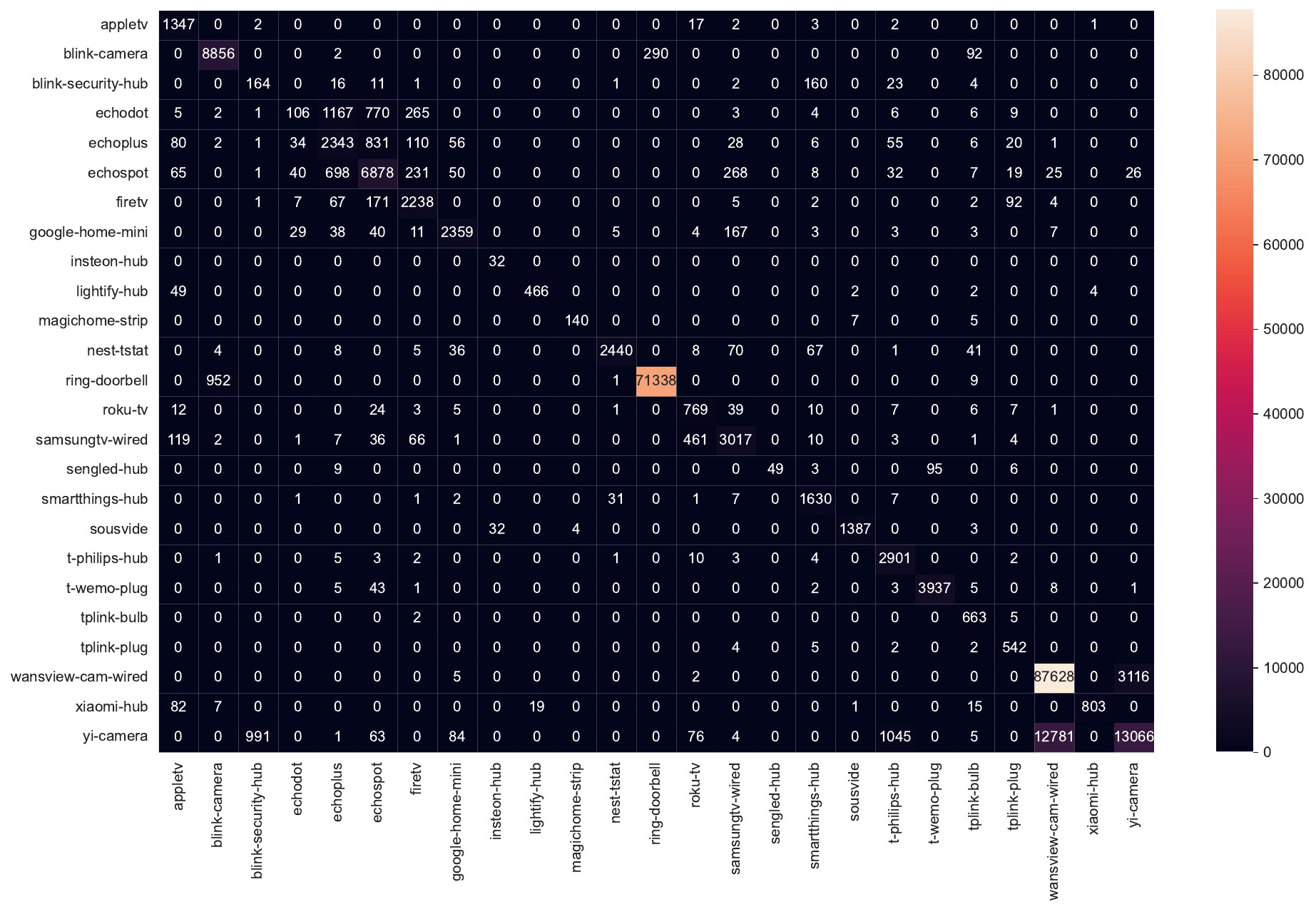}}\\
	\subfloat[\label{fig:usv-uk}USA-VPN$|$UK]{\includegraphics[width=80mm]{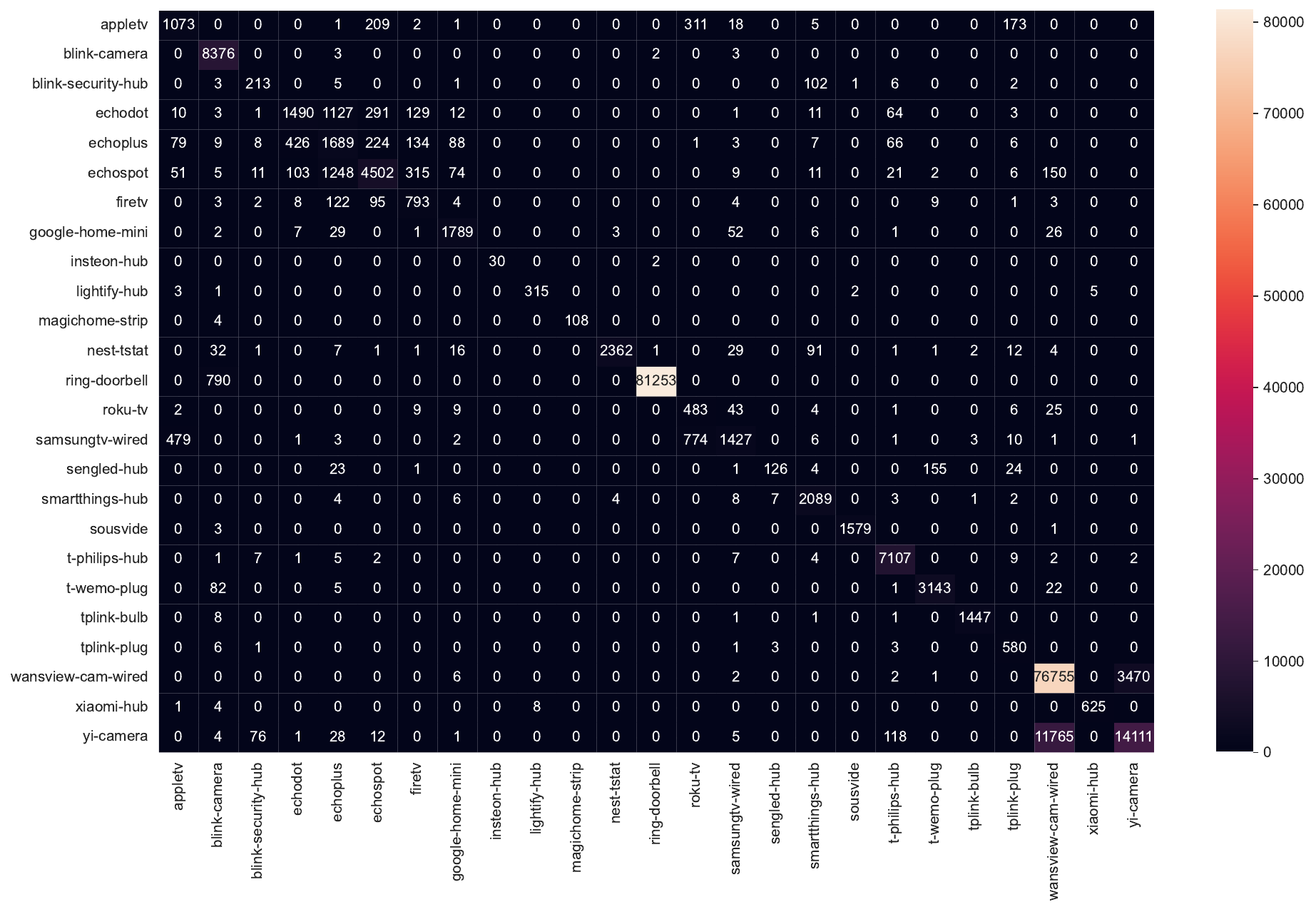}}
	\subfloat[\label{fig:usv-ukv}USA-VPN$|$UK-VPN]{\includegraphics[width=80mm]{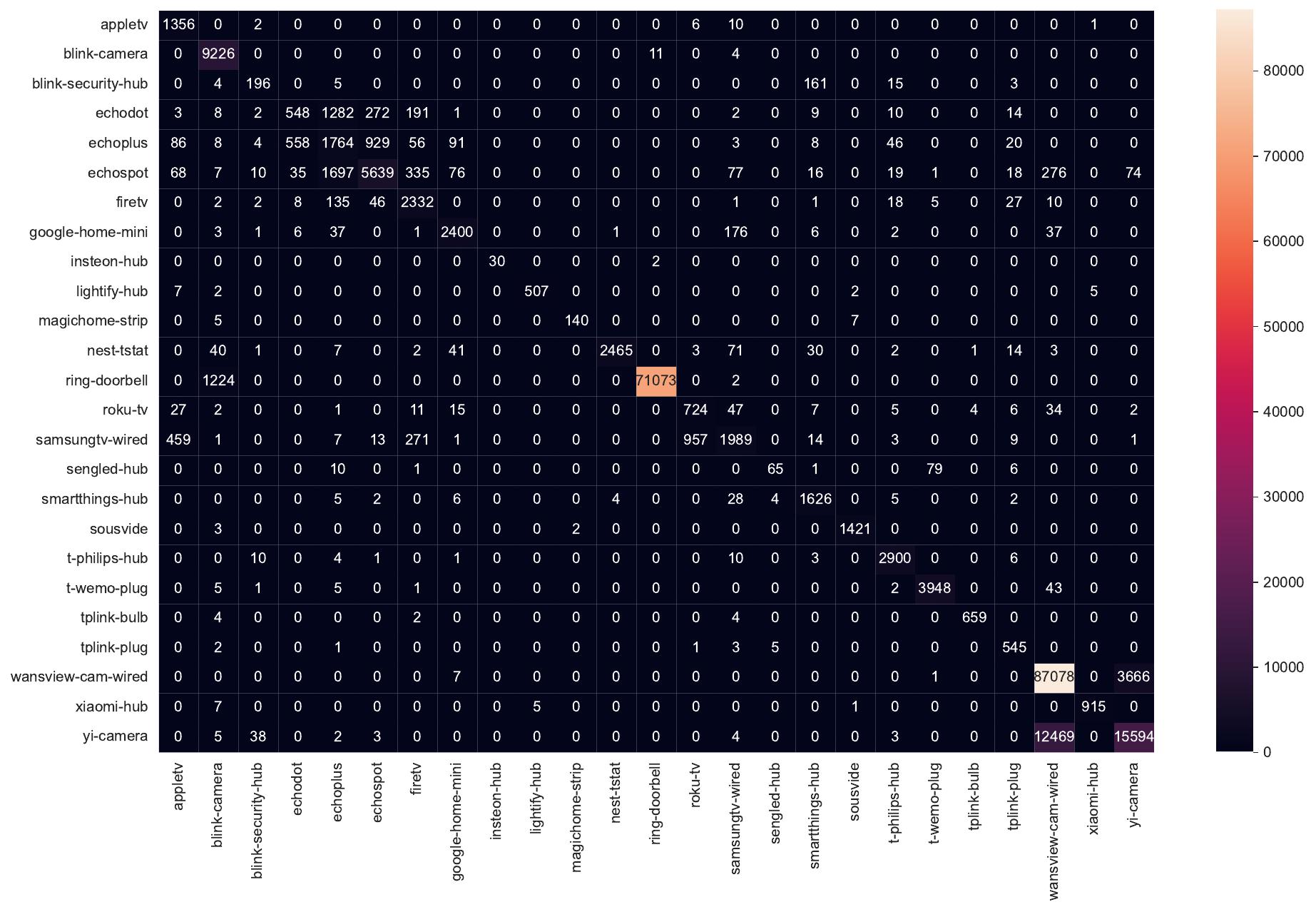}}\\
	\caption[Confusion matrices of MonIoTr dataset (USA vs UK case)]{Confusion matrices depict scenarios where distinct laboratories were employed in the MonIoTr dataset. USA data is utilized for training, while UK data is utilized for testing.}
	\label{fig:mon_data_us2uk}
\end{figure}

-----------------------------

\subsection{Other Supporting Tables}
\begin{table}
	\centering
	\caption{Devices common to UK and USA sites and their packet counts, with green indicating higher and red indicating lower packet numbers.}
	\begin{tabular}{lcccc}
		\toprule
		\multirow{2}[4]{*}{Device Names} & \multicolumn{4}{c}{Packet (Sample) Number} \\
		\cmidrule{2-5}          & USA    & USA-VPN & UK    & UK-VPN \\
		\midrule
		appletv & \cellcolor[rgb]{ .98,  .808,  .816}13319 & \cellcolor[rgb]{ .984,  .847,  .859}14685 & \cellcolor[rgb]{ .984,  .945,  .957}17977 & \cellcolor[rgb]{ .98,  .82,  .831}13775 \\
		blink-camera & \cellcolor[rgb]{ .984,  .988,  .996}30560 & \cellcolor[rgb]{ .965,  .98,  .98}62151 & \cellcolor[rgb]{ .949,  .973,  .969}83855 & \cellcolor[rgb]{ .945,  .973,  .965}92434 \\
		blink-security-hub & \cellcolor[rgb]{ .973,  .498,  .506}2983 & \cellcolor[rgb]{ .973,  .522,  .529}3771 & \cellcolor[rgb]{ .973,  .51,  .518}3352 & \cellcolor[rgb]{ .973,  .525,  .533}3863 \\
		echodot & \cellcolor[rgb]{ .98,  .765,  .776}11955 & \cellcolor[rgb]{ .965,  .98,  .98}63302 & \cellcolor[rgb]{ .984,  .988,  .996}31471 & \cellcolor[rgb]{ .988,  .988,  1}23483 \\
		echoplus & \cellcolor[rgb]{ .976,  .984,  .988}43742 & \cellcolor[rgb]{ .973,  .98,  .984}51061 & \cellcolor[rgb]{ .984,  .988,  .996}27440 & \cellcolor[rgb]{ .98,  .984,  .992}35799 \\
		echospot & \cellcolor[rgb]{ .976,  .984,  .988}43504 & \cellcolor[rgb]{ .969,  .98,  .984}53605 & \cellcolor[rgb]{ .961,  .98,  .976}65150 & \cellcolor[rgb]{ .949,  .973,  .969}83536 \\
		firetv & \cellcolor[rgb]{ .988,  .988,  1}19479 & \cellcolor[rgb]{ .988,  .988,  1}20274 & \cellcolor[rgb]{ .98,  .722,  .733}10487 & \cellcolor[rgb]{ .984,  .988,  1}25935 \\
		google-home-mini & \cellcolor[rgb]{ .98,  .827,  .835}13992 & \cellcolor[rgb]{ .984,  .89,  .902}16102 & \cellcolor[rgb]{ .984,  .98,  .992}19209 & \cellcolor[rgb]{ .984,  .988,  1}26735 \\
		insteon-hub & \cellcolor[rgb]{ .984,  .859,  .871}15067 & \cellcolor[rgb]{ .984,  .937,  .949}17724 & \cellcolor[rgb]{ .973,  .412,  .42}33 & \cellcolor[rgb]{ .973,  .412,  .42}33 \\
		lightify-hub & \cellcolor[rgb]{ .976,  .643,  .651}7837 & \cellcolor[rgb]{ .976,  .647,  .655}7916 & \cellcolor[rgb]{ .973,  .506,  .514}3274 & \cellcolor[rgb]{ .976,  .565,  .573}5245 \\
		magichome-strip & \cellcolor[rgb]{ .973,  .518,  .525}3667 & \cellcolor[rgb]{ .973,  .51,  .518}3336 & \cellcolor[rgb]{ .973,  .443,  .451}1138 & \cellcolor[rgb]{ .973,  .455,  .463}1521 \\
		nest-tstat & \cellcolor[rgb]{ .98,  .812,  .82}13438 & \cellcolor[rgb]{ .98,  .812,  .824}13508 & \cellcolor[rgb]{ .988,  .988,  1}25651 & \cellcolor[rgb]{ .984,  .988,  .996}26854 \\
		ring-doorbell & \cellcolor[rgb]{ .729,  .886,  .776}441388 & \cellcolor[rgb]{ .671,  .859,  .725}540796 & \cellcolor[rgb]{ .498,  .792,  .576}820444 & \cellcolor[rgb]{ .557,  .816,  .627}723017 \\
		roku-tv & \cellcolor[rgb]{ .98,  .725,  .737}10606 & \cellcolor[rgb]{ .98,  .71,  .722}10095 & \cellcolor[rgb]{ .976,  .584,  .592}5876 & \cellcolor[rgb]{ .976,  .675,  .682}8900 \\
		samsungtv-wired & \cellcolor[rgb]{ .953,  .973,  .969}82579 & \cellcolor[rgb]{ .98,  .988,  .992}34790 & \cellcolor[rgb]{ .984,  .988,  .996}27148 & \cellcolor[rgb]{ .98,  .984,  .992}37328 \\
		sengled-hub & \cellcolor[rgb]{ .973,  .541,  .549}4464 & \cellcolor[rgb]{ .976,  .557,  .565}4967 & \cellcolor[rgb]{ .973,  .51,  .518}3370 & \cellcolor[rgb]{ .973,  .459,  .467}1642 \\
		smartthings-hub & \cellcolor[rgb]{ .984,  .988,  1}25970 & \cellcolor[rgb]{ .98,  .988,  .996}33576 & \cellcolor[rgb]{ .988,  .988,  1}21271 & \cellcolor[rgb]{ .984,  .914,  .922}16856 \\
		sousvide & \cellcolor[rgb]{ .973,  .412,  .42}73 & \cellcolor[rgb]{ .973,  .412,  .42}69 & \cellcolor[rgb]{ .984,  .882,  .894}15842 & \cellcolor[rgb]{ .98,  .835,  .847}14271 \\
		t-philips-hub & \cellcolor[rgb]{ .898,  .953,  .922}167841 & \cellcolor[rgb]{ .918,  .961,  .941}136289 & \cellcolor[rgb]{ .957,  .976,  .973}71515 & \cellcolor[rgb]{ .984,  .988,  .996}29364 \\
		t-wemo-plug & \cellcolor[rgb]{ .976,  .984,  .992}40616 & \cellcolor[rgb]{ .976,  .984,  .992}40155 & \cellcolor[rgb]{ .98,  .988,  .996}32546 & \cellcolor[rgb]{ .976,  .984,  .992}40069 \\
		tplink-bulb & \cellcolor[rgb]{ .988,  .988,  1}19753 & \cellcolor[rgb]{ .973,  .984,  .984}49492 & \cellcolor[rgb]{ .984,  .843,  .855}14580 & \cellcolor[rgb]{ .976,  .608,  .62}6715 \\
		tplink-plug & \cellcolor[rgb]{ .976,  .612,  .62}6812 & \cellcolor[rgb]{ .984,  .914,  .925}16968 & \cellcolor[rgb]{ .976,  .588,  .596}5965 & \cellcolor[rgb]{ .976,  .576,  .584}5588 \\
		wansview-cam-wired & \cellcolor[rgb]{ .388,  .745,  .482}996276 & \cellcolor[rgb]{ .498,  .788,  .576}823173 & \cellcolor[rgb]{ .51,  .796,  .588}802384 & \cellcolor[rgb]{ .443,  .769,  .529}907544 \\
		yi-camera & \cellcolor[rgb]{ .804,  .914,  .839}325660 & \cellcolor[rgb]{ .831,  .925,  .867}277292 & \cellcolor[rgb]{ .843,  .929,  .875}261276 & \cellcolor[rgb]{ .827,  .925,  .863}281210 \\
		xiaomi-hub & \cellcolor[rgb]{ .973,  .494,  .502}2897 & \cellcolor[rgb]{ .973,  .486,  .494}2588 & \cellcolor[rgb]{ .976,  .6,  .608}6394 & \cellcolor[rgb]{ .976,  .686,  .698}9298 \\
		\midrule
		Average & 93779 & 91907 & 95106 & 96841 \\
		\bottomrule
	\end{tabular}%
	\label{tab:numberofpackets}%
\end{table}%



\begin{table}[htbp]
  \centering
  \caption{GeMID vs IoTDevID on the Aalto dataset with overall scores}
    \begin{tabular}{lllll}
    \toprule
          & \multicolumn{1}{c}{Accuracy} & \multicolumn{1}{c}{Precision} & \multicolumn{1}{c}{Recall} & \multicolumn{1}{c}{F1 Score} \\
    \midrule
    GEMID & 0.729±0.003 & 0.705±0.004 & 0.592±0.002 & 0.624±0.003 \\
    IoTDevID & 0.686+0.000 & 0.730+0.005 & 0.665±0.000 & 0.683±0.002 \\
    \bottomrule
    \end{tabular}%
  \label{tab:aalto}%
\end{table}%


\newpage
\subsection{Results from Aggregation Algorithm}

The aggregation algorithm proposed in the IoTDevID study is designed to enhance performance by consolidating the machine learning results of individual packets originating from the same source.

\begin{table}[htbp]
  \centering
  \caption{Comparison of DI methods on MonIoTr dataset with aggregation algorithm. F1 scores.}
    \begin{tabular}{clccc}
    \toprule
          & \textbf{Dataset} & \textbf{GeMID} & \textbf{IoTDevID} & \textbf{Kitsune} \\
    \midrule
    \multirow{5}[2]{*}{\begin{sideways}\textbf{CV}\end{sideways}} & UK    & 1.000 & 0.970 & 1.000 \\
          & UKVPN & 1.000 & 0.976 & 0.997 \\
          & USA    & 0.998 & 0.988 & 0.960 \\
          & USAVPN & 0.996 & 0.995 & 1.000 \\
          & \cellcolor[rgb]{ .851,  .851,  .851}Mean & \cellcolor[rgb]{ .851,  .851,  .851}\textbf{0.998} & \cellcolor[rgb]{ .851,  .851,  .851}0.982 & \cellcolor[rgb]{ .851,  .851,  .851}0.989 \\
    \midrule
    \multirow{5}[2]{*}{\begin{sideways}\textbf{SS}\end{sideways}} & UK|UKVPN & 0.962 & 0.943 & 0.894 \\
          & UKVPN|UK & 0.951 & 0.903 & 0.869 \\
          & USA|USAVPN & 0.925 & 0.911 & 0.896 \\
          & USAVPN|USA & 0.968 & 0.978 & 0.895 \\
          & \cellcolor[rgb]{ .851,  .851,  .851}Mean & \cellcolor[rgb]{ .851,  .851,  .851}\textbf{0.952} & \cellcolor[rgb]{ .851,  .851,  .851}0.934 & \cellcolor[rgb]{ .851,  .851,  .851}0.889 \\
    \midrule
    \multirow{9}[2]{*}{\begin{sideways}\textbf{DD}\end{sideways}} & UK|USA  & 0.894 & 0.851 & 0.638 \\
          & UK|USAVPN  & 0.885 & 0.815 & 0.536 \\
          & UKVPN|USA  & 0.882 & 0.804 & 0.618 \\
          & UKVPN|USAVPN  & 0.838 & 0.862 & 0.637 \\
          & USA|UK  & 0.916 & 0.940 & 0.723 \\
          & USA|UKVPN  & 0.889 & 0.856 & 0.605 \\
          & USAVPN|UK  & 0.931 & 0.880 & 0.616 \\
          & USAVPN|UKVPN  & 0.900 & 0.846 & 0.643 \\
          & \cellcolor[rgb]{ .851,  .851,  .851}Mean & \cellcolor[rgb]{ .851,  .851,  .851}\textbf{0.892} & \cellcolor[rgb]{ .851,  .851,  .851}0.857 & \cellcolor[rgb]{ .851,  .851,  .851}0.627 \\
    \bottomrule
    \end{tabular}%
  \label{tab:aggre1}%
\end{table}%

\begin{table}[htbp]
  \centering
  \caption{GeMDID F1 scores per device in all 8 DD cases for MonIoTr data with aggregation algorithm.}
    \resizebox{0.9\columnwidth}{!}{\begin{tabular}{llrrrrrrrr}
    \toprule
    \multirow{2}[2]{*}{Devices} & \multicolumn{1}{r}{Train→} & \multicolumn{1}{c}{UK} & \multicolumn{1}{c}{UK} & \multicolumn{1}{c}{UKVPN} & \multicolumn{1}{c}{UKVPN} & \multicolumn{1}{c}{USA} & \multicolumn{1}{c}{USA} & \multicolumn{1}{c}{USAVPN} & \multicolumn{1}{c}{USAVPN} \\
          & \multicolumn{1}{r}{Test→} & \multicolumn{1}{c}{USA} & \multicolumn{1}{c}{USAVPN} & \multicolumn{1}{c}{USA} & \multicolumn{1}{c}{USAVPN} & \multicolumn{1}{c}{UK} & \multicolumn{1}{c}{UKVPN} & \multicolumn{1}{c}{UK} & \multicolumn{1}{c}{UKVPN} \\
    \midrule
    \multicolumn{2}{l}{appletv} & 0.999 & 1.000 & 1.000 & 1.000 & 0.987 & 1.000 & 0.950 & 0.978 \\
    \multicolumn{2}{l}{blink-camera} & 1.000 & 1.000 & 1.000 & 1.000 & 1.000 & 1.000 & 1.000 & 1.000 \\
    \multicolumn{2}{l}{blink-security-hub} & 1.000 & 1.000 & 0.660 & 0.952 & 0.914 & 0.697 & 0.940 & 0.802 \\
    \multicolumn{2}{l}{echodot} & 0.025 & 0.000 & 0.000 & 0.000 & 0.621 & 0.001 & 0.788 & 0.140 \\
    \multicolumn{2}{l}{echoplus} & 0.141 & 0.320 & 0.559 & 0.116 & 0.587 & 0.785 & 0.814 & 0.675 \\
    \multicolumn{2}{l}{echospot} & 0.699 & 0.510 & 0.741 & 0.502 & 0.913 & 0.961 & 0.990 & 0.956 \\
    \multicolumn{2}{l}{firetv} & 0.969 & 0.992 & 0.977 & 0.992 & 0.979 & 0.995 & 1.000 & 0.996 \\
    \multicolumn{2}{l}{google-home-mini} & 0.999 & 1.000 & 0.997 & 1.000 & 1.000 & 1.000 & 1.000 & 1.000 \\
    \multicolumn{2}{l}{insteon-hub} & 0.986 & 0.839 & 0.725 & 0.273 & 1.000 & 1.000 & 1.000 & 1.000 \\
    \multicolumn{2}{l}{lightify-hub} & 1.000 & 1.000 & 0.999 & 1.000 & 0.965 & 1.000 & 1.000 & 1.000 \\
    \multicolumn{2}{l}{magichome-strip} & 0.969 & 0.969 & 0.999 & 1.000 & 1.000 & 1.000 & 1.000 & 1.000 \\
    \multicolumn{2}{l}{nest-tstat} & 0.999 & 1.000 & 1.000 & 0.999 & 1.000 & 1.000 & 1.000 & 1.000 \\
    \multicolumn{2}{l}{ring-doorbell} & 1.000 & 1.000 & 1.000 & 1.000 & 1.000 & 1.000 & 1.000 & 1.000 \\
    \multicolumn{2}{l}{roku-tv} & 0.949 & 0.984 & 1.000 & 0.997 & 0.988 & 0.993 & 0.744 & 0.787 \\
    \multicolumn{2}{l}{samsungtv-wired} & 0.994 & 0.997 & 1.000 & 1.000 & 0.998 & 0.998 & 0.912 & 0.917 \\
    \multicolumn{2}{l}{sengled-hub} & 0.768 & 0.869 & 1.000 & 1.000 & 0.450 & 0.261 & 0.491 & 0.561 \\
    \multicolumn{2}{l}{smartthings-hub} & 1.000 & 0.999 & 0.979 & 0.993 & 0.989 & 0.950 & 0.991 & 0.964 \\
    \multicolumn{2}{l}{sousvide} & 0.905 & 0.723 & 0.441 & 0.186 & 1.000 & 1.000 & 1.000 & 1.000 \\
    \multicolumn{2}{l}{t-philips-hub} & 1.000 & 1.000 & 1.000 & 1.000 & 1.000 & 1.000 & 1.000 & 1.000 \\
    \multicolumn{2}{l}{t-wemo-plug} & 0.987 & 0.981 & 0.999 & 0.975 & 0.966 & 0.983 & 0.967 & 0.988 \\
    \multicolumn{2}{l}{tplink-bulb} & 1.000 & 1.000 & 0.998 & 1.000 & 0.993 & 1.000 & 1.000 & 1.000 \\
    \multicolumn{2}{l}{tplink-plug} & 0.996 & 0.997 & 1.000 & 1.000 & 0.989 & 1.000 & 0.990 & 1.000 \\
    \multicolumn{2}{l}{wansview-cam-wired} & 0.998 & 0.983 & 0.998 & 0.991 & 0.928 & 0.930 & 0.942 & 0.950 \\
    \multicolumn{2}{l}{xiaomi-hub} & 0.972 & 1.000 & 0.972 & 1.000 & 0.955 & 1.000 & 1.000 & 1.000 \\
    \multicolumn{2}{l}{yi-camera} & 0.995 & 0.954 & 0.993 & 0.975 & 0.686 & 0.678 & 0.767 & 0.796 \\
    \midrule
    \multicolumn{2}{l}{accuracy} & 0.975 & 0.941 & 0.978 & 0.939 & 0.934 & 0.932 & 0.950 & 0.945 \\
    \multicolumn{2}{l}{macro avg} & 0.894 & 0.885 & 0.882 & 0.838 & 0.916 & 0.889 & 0.931 & 0.900 \\
    \bottomrule
    \end{tabular}}%
  \label{tab:aggre2}%
\end{table}%


\subsection{Hyperparameter Optimization}\label{Hyperparameter}

This section presents the hyperparameter ranges and the selected values for both classical machine learning (ML) and neural network-based models. Tables~\ref{tab:hyperparametersml} and~\ref{tab:hyperparametersdl} outline the hyperparameter tuning process, highlighting the optimal parameter settings identified during experimentation to achieve the best model performance.

\begin{table}[h!]
\centering
\renewcommand{\arraystretch}{1.2} 
\setlength{\tabcolsep}{5pt} 
\caption{Hyperparameter Selection - value ranges and selected parameters for classical ML algorithms}
\begin{tabular}{|l|l|p{2.5cm}|l|l|l|l|}
\hline
    \begin{turn}{90}\textbf{Method}    \end{turn} & \begin{turn}{90}\textbf{Parameter}\end{turn}       & \begin{turn}{90}\textbf{Value Range}\end{turn}    & \begin{turn}{90}\textbf{GEMID}\end{turn} & \begin{turn}{90}\textbf{CICFwMtr}\end{turn} & \begin{turn}{90}\textbf{IoTDevID}\end{turn} & \begin{turn}{90}\textbf{Kitsune}\end{turn} \\ \hline
\multirow{4}{*}{DT} & Criterion   & gini, entropy       & Entropy        & Entropy           & Entropy           & Entropy          \\ 
& Max Depth   & 1 - 32      & 14 & 18    & 19    & 16   \\ 
& Max Features& 1-Feature num.   & 21 & 6     & 24    & 33   \\ 
& Min Samples Split       & 2 - 10      & 5  & 6     & 5     & 5    \\ \hline
\multirow{5}{*}{RF} & Bootstrap   & True, False         & False          & False & False & False           \\
& Criterion   & gini, entropy       & Entropy        & Gini  & Gini  & Entropy          \\ 
& Max Depth   & 1 - 32      & 17 & 29    & 31    & 20   \\ 
& Max Features& 1 - 11      & 3  & 6     & 10    & 2    \\ 
& Min Samples Split       & 2 - 11      & 5  & 5     & 2     & 2    \\ 
& N Estimators& 1 - 200     & 71 & 88    & 136   & 79   \\ \hline
\multirow{4}{*}{XGB} & N Estimators           & 100, 500, 900, \newline 1100, 1500  & 900& 1100  & 1500  & 1100 \\ 
& Max Depth   & 2, 3, 5, 10, 15     & 3  & 3    & 3     & 2    \\ 
& Learning Rate           & 0.05, 0.1, 0.15, \newline 0.20       & 0.15           & 0.15  & 0.15  & 0.2  \\ 
& Min Child Weight        & 1, 2, 3, 4          & 1  & 1     & 1     & 1   \\ \hline
\multirow{4}{*}{KNN} & Algorithm  & auto, ball\_tree, \newline kd\_tree, brute       & kd\_tree       & ball\_tree        & ball\_tree        & ball\_tree       \\ 
& Leaf Size   & 1 - 50      & 39 & 44    & 35    & 44   \\ 
& N Neighbors & 1 - 64      & 5  & 4     & 13    & 2    \\ 
& Weights     & uniform, \newline distance   & Distance       & Distance          & Distance          & Distance         \\ \hline
NB      & Var Smoothing           & $10^0$ - $10^{-9}$      & $1.52 \times 10^{-6}$ & $8.11 \times 10^{-9}$ & $2.31 \times 10^{-6}$ & $8.11 \times 10^{-9}$ \\ \hline
\multirow{3}{*}{LR} & C   & $10^{-5}$ - 100 \newline (log-uniform)   & 0.0809264      & 0.10342           & 0.689136          & 0.071026         \\ 
& Penalty     & none, L1, L2, \newline elasticnet       & L2 & L1    & L1    & L1   \\ 
& Solver      & newton-cg, lbfgs, liblinear    & newton-cg      & liblinear         & liblinear         & liblinear        \\ \hline
\multirow{2}{*}{SVM} & Gamma      & 0.001, 0.01, 0.1, 1 & 0.001          & 0.001 & 1     & 0.001\\ 
& C   & 0.001, 0.01, 0.1, 1, 10     & 1  & 1     & 10    & 10   \\ \hline
\end{tabular}

\label{tab:hyperparametersml}
\end{table}

\begin{table}[h]
    \centering
        \caption{Hyperparameter Selection - value ranges and selected parameters for neural network neural network models}
    \begin{tabular}{|l|l|l|l|l|l|l|l|}
        \hline
        \begin{sideways}\textbf{Method}\end{sideways} & 
        \begin{sideways}\textbf{Parameter}\end{sideways} & 
        \begin{sideways}\textbf{Value Range}\end{sideways} & 
        \begin{sideways}\textbf{GEMID}\end{sideways} & 
        \begin{sideways}\textbf{CICFwMtr}\end{sideways} & 
        \begin{sideways}\textbf{IoTDevID}\end{sideways} & 
        \begin{sideways}\textbf{Kitsune}\end{sideways} \\         \hline
        CNN & Filters & [32, 64, 96, 128] & 64 & 32 & 32 & 32 \\
& Kernel Size & [3, 4, 5] & 3 & 3 & 3 & 3 \\
& Num Dense Layers & [1, 2, 3] & 1 & 1 & 1 & 1 \\
& Dense Units & [32, 64, 96, 128] & 32 & 128 & 96 & 32 \\
& Dropout & [0.0, 0.1, 0.2, 0.3, 0.4, 0.5] & 0.0 & 0.0 & 0.0 & 0.0 \\
& Learning Rate & [1e-2, 1e-3, 1e-4] & 0.01 & 0.01 & 0.01 & 0.01 \\
& Epochs & [10] & 10 & 10 & 10 & 10 \\
&Batch Size & [32] & 32 & 32 & 32 & 32 \\
        \hline
        LSTM & Units & [32, 64, 96, 128] & 64 & 128 & 32 & 32 \\
 & Dropout 1 & [0.0, 0.1, 0.2, 0.3, 0.4, 0.5] & 0.1 & 0.3 & 0.2 & 0.1 \\
 & Num Dense Layers& [1, 2, 3] & 1 & 2 & 1 & 1 \\
 & LSTM Units & [32, 64, 96, 128] & 64 & 96 & 96 & 128 \\
 & Dropout 2 & [0.0, 0.1, 0.2, 0.3, 0.4, 0.5] & 0.3 & 0.4 & 0.2 & 0.4 \\
 & Units Last & [32, 64, 96, 128] & 128 & 128 & 128 & 96 \\
 & Dropout Last & [0.0, 0.1, 0.2, 0.3, 0.4, 0.5] & 0.4 & 0.4 & 0.2 & 0.3 \\
 & Learning Rate & [1e-2, 1e-3, 1e-4] & 0.01 & 0.01 & 0.01 & 0.0001 \\
 & Epochs & [10] & 10 & 10 & 10 & 10 \\
 &Batch Size & [32] & 32 & 32 & 32 & 32 \\
        \hline
        ANN & Units Input & [32, 64, 96, 128] & 128 & 128 & 32 & 32 \\
& Num Dense Layers& [1, 2, 3] & 1 & 3 & 2 & 1 \\
& Units & [32, 64, 96, 128] & 96 & 128 & 128 & 32 \\
& Dropout & [0.0, 0.1, 0.2, 0.3, 0.4, 0.5] & 0.0 & 0.1 & 0.0 & 0.0 \\
& Learning Rate & [1e-2, 1e-3, 1e-4] & 0.01 & 0.01 & 0.01 & 0.001 \\
& Epochs & [10] & 10 & 10 & 10 & 10 \\
&Batch Size & [32] & 32 & 32 & 32 & 32 \\
& Validation Split & [0.3] & 0.3 & 0.3 & 0.3 & 0.3 \\
        \hline
        BERT & Batch Size & [16, 32, 64] & 16 & 32 & 16 & 16 \\
 & Learning Rate & [1e-5, 3e-5, 5e-5] & 1e-5 & 3e-5 & 1e-5 & 3e-5 \\
 & Epochs & [3, 5, 7] & 7 & 3 & 3 & 7 \\
 & Weight Decay & [0, 0.01, 0.1] & 0 & 0.01 & 0 & 0.01 \\
        \hline
    \end{tabular}

    \label{tab:hyperparametersdl}
\end{table}

\begin{table}[h!]
    \centering
    \caption{Specifications of the experimental platform}
    \begin{tabular}{l|l}
        \hline
        \textbf{Component} & \textbf{Details} \\ \hline
        Processor          & 12th Gen Intel(R) Core(TM) i7-12700H 2.30 GHz \\ 
        Installed RAM      & 16.0 GB (15.7 GB usable) \\ 
        System Type        & 64-bit operating system, x64-based processor \\ 
        Edition            & Windows 11 Home / Version 23H2 \\ \hline
    \end{tabular}
\end{table}

\begin{table}[h]
    \centering
    \caption{Description of non-core python libraries used in the project}
    \begin{tabular}{@{}ll@{}}
        \toprule
        \textbf{Library} & \textbf{Description} \\ \midrule
        numpy            & Fundamental package for numerical computing in Python. \\
        tqdm             & Library for adding progress bars to Python code. \\
        zipfile          & Module to read and write ZIP files (standard library). \\
        scapy            & Library for packet manipulation and network analysis. \\
        pandas           & Data manipulation and analysis tool for structured data. \\
        pickle           & Module for serializing and de-serializing Python objects. \\
        scikit-learn     & Machine learning library for Python. \\
        tabulate         & Library for creating nicely formatted tables in Python. \\
        pyximport        & Module for importing Python modules written in C/C++. \\
        seaborn          & Statistical data visualization library based on Matplotlib. \\
        transformers     & Library for natural language processing with pre-trained models. \\
        torch            & Deep learning framework providing GPU acceleration. \\
        keras\_tuner     & Library for hyperparameter tuning in Keras models. \\
        tensorflow       & End-to-end open-source platform for machine learning. \\ \bottomrule
    \end{tabular}
\end{table}
\newpage
\clearpage

\section{Adapting Neural Network-Based Learning Models for IoT Device Identification}

This section provides an in-depth explanation of how specific machine learning models—Convolutional Neural Networks (CNN), Long Short-Term Memory (LSTM), and Bidirectional Encoder Representations from Transformers (BERT)—were adapted to classify IoT devices. While these models are commonly applied to distinct data types (e.g., image data for CNNs, sequential data for LSTMs, and natural language processing for BERT), their adaptability to IoT device data required tailored preprocessing and structuring. Here, we detail the approach taken for each model, including preprocessing techniques and model architecture adjustments, to 
fit them to the IoT device identification task.

\subsection{Input Representation in BERT-based Model}

\subsubsection{Tabular Data and Feature Engineering} 
Our input features consist of numerical and categorical data (e.g., port numbers, IP flags, TCP header lengths), commonly used in IoT traffic analysis. Unlike BERT’s traditional text-tokenized inputs, these features require a tailored embedding approach.

\subsubsection{Customized Input Handling} 
To adapt these tabular features for BERT, we input them directly into the model through a fully connected layer (\texttt{fc1}, implemented as \texttt{torch.nn.Linear}). This layer projects feature vectors from the original input space into a 768-dimensional embedding space, aligning with the BERT model's expected hidden layer size.

\begin{verbatim}
x = torch.relu(self.fc1(x)) 
x = self.transformer(inputs_embeds=x.unsqueeze(1)).last_hidden_state 
\end{verbatim}

\subsubsection{Utilizing BERT as a Transformer Backbone} 
We utilize \texttt{inputs\_embeds} instead of tokenized input IDs, allowing embedded numerical features to pass directly into BERT.

\begin{verbatim} 
x = self.transformer(inputs_embeds=x.unsqueeze(1)).last_hidden_state 
\end{verbatim}

Typically, BERT processes sequences of token embeddings. By passing our feature embeddings via \texttt{inputs\_embeds}, we adapt tabular data as sequences of length one. Although unconventional, each row in the IoT dataset becomes a distinct input sequence.

\subsubsection{Pooling and Prediction} 
After processing the features through BERT, we apply mean pooling to aggregate hidden states across the input sequence. The pooled output is then passed through a classification layer (\texttt{fc2}) to generate predictions.

\subsection{Contrasting Standard and Custom BERT Workflows} 
\textbf{Standard BERT workflow:} 
\begin{verbatim}
Input text → Tokenization → Token Embeddings → Transformer → Classification Layer
\end{verbatim}

\noindent\textbf{Custom Workflow in This Study:} 
\begin{verbatim}
Tabular Data → Linear Embedding → Transformer → Pooling → Classification Layer
\end{verbatim}

\subsection{Input Representation in CNN-based Model}

While Convolutional Neural Networks (CNNs) are typically used for image classification, they can also be applied to other data formats, such as time-series or tabular data. In this study, a \textbf{1D CNN} was used for classifying IoT devices based on network traffic features.

Tabular IoT data usually comprises numerical features (e.g., packet length, header flags). Although CNNs are generally applied to spatial data (like images), they are also capable of capturing \textbf{local patterns} in sequential or structured data, making them useful for:
\begin{itemize}
    \item \textbf{Time-series analysis}, where the sequence of inputs matters.
    \item \textbf{Feature correlations} in structured datasets, like network traffic logs.
\end{itemize}

\subsubsection{Input Shape for 1D CNN}
Each row in the dataset corresponds to one instance (e.g., an intercepted traffic session) with multiple features. Let:
\begin{itemize}
    \item $n$: Number of instances (rows).
    \item $m$: Number of features (columns).
\end{itemize}

The input shape for a 1D CNN is transformed to: 
\[
(n, m, 1)
\]
where 1 is the \textbf{channel dimension}, analogous to the RGB channels in image data but here signifying a \textbf{single feature channel}.

\subsubsection{Steps to Prepare Data for CNN}
Before feeding the data into the model, the following steps are performed:
\begin{enumerate}
    \item \textbf{Data Scaling:} A Min-Max scaler is applied to normalize the feature values between 0 and 1:
    \[
    X_{\text{scaled}} = \frac{X - X_{\text{min}}}{X_{\text{max}} - X_{\text{min}}}
    \]
    \item \textbf{Reshaping:} The scaled data is reshaped to:
    \[
    X_{\text{train}} \in \mathbb{R}^{n_{\text{train}} \times m \times 1}, \quad X_{\text{test}} \in \mathbb{R}^{n_{\text{test}} \times m \times 1}
    \]
\end{enumerate}

\subsubsection{CNN Architecture for IoT Device Classification}
The CNN model is defined as follows:

\begin{verbatim}
model = Sequential()
model.add(Conv1D(filters=32, kernel_size=3, activation='relu', 
     input_shape=(m, 1)))
model.add(MaxPooling1D(pool_size=2))
model.add(Flatten())
model.add(Dense(32, activation='relu'))
model.add(Dropout(0.0))
model.add(Dense(96, activation='relu'))
model.add(Dropout(0.0))
model.add(Dense(21, activation='softmax'))
\end{verbatim}

\begin{itemize}
    \item \textbf{Conv1D Layer:} Applies convolution over the feature dimension (of size $m$) to extract \textbf{local patterns} between adjacent features.
    \item \textbf{MaxPooling1D Layer:} Reduces the dimensionality of the output and helps prevent overfitting.
    \item \textbf{Flatten Layer:} Converts the 2D tensor output of the previous layer into a 1D vector for further processing.
    \item \textbf{Dense Layers:} Fully connected layers for learning non-linear combinations of the extracted features.
    \item \textbf{Softmax Layer:} The final dense layer with 21 units corresponds to the 21 device classes, and the \texttt{softmax} activation outputs class probabilities.
\end{itemize}

\subsubsection{Feeding the Input to the CNN}
The \texttt{input\_shape} argument of the first \texttt{Conv1D} layer is defined as:
\[
\texttt{input\_shape} = (m, 1)
\]
This indicates that each input sample consists of $m$ features arranged along the feature axis, with 1 channel per feature.

During training, batches of data with shape:
\[
\text{batch\_size} \times m \times 1
\]
are fed into the CNN. For example, if the batch size is 32, the input will have the shape:
\[
(32, m, 1)
\]

\subsection{Input Representation in LSTM-based Model}

This appendix details the key steps and design choices made to adapt a Long Short-Term Memory (LSTM) neural network model for IoT device identification. While LSTMs are traditionally used in text and sequential data analysis, their ability to capture temporal dependencies makes them a viable option for analyzing IoT traffic patterns and device identification.

\subsubsection{Data Preparation} 
The data preprocessing workflow involves several key steps to ensure compatibility with the LSTM model’s input requirements: 
\begin{itemize} 
    \item \textbf{Data Loading:} Device feature data and corresponding labels are imported from CSV files, enabling a structured format for the model input. 
    \item \textbf{Feature Normalization:} The \texttt{MinMaxScaler} from the scikit-learn library is applied to scale feature values to a standardized range, which improves convergence and stability during training. 
    \item \textbf{Input Reshaping:} To align with the LSTM’s expected 3D input shape (samples, time\_steps, features), each 1D feature vector is reshaped with time\_steps = 1, accommodating the sequential nature of LSTM processing. 
\end{itemize}

\subsubsection{LSTM Model Architecture} 
The architecture design leverages the LSTM’s capabilities for temporal feature extraction and sequence learning: 
\begin{itemize} 
    \item \textbf{LSTM Layers:} The model comprises multiple LSTM layers with varying units (32, 64, 96), structured to progressively extract higher-level temporal features. Each layer uses \texttt{return\_sequences=True}, enabling the passage of sequential outputs across layers. 
    \item \textbf{Dropout Regularization:} Dropout layers interspersed between LSTM layers help reduce overfitting, improving model generalization on unseen data. 
    \item \textbf{Output Layer:} A Dense output layer with 21 units and a softmax activation function produces a probability distribution over the device classes, effectively handling the multi-class classification problem. 
\end{itemize}

------

\section{Genetic Algorithm Implementation Explanation} \label{ga-exp}

\subsection{Algorithm Overview}
The genetic algorithm (GA) implemented for feature selection follows a standard evolutionary approach adapted for binary feature selection problems. The algorithm evolves a population of binary chromosomes, where each chromosome represents a subset of features from the original feature space.

\subsection{Population Initialization}
The initial population is created using a controlled random approach:
\begin{itemize}
    \item Each chromosome is initialized as a boolean array of length equal to the number of features
    \item 30\% of features are initially set to False (not selected)
    \item 70\% of features are set to True (selected)
    \item The boolean array is randomly shuffled to ensure random distribution of selected features
\end{itemize}

This initialization strategy ensures a reasonable starting point while maintaining diversity in the initial population.

\subsection{Fitness Evaluation Process}
The fitness evaluation follows these steps:
\begin{enumerate}
    \item For each chromosome in the population:
    \begin{itemize}
        \item Extract features corresponding to True values in the chromosome
        \item Train a Decision Tree Classifier on the selected features using training data
        \item Make predictions on the test set using the same feature subset
        \item Calculate macro F1-score as the fitness value
    \end{itemize}
    \item Sort population by fitness scores in descending order
    \item Return sorted fitness scores and corresponding chromosomes
\end{enumerate}

\subsection{Selection Strategy}
The algorithm employs elitist selection:
\begin{itemize}
    \item Top 120 chromosomes (60\% of population) are selected as parents
    \item This ensures that high-quality solutions are preserved across generations
    \item The selection pressure is moderate, maintaining population diversity
\end{itemize}

\subsection{Crossover Implementation}
A fixed-point crossover is implemented:
\begin{itemize}
    \item Each parent undergoes crossover with the next parent in the sorted list.
    \item Genes at positions 3-6 (indices 3, 4, 5, 6) are exchanged between parents.
    \item Each parent produces one offspring, doubling the population from 120 to 240 (parents plus offspring).
    \item This creates offspring that combine characteristics of high-performing parents.
\end{itemize}

\subsection{Mutation Process}
Bit-flip mutation is applied:
\begin{itemize}
    \item Each gene in each chromosome has a 5\% probability of mutation
    \item Mutation flips the boolean value (True - False)
    \item This maintains genetic diversity and prevents premature convergence
\end{itemize}

\subsection{Algorithmic Limitations and Considerations}
Several aspects of the implementation could be enhanced for better reproducibility and performance:

\begin{itemize}
    \item \textbf{Fixed crossover points}: The crossover always occurs at positions 3-7, which may not be optimal for all feature sets
    \item \textbf{No convergence criteria}: The algorithm runs for exactly 25 generations regardless of fitness improvement
    \item \textbf{Limited diversity maintenance}: No explicit mechanisms to prevent population homogenization
    \item \textbf{Fitness evaluation on test set}: Using test data for fitness evaluation may lead to overfitting
\end{itemize}

\subsection{Reproducibility Parameters}
For complete reproducibility, the following parameters should be documented:
\begin{itemize}
    \item Random seed for population initialization and mutation
    \item Decision Tree Classifier hyperparameters (max\_depth, criterion, etc.)
    \item Cross-validation strategy for fitness evaluation
    \item Hardware specifications and execution environment
\end{itemize}

\begin{table}[htbp]
\centering
\caption{Genetic Algorithm Implementation Details for Feature Selection}
\label{tab:ga_implementation}
\begin{tabular}{p{4cm}p{3cm}p{6cm}}
\toprule
\textbf{Parameter/Component} & \textbf{Value/Method} & \textbf{Description} \\
\midrule
\multicolumn{3}{l}{\textit{Population Parameters}} \\
Population Size & 200 & Number of chromosomes in each generation \\
Number of Generations & 25 & Maximum iterations for evolution process \\
Number of Parents & 120 & Top-performing chromosomes selected for reproduction \\
\midrule
\multicolumn{3}{l}{\textit{Genetic Operators}} \\
Mutation Rate & 0.05 (5\%) & Probability of bit-flip mutation per gene \\
Selection Method & Elitist Selection & Top 60\% of population based on fitness scores \\
Crossover Method & Fixed-point & Genes 3-6 exchanged between consecutive parents \\
Crossover Rate & 100\% & All selected parents produce offspring, doubling population to 240 before next generation \\

\midrule
\multicolumn{3}{l}{\textit{Chromosome Representation}} \\
Encoding & Binary & Boolean array where True = feature selected \\
Initial Feature Ratio & 70\% & 30\% of features initially set to False \\
Chromosome Length & Variable & Equal to total number of input features \\
\midrule
\multicolumn{3}{l}{\textit{Fitness Evaluation}} \\
Fitness Function & Macro F1-score & \begin{tabular}[t]{@{}p{5.5cm}@{}}
Model trained on selected features, \\
evaluated on test set using \\
Decision Tree Classifier
\end{tabular} \\
Model & Decision Tree & sklearn.DecisionTreeClassifier \\
Performance Metric & F1-score (macro) & Handles class imbalance effectively \\
\midrule
\multicolumn{3}{l}{\textit{Stopping Criteria}} \\
Primary Criterion & Fixed Generations & Algorithm terminates after 25 generations \\
Convergence Check & None implemented & No early stopping based on fitness plateau \\
\midrule
\multicolumn{3}{l}{\textit{Output and Selection}} \\
Best Chromosome & Final generation & Highest fitness score in last generation \\
Feature Subset & Binary mask & Applied to original feature set \\
\bottomrule
\end{tabular}
\end{table}

\begin{table}[htbp]
  \centering
  \caption{Detailed Generalization Gap ($\Delta F{1}$) Analysis in the DD Evaluation Context. This table presents a detailed breakdown of the generalization gap, calculated as 	$\Delta F1 = F1_{source} - F1_{target}$, for each method on a per-device basis across all eight DD test cases. These results provide a granular view of model performance degradation and highlight the superior consistency and lower generalization gap of the GeMID method across a wide range of devices.}
    \begin{tabular}{@{}clrrrr@{}}
    \toprule
    \multicolumn{1}{l}{$\Delta F1$} & Dataset & \multicolumn{1}{c}{GeMID} & \multicolumn{1}{c}{CICFlwM} & \multicolumn{1}{c}{IoTDevID} & \multicolumn{1}{c}{Kitsune} \\
    \midrule
    \multirow{6}[4]{*}{\begin{sideways}CV→SS Gap\end{sideways}} & UK→UK|UKVPN & 0.065 & 0.213 & 0.076 & 0.190 \\
          & UKVPN→UKVPN|UK & 0.088 & 0.298 & 0.106 & 0.214 \\
          & USA→USA|USAVPN & 0.104 & 0.298 & 0.089 & 0.191 \\
          & USAVPN→USAVPN|USA & 0.060 & 0.288 & 0.084 & 0.257 \\
\cmidrule{2-6}          & \cellcolor[rgb]{ .851,  .851,  .851}Mean & \cellcolor[rgb]{ .851,  .851,  .851}0.079 & \cellcolor[rgb]{ .851,  .851,  .851}0.274 & \cellcolor[rgb]{ .851,  .851,  .851}0.089 & \cellcolor[rgb]{ .851,  .851,  .851}0.213 \\
          & \cellcolor[rgb]{ .851,  .851,  .851}Std Dev. & \cellcolor[rgb]{ .851,  .851,  .851}0.021 & \cellcolor[rgb]{ .851,  .851,  .851}0.041 & \cellcolor[rgb]{ .851,  .851,  .851}0.013 & \cellcolor[rgb]{ .851,  .851,  .851}0.031 \\
    \midrule
    \multirow{10}[4]{*}{\begin{sideways}CV→DD Gap\end{sideways}} & UK→UK|USA & 0.200 & 0.382 & 0.197 & 0.467 \\
          & UK→UK|USAVPN & 0.191 & 0.367 & 0.234 & 0.496 \\
          & UKVPN→UKVPN|USA & 0.217 & 0.337 & 0.254 & 0.532 \\
          & UKVPN→UKVPN|USAVPN & 0.219 & 0.399 & 0.199 & 0.480 \\
          & USA→USA|UK & 0.187 & 0.450 & 0.158 & 0.362 \\
          & USA→USA|UKVPN & 0.185 & 0.387 & 0.216 & 0.448 \\
          & USAVPN→USAVPN|UK & 0.143 & 0.389 & 0.213 & 0.505 \\
          & USAVPN→USAVPN|UKVPN & 0.138 & 0.440 & 0.192 & 0.423 \\
\cmidrule{2-6}          & \cellcolor[rgb]{ .851,  .851,  .851}Mean & \cellcolor[rgb]{ .851,  .851,  .851}0.185 & \cellcolor[rgb]{ .851,  .851,  .851}0.394 & \cellcolor[rgb]{ .851,  .851,  .851}0.208 & \cellcolor[rgb]{ .851,  .851,  .851}0.464 \\
          & \cellcolor[rgb]{ .851,  .851,  .851}Std Dev. & \cellcolor[rgb]{ .851,  .851,  .851}0.030 & \cellcolor[rgb]{ .851,  .851,  .851}0.037 & \cellcolor[rgb]{ .851,  .851,  .851}0.029 & \cellcolor[rgb]{ .851,  .851,  .851}0.053 \\
    \midrule
    \multirow{10}[4]{*}{\begin{sideways}SS→DD Gap\end{sideways}} & UK|UKVPN→UK|USA & 0.135 & 0.169 & 0.121 & 0.277 \\
          & UK|UKVPN→UK|USAVPN & 0.126 & 0.154 & 0.158 & 0.306 \\
          & UKVPN|UK→UKVPN|USA & 0.129 & 0.039 & 0.148 & 0.318 \\
          & UKVPN|UK→UKVPN|USAVPN & 0.131 & 0.101 & 0.093 & 0.266 \\
          & USA|USAVPN→USA|UK & 0.083 & 0.152 & 0.069 & 0.171 \\
          & USA|USAVPN→USA|UKVPN & 0.081 & 0.089 & 0.127 & 0.257 \\
          & USAVPN|USA→USAVPN|UK & 0.083 & 0.101 & 0.129 & 0.248 \\
          & USAVPN|USA→USAVPN|UKVPN & 0.078 & 0.152 & 0.108 & 0.166 \\
\cmidrule{2-6}          & \cellcolor[rgb]{ .851,  .851,  .851}Mean & \cellcolor[rgb]{ .851,  .851,  .851}0.106 & \cellcolor[rgb]{ .851,  .851,  .851}0.120 & \cellcolor[rgb]{ .851,  .851,  .851}0.119 & \cellcolor[rgb]{ .851,  .851,  .851}0.251 \\
          & \cellcolor[rgb]{ .851,  .851,  .851}Std Dev. & \cellcolor[rgb]{ .851,  .851,  .851}0.026 & \cellcolor[rgb]{ .851,  .851,  .851}0.044 & \cellcolor[rgb]{ .851,  .851,  .851}0.029 & \cellcolor[rgb]{ .851,  .851,  .851}0.056 \\
    \bottomrule
    \end{tabular}%
  \label{tab:deltaf1}%
\end{table}%

\end{document}